\documentclass[12pt,twoside,a4paper]{report}

\usepackage[utf8]{inputenc}
\usepackage[T1]{fontenc}
\usepackage{lmodern}

\usepackage[a4paper,width=150mm,top=25mm,bottom=25mm, bindingoffset=6mm]{geometry}

\usepackage[ngerman,KeepShorthandsActive,british]{babel}
\usepackage{graphicx}

\usepackage{eurosym}
\DeclareTextCommandDefault{\texteuro}{\euro}
\DeclareUnicodeCharacter{20AC}{\euro}

\defineshorthand{"`}{„}
\defineshorthand{"'}{“}

\usepackage{fancyhdr}
\usepackage[hyphens]{url} 
\usepackage{listings}

\usepackage{graphicx}
\graphicspath{{figures/}}

\usepackage[export]{adjustbox}
\usepackage{subfig}
\usepackage{color}
\usepackage[table]{xcolor}
\usepackage{multirow}

\usepackage{caption} 
\usepackage{float}
\usepackage{verbatim}
\usepackage{tabu}

\usepackage{titlesec}
\usepackage{tabto} 

\usepackage[bottom,hang,flushmargin]{footmisc}
\usepackage{tabularx}
\usepackage{array}
\newcolumntype{Y}{>{\raggedright\arraybackslash}X}
\usepackage{rotating} 

\usepackage{listings}
\usepackage{xcolor}
\lstdefinestyle{prompt}{
  basicstyle=\ttfamily\small,
  breaklines=true,
  columns=fullflexible,
  frame=single,
  backgroundcolor=\color{gray!7},
  numbers=left,
  numberstyle=\footnotesize,
  numbersep=8pt,          
  xleftmargin=3.2em,      
  framexleftmargin=3.2em, 
}

\usepackage[toc,page]{appendix}

\usepackage[hidelinks=true,breaklinks=true]{hyperref}
\usepackage{listings}
\usepackage{amsthm}
\theoremstyle{definition}
\newtheorem{definition}{Definition}[section]
\usepackage{amsmath}
\usepackage{parskip} 

\usepackage{array} 
\usepackage{amssymb} 

\usepackage{enumitem} 
\usepackage{xcolor}
\usepackage{tablefootnote}

\usepackage{pdflscape} 

\usepackage[colorinlistoftodos,textsize=small]{todonotes}

\newcommand{\tagbadge}[1]{\fcolorbox{black!20}{black!5}{\footnotesize\ttfamily #1}}

\begin{document}
\pagenumbering{roman}
\begin{titlepage}
  \begin{center}
    \vspace*{1cm}
    \begin{figure}
      \centering        
      \includegraphics[scale=0.4]{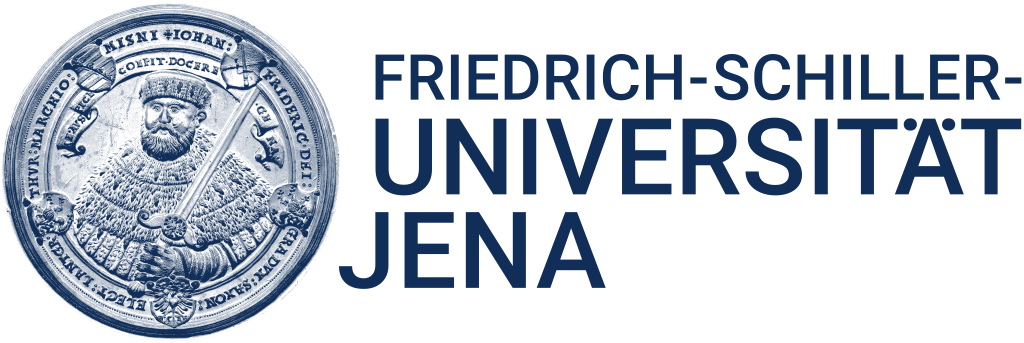}
    \end{figure}
    \LARGE
    \textbf{From Metadata to Meaning:\\
    A Semantic Units Knowledge Graph for the Biodiversity Exploratories}\\*[2.5cm]
    \large
    \textbf{Masterarbeit\\
            zur Erlangung des akademischen Grades\\
            Master of Science (M. Sc.)\\
            im Studiengang Informatik}\\*[2.5cm]
    FRIEDRICH-SCHILLER-UNIVERSITÄT JENA\\
    Fakultät für Mathematik und Informatik\\
    
    \vspace{0.8cm}        
    {eingereicht von Tarek Al Mustafa\\
     geboren am 20.09.1999 in Immenstadt im Allgäu\\
     \vspace{0.8cm}    
     Themenverantwortliche: Prof. Dr. Birgitta König-Ries\\
     \vspace{0.8cm}    
     Jena, den 15.09.2025
    }
    \vfill
      
  \end{center}
\end{titlepage}

\chapter*{\centering Abstract}
\textit{Knowledge Graphs} (KGs) bear great potential for ecology and biodiversity researchers in their ability to support synthesis and integration efforts, meta-analyses, reasoning tasks, and overall machine interoperability of research data.
However, this potential is yet to be realized as KGs are notoriously difficult to interact with via their query language \textit{SPARQL} for many user groups alike.
Additionally, a further hindrance for user--KG interaction is the fundamental disconnect between user requirements and requirements KGs have to fulfill regarding machine-interoperability, reasoning tasks, querying, and further technical requirements.
Thus, many statements in a KG are of no semantic significance for end users.
In this work, we investigate a potential remedy for this challenge: \textit{Semantic Units} (SUs) are semantically significant, named subgraphs in a KG with the goal to enhance \textit{cognitive interoperability} for users, and to provide responses to common KG modelling challenges.
We model and construct a KG from publication and dataset metadata of the \textit{Biodiversity Exploratories} (BE), a research platform for functional biodiversity research across research plots in Germany to contribute to biodiversity research from the perspective of computer science.
We contribute further by delivering the first implementation of semantic units on a knowledge graph and investigate how SUs impact KG querying.
Finally, we present two implementations of tasks that show how \textit{large language models} (LLMs) can be used to extract structured metadata categories from publication and dataset titles and abstracts, and how embedding models can be used to enrich metadata with latent information, in an effort to support the creation of structured and \textit{FAIR} (findable, accessible, interoperable, and reusable) metadata.
\begin{otherlanguage}{ngerman}
\chapter*{\centering Zusammenfassung}
\textit{Wissensgraphen} bieten großes Potenzial für Ökologie- und Biodiversitätsforschung durch ihre Fähigkeiten, Datensynthese, Datenintegration, Meta-Analysen, logische Schlussfolgerungsanwendungen und die Maschinen-Interoperabilität von Daten zu unterstützen.
Derzeit wird dieses Potenzial jedoch nicht ausgeschöpft, da die Interaktion zwischen Menschen und Wissensgraphen bekanntermaßen schwierig ist. 
Besonders die Abfragesprache \textit{SPARQL} ist für viele Gruppen von Nutzenden nur schwer nutzbar.
Ein weiteres Hindernis für diese Interaktion ist die grundlegende Spannung zwischen den Anforderungen von Nutzenden für einfachere Nutzbarkeit und den Anforderungen, die die Ermöglichung von Maschineninteroperabilität, logischen Schlussfolgerungen, Abfragen und weiteren technische Anforderungenn an Wissensgraphen stellen, weswegen viele Aussagen innerhalb eines Wissensgraphen keinen semantischen Mehrwert für Nutzende darstellen.
In dieser Arbeit befassen wir uns mit einer möglichen Lösung für dieses Problem in der Form von \textit{Semantischen Einheiten}.
Diese sind semantisch signifikante, benannte Subgraphen innerhalb eines Wissensgraphen, die die \textit{kognitive Interoperabilität} von Nutzenden unterstützen und weitere Lösungen für geläufige Modellierungsprobleme von Wissensgraphen liefern sollen.
Wir modellieren und konstruieren einen Wissensgraphen über Publikations- und Datensatzmetadaten der \textit{Biodiversitäts-Exploratorien}, einer Forschungsplattform für funktionale Biodiversitätsforschung auf Forschungsflächen in Deutschland, um diese Domäne aus der Perspektive der Informatik zu unterstützen.
Einen weiteren Beitrag dieser Arbeit stellt die erste Implementierung von semantischen Einheiten auf einem Wissensgraphen dar und wir untersuchen zusätzlich, wie sich semantische Einheiten auf die Abfrage des Wissensgraphen auswirken.
Zuletzt präsentieren wir zwei Anwendungen, die das Potential von \textit{großen Sprachmodellen} für das Extrahieren von strukturierten Metadatenkategorien von Publikations- und Datensatztiteln und Zusammenfassungen aufzeigen und wie Einbettungsmodelle Metadaten mit latenten Informationen aufbereiten können, um die Erstellung von strukturierten und \textit{FAIR}en (auffindbare, zugängliche, interoperable und wiederverwendbare) Metadatensätze zu erleichtern.
\end{otherlanguage}

\chapter*{\centering Acknowledgments}

I would like to extend my gratitude to Birgitta and Tina for their supervision of this thesis and their continued mentorship and guidance over the past few years.

Further, we thank Lars Vogt for his guidance over the course of this work.
Modelling and implementing Semantic Units would not have been possible without help directly from the source.

We also extend our gratitude to the BExIS team, especially Franziska Zander, Andreas Ostrowski, Cornelia Fürstenau, and Eleonora Petzold, for their advice and insights.

Further, we thank the interdisciplinary research project that started as a resident group \textit{Mapping Evidence to Theory in Ecology: Addressing the Challenges of Generalization and Causality} at the Center for Interdisciplinary Research (ZiF: Zentrum für interdisziplinäre Forschung)\footnote{\url{https://www.uni-bielefeld.de/einrichtungen/zif/}} at Bielefeld University and now continues as EcoWeaver\footnote{\url{https://ecoweaver.github.io/}}.

Finally, I extend my deepest gratitude to my friends and family, Without your support I would not be in the position I am in today.        
\tableofcontents 
\chapter*{\centering List of Abbreviations I}

\begin{tabularx}{\linewidth}{@{}r Y@{}}
    KG & Knowledge Graph \\
    LLM & Large Language Model \\
    SU & Semantic Unit \\
    FAIR & The FAIR Principles: Findability, accessibility, interoperability, and reusability \\
    DFG & Deutsche Forschungsgemeinschaft (German Research Federation \\
    GBIF & Global Biodiversity Information Facility \\
    RQ & Research Question \\
    SQ & Sub-question\\
    CQ & Competency Question \\
    BE&Biodiversity Exploratories\\
    BExIS&Biodiversity Exploratories Information System \\
    DOI&Digital Object Identifier\\
    RDF&Resource Description Framework\\
    OWL&Web Ontology Language\\
    BFO&Basic Formal Ontology\\
    SKOS&Simple Knowledge Organization System\\
    UFO&Unified Foundational Ontology\\
    GFO&General Formal Ontology\\
    T-Box&Terminology Box\\
    A-Box&Assertion Box\\
    W3C&World Wide Web Consortium\\
    Def&Definition\\
    GUPRI&Globally Unique Persistent and Resolvable Identifier\\
    UPRI&Unique Persistent and Resolvable\\
    OBI&Ontology of Biomedical Investigations\\
    SHACL&Shapes Constraint Language\\
    XSD& XML Schema Datatypes\\
    GEO&Geospatial Ontology\\
    PCO&Population and Community Ontology\\
    NCBI&National Center for Biotechnology Information Organismal Classification\\
    FABIO&FRBR-aligned Bibliographic Ontology\\
    VIVO& VIVO Ontology\\
    VCARD&vCard Ontology\\
    DCAT&Data Catalog Vocabulary\\
\end{tabularx}
\chapter*{\centering List of Abbreviations II}

\begin{tabularx}{\linewidth}{@{}r Y@{}}
    ALB&Schwäbische Alb\\
    HAI&Hainich-Dün\\
    SCH&Schorfheide Chorin\\
    GP&Grid Plot\\
    EP&Experimental Plots\\
    MIP&Medium Intensive Plots\\
    VIP&Very Intensive Plots\\
    REX&Reduced land-use intensity experiment\\
    LUX&Land-use experiment\\
    FOX&Forest gap experiment\\
    PROV&Provenance\\
    NCIT&National Cancer Institute Thesaurus\\
    RO&Relation Ontology\\
    SIO&Semanticscience Integrated Ontology\\
    EnvThes&Environmental Thesaurus Ontology\\
    ENVO&Environment Ontology\\
    AGRO&Agronomy Ontology\\
    MESH&Medical Subject Headings Ontology\\
    METPO&Ontology of ecophysiological traits of microbes\\
    DCTERMS&Dublin Core Terms Ontology\\
    IAO&Information Artifact Ontology\\
    FOAF&Friend-Of-A-Friend Ontology\\
    EFO&Experimental Factor Ontology\\
    ADDICTO&Addiction Ontology\\
    OPMI&Ontology of Precision Medicine and Investigation\\
    TTL&Turtle Syntax\\
    TRIG&TriG Syntax\\
    IRI& Internationalized Resource Identifier\\
    INRAE&French National Research Institute Ontology\\
    CITO&Citation Typing Ontology\\
    STATO&Ontology of Statistical Methods\\
    OMO& OMO Ontology\\
\end{tabularx}
\clearpage           
\pagenumbering{arabic} 
\chapter{Introduction} \label{chapter1}
\section{Motivation}
\begin{sloppypar}
\textit{Knowledge graphs} (KGs) have received ever growing attention in recent decades and are used in many applications in industry and science \cite{peng2023knowledge}.
Even though their usage is as wide-spread as it is, knowledge graphs are notoriously difficult to query, even for users with a background in technical fields \cite{warren2020comparison, vargas2020user, mishra2022natural, li2023knowledge}. 
Therefore, they oftentimes stay hidden behind user interfaces for larger applications, and as a result, most users are never asked to interact with one directly. 
\end{sloppypar}
This work, in line with previous efforts, continues to investigate how the interaction between users and KGs might be simplified to further support users in leveraging the advantages of integrated semantic web technologies.
To this end, we view it as the responsibility of computer scientists to deliver solutions for making the systems we develop as useful and usable as possible for different user groups.
We focus specifically on interdisciplinary research between computer scientists and researchers from the domains of ecology and biodiversity science, aiming to support researchers in leveraging semantic technologies from computer science in their daily work.

As such, the source data used as the starting point of this thesis consists of publication and dataset metadata of the \textit{Biodiversity Exploratories} (BE)\footnote{\url{https://www.biodiversity-exploratories.de/en/}}, a DFG-funded Infrastructure Priority Programme (SPP 1374) that presents a research platform for functional biodiversity research on selected research plots across Germany \cite{fischer2010implementing}. 

We estimate that the BE would benefit greatly if this metadata were to be made available in a knowledge graph for several reasons.
First, the process of semantification that supplies well defined semantic meanings and definitions to terms in a domain that is inherent in KG creation workflows, bears potential for enriching the metadata schemata for publications and datasets of the BE.
This would be beneficial both for researchers uploading metadata about their work and for users interested in reuse.
Second, if such metadata were made available in a knowledge graph, it would be more machine-actionable than it currently is.
This could enable a variety of KG specific applications that support the \textit{integration} of BE data to other knowledge sources:
Connect taxa observed in BE studies to \textit{GBIF}\footnote{\url{https://www.gbif.org/}} and gain access to trait and occurrence information,
link datasets to \textit{DOIs} (Digital Object Identifier), disambiguate authors and organizations with \textit{ORCID}\footnote{\url{https://orcid.org/}} and \textit{ROR}\footnote{\url{https://ror.org/}} identifiers, and retrieve additional metadata about affiliations or citations.
Third, the Biodiversity Exploratories have put much effort into synthesis (the tenth core project of the BE)\footnote{\url{https://www.biodiversity-exploratories.de/en/projects/core-project-10-biodiversity-synthesis/}} in recent years, aiming to create new insights from a combination of multiple information sources.
As we have outlined before, integrated publications and datasets may contribute to meta-analyses. \textit{'These analyses rely on the synthesis (or in computer science terms: integration) of numerous datasets. Today, this is a labor-intensive task, often requiring months if not years of PostDoc time'} \cite{al2025semantic}

However, with the potential upsides a knowledge graph may deliver, some challenges arise for users and their interaction with KGs, leading us to formulate the problem statement investigated in this work in the next section.

\section{Problem Statement and Contribution}
The upsides of knowledge graphs stem from rigorous semantic modelling and the adherence to their underlying data model, the \textit{Resource Description Framework} (RDF) \cite{miller1998introduction, pan2009resource}, in which data is represented as triples:

\textbf{<subject> <predicate> <object>}

With these triples, also called \textit{facts}, it becomes possible to represent minute units of information, such that complex statements about entities and their relationships can be expressed in detail.
As a result, simple natural language statements such as \textit{'Apple X has a weight of 204.56 grams'} 
require many triples to enable reasoning capabilities, query-functionality, machine interpretability, and \textit{FAIR} (findable, reusable, interoperable, and reusable) \cite{wilkinson2016fair} representation.
This leads to many graph triples bearing no semantic significance for users, which poses a barrier for user--KG interaction as graph schemata become more difficult to understand and query, which constitutes the main research problem we investigate in this work:

\textbf{Problem Statement:}
\begin{itemize}
    \item \textit{There is a fundamental mismatch between the requirements that ontology and knowledge graph engineers must satisfy to comply with best practices regarding data governance, technical requirements, and enabling machine interoperability, and the requirements that users of these technologies impose on their ability to provide semantically meaningful statements to convey information.}
\end{itemize}

\begin{sloppypar}
To address this problem, we investigate the theoretical background behind \textit{Semantic Units} \cite{vogt2024asemantic}, a novel approach for semantically meaningful, named subgraphs in a knowledge graph, and contribute by presenting their first technical implementation. As the foundation of this implementation, we model and construct a knowledge graph for publication and dataset metadata of the Biodiversity Exploratories.

The metadata used as basis for the knowledge graph is available on BExIS, the \textit{Biodiversity Exploratories Information System}\footnote{\url{https://www.bexis.uni-jena.de/}}, where it exists in a semi-structured format, consisting of both structured metadata categories and unstructured free-text fields.
As some complexity of knowledge graph creation is alleviated when source data is structured, we additionally investigate \textit{Large Language Model} (LLM) facilitated approaches to extract structured metadata categories from publication and dataset titles and abstracts.
This could not only support KG creation workflows, but also aid researchers in creating FAIR-aligned datasets with less effort.

Finally, we investigate a second LLM application that may extract latent information from BE metadata: 
Using embedding models, we aim to extract latent topics that publications and datasets cluster around, categorize documents by providing topic hierarchies as anchors in an embedding space, and link documents to the overarching research goals of the Biodiversity Exploratories to further enrich BE metadata.
\end{sloppypar}

\section{Thesis Overview}
In the last section of this introductory chapter, we present an overview of the contents of this thesis:

The work is split into three background an related work chapters, three method chapters, and three further chapters that present results, discuss them, and conclude the thesis.

In Ch. \ref{chapter2}, we present core concepts of the semantic web including knowledge graphs, ontologies, description logics and reasoning, the FAIR Guiding Principles, the concept of cognitive interoperability, and review the process of KG and ontology engineering.
In the second related work chapter, Ch. \ref{chapter3}, we provide context for the user perspective of knowledge graph interaction, review common modelling challenges in KG knowledge representation, and present an introduction to semantic units that aim to provide solutions to these challenges.
The final related work chapter, Ch. \ref{chapter4}, covers other approaches for simplified knowledge graph interaction and presents literature on the interplay between knowledge graphs and large language models.

To lead into the three method chapters, we first present the Biodiversity Exploratories, their components and research design, BExIS, and the source data used in this work in detail in Ch. \ref{chapter4-5}.
Afterwards, we state the research questions investigated in this work and present the two parts of our method and their contents in Ch. \ref{chapter5}.
In the second method chapter, Ch.\ref{chapter6}, we present our modelling approach, state competency questions to evaluate the finished KG, describe the modelling process in detail, present our final knowledge graph schema, discuss modelling challenges, provide implementation details, and describe and approach for schema validation.
The final method chapter, Ch. \ref{chapter7}, presents our approach for two LLM applications: Extracting metadata categories from publication and dataset titles and abstracts, and embedding methods for latent information extraction.

Following the method chapters, we present our results in Ch. \ref{chapter8}, in which we detail how to write SPARQL queries for semantic units, evaluate our approach using competency questions, and present semantic unit visualizations.
Ch. \ref{chapter9} contains detailed discussions of the contents and results of this work, addresses our main research questions, lists limitations to our approach, and presents directions for future work.
Finally, we conclude the thesis in Ch. \ref{chapter11}.

\chapter{The Semantic Web} \label{chapter2}

To lay a proper foundation for the contents of this work, in this chapter, we provide background information on concepts of and related to the \textit{Semantic Web}.
Topics covered range from clarifications on ontology and knowledge graph definitions, logical frameworks and reasoning, to standards like the \textit{FAIR Guiding Principles} \cite{wilkinson2016fair} and an overview of concepts for knowledge graph and ontology engineering.

\section{Knowledge Graphs and Ontologies}
One of the core parts of this work is the application of advanced modelling techniques for ontologies and knowledge graphs. 
As a preliminary, we will first define the meaning of these terms in the context of this work. 
This is necessary as both terms are frequently used in the literature to describe ideas that point in similar directions, but differ significantly upon closer inspection.
Therefore, we will first present common definitions of both ontologies and knowledge graphs before settling on those most appropriate for our purposes.

\begin{sloppypar}

The term ontology is not a new invention by computer scientists. Consisting of the Greek terms \textit{ontos} and \textit{logos}, translated as \textit{being} and \textit{word} respectively \cite{Breitman2007}, its first mention came in the year of 1613 by two philosophers, \textit{Rudolf Göckel} and \textit{Jacob Lorhard}, who both independently used it in their works \textit{Lexicon philosophicum} and \textit{Theatrum philosophicum} \cite{Smith2012-SMIOJD}. 
As such, \textit{'Ontology as a branch of philosophy is the science of what is, of the kinds and structures of objects, properties, events, processes[,] and relations in every area of reality'} \cite{Smith2012-SMIOJD}.    
\end{sloppypar}
This philosophical definition already encapsulates the essence of what an ontology in computer science aims to achieve: Defining objects and properties that can exist, and what structure they exist in.
Thomas Gruber's definition motivates this concept further: \textit{'An ontology is an explicit specification of a conceptualization'} \cite{gruber1993translation}, which Studer et al. later specified as \textit{'a formal, explicit specification of a shared conceptualization'} \cite{studerKnowledgeEngineeringPrinciples1998}. 
From this definition it follows that in practice, one goal of an ontology is to explicitly state the shared understanding of people working in the same area about the concepts that exist and how they relate to each other, such that common definitions can be referenced.

\begin{sloppypar}
However, not only domain-specific ontologies exist. 
An ontology may also try to define existing entities and relationships across all fields, or supply definitions of commonly used notions or highly general, domain independent categories such as \textit{continuants} and \textit{occurrents} in \textit{BFO}, the Basic Formal Ontology \cite{otte2022bfo}.
In this case, we speak of \textit{upper-level ontologies} or \textit{top-level ontologies}, that can be used as a domain-agnostic backbone or starting point to build domain-specific terms on top of \cite{Smith2012-SMIOJD}.
Further examples of top-level ontologies include \textit{SKOS}, the Simple Knowledge Organization System \cite{miles2009skos}, \textit{UFO}, the Unified Foundational Ontology \cite{guizzardi2015towards, guizzardi2022ufo}, and \textit{GFO}, the General Formal Ontology \cite{herre2010general}. For a comprehensive review of upper-level ontologies and their ontological commitments we refer to \cite{partridge2020survey}.
\end{sloppypar}

Using these definitions of ontologies as basis, we derive the definition of knowledge graphs as the term is used in this work.
Ontologies define the concepts and terms that exist in a domain, and thus represent the \textit{terminology box} (T-Box or TBox) by expressing universal statements.
A knowledge graph, however, contains assertions about specific individuals that are instances of ontology classes, and thus also contains the \textit{assertion box} (A-Box or ABox).
Therefore, we define a knowledge graph as the combination of a T-Box and A-Box, a definition that is adapted widely across the literature \cite{paulheim2016knowledge, ma2022knowledge, simsek2023knowledge, callahan2024open, vogt2025framework}.

However, many other definitions of knowledge graphs exist that focus on other concepts than the T-Box and A-Box:

Ehrlinger and Wöß collect a variety of KG definitions and define a KG as follows: \textit{'a KG acquires and integrates information into an ontology and applies a reasoner to derive new knowledge'} \cite{ehrlinger2016towards}.

In a 2021 survey, KGs are defined as \textit{'G = \{E, R, F\},
where E, R and F are sets of entities, relations and facts,
respectively. A fact is denoted as a triple $(h, r, t) \in F$'} \cite{ji2021survey}.

Hogan et al. define KGs as \textit{'graph[s] of
data intended to accumulate and convey knowledge of the real world, whose nodes represent entities
of interest and whose edges represent potentially different relations between these entities. The graph of data (a.k.a. data graph) conforms to a graph-based data model, which may be a directed edge-labelled graph, a heterogeneous graph, a property graph, and so on'} \cite{hogan2021knowledge} .

\section{Logical Frameworks and Reasoning}
\begin{sloppypar}
Ontologies used in this work are based on \textit{The Web Ontology Language} \cite{mcguinness2004owl}, OWL for short, using RDF, \textit{The Resource Description Framework} \cite{miller1998introduction, pan2009resource} to represent data.
As stated in this excerpt from the introduction of OWL's \textit{W3C} (World Wide Web Consortium) recommendation: 
\end{sloppypar}

\textit{'The first level above RDF required for the Semantic
Web is an ontology language what can formally describe the meaning of terminology used in Web
documents. If machines are expected to perform useful reasoning tasks on these documents, the
language must go beyond the basic semantics of RDF Schema'} \cite{mcguinness2004owl}.

The key point is that OWL provides a formal way for description, and aims to perform reasoning tasks on data.
In OWL, the logical framework used is that of \textit{Description Logics} \cite{baader2003basic, baader2008description}, to check decidability and solve inference problems \cite{horrocks2005owl}.
These include query-rewriting and rule-based inferences from \textit{owl:sameAs}\footnote{\url{https://www.w3.org/TR/owl-ref/\#sameAs-def}} \cite{polleres2013rdfs}, reasoning over transitive properties, \textit{'concept satisfiability, class subsumption, class consistency, and instance checking'} \cite{wang2004ontology}. 
Also, reasoning can be used when answering queries by expanding over explicit terms written in the query \cite{fikes2004owl}.

\section{Cognitive Interoperability}
A core motivation of this work is to investigate whether semantic units may enhance the \textit{cognitive interoperability} between users and knowledge graphs.
We believe that this term is explained well by presenting definitions of related concepts that cognitive interoperability can be derived.
A starting point that is fitting for this purpose are the \textit{FAIR Guiding Principles} \cite{wilkinson2016fair} as they represent perhaps the most commonly known guidelines for scientific data management. 
(Meta)data should be findable, accessible, interoperable, and reusable to allow for better machine-readability, -interpretability, and -actionability.
Weiland et al. define these terms as follows:

\textbf{Machine-readability.} Machine-readable \textit{“are those elements in bit-sequences that are clearly defined by structural specifications”} \cite{weiland2022fdo}.

\textbf{Machine-interpretability.} Machine-interpretable \textit{“are those elements that are machine-readable and can be related with semantic artifacts in a given context and therefore have a defined purpose”}\cite{weiland2022fdo}.

\textbf{Machine-actionability.}     Machine-actionable \textit{“are those elements in bit-sequences that are machine-interpretable and belong to a type for which operations have been specified in symbolic grammar”} \cite{weiland2022fdo}.

Vogt et al. develop these ideas further and suggest extensions for the FAIR principles to enhance \textit{semantic interoperability} \cite{vogt2025suggestions}, which \textit{'requires systems to agree on a shared standard of representing information to enable them to efficiently communicate information without human intervention'} \cite{maciel2024systems}.
Next to these machine centric descriptors, the concept of cognitive interoperability is mentioned in \cite{vogt2023fair} and defined in \cite{vogt2024bsemantic}:

\textbf{Cognitive Interoperability.} \textit{'A critical characteristic of information technology systems and research data infrastructures that plays a central role in facilitating the efficient communication of data and metadata to human users. By providing intuitive tools and functions, these systems enable users to gain an overview of data, locate data of interest, and explore related data points in semantically meaningful and intuitive ways. Thereby, the systems must take into account the general cognitive conditions of human beings, not only in terms of how humans prefer to interact with technology (human-computer interaction), but also how they prefer to interact with information (human information
interaction)'} \cite{vogt2024bsemantic}.

The concept of cognitive interoperability is the core of what we aim to investigate in this thesis:
Using advanced modelling techniques like semantic units, can cognitive interoperability for users of a knowledge graph be enhanced?
Before we present more background and information on semantic units in the next chapter, we present advantages of knowledge graphs and give an overview of knowledge graph and ontology engineering in the next section.

\section{Knowledge Graph and Ontology Engineering}\label{sec:KGENG}
In the final section of this chapter, we give an overview of knowledge graph and ontology engineering approaches.
We will return to concepts mentioned in this section in Ch. \ref{chapter6}, the chapter in which we describe our own modelling approach for the knowledge graph developed in this work.


\textbf{The modelling process.} Studer et al. present one perspective in which knowledge engineering has at its center the modelling process. 
A model is only an approximation of reality, and modelling an incessant activity with the goal of reaching an intended behavior.
The process is also cyclic, meaning new observations lead to further model refinement. 
Finally, the modelling process is also highly subjective as to the interpretations of the engineers \cite{studerKnowledgeEngineeringPrinciples1998}.

\textbf{Problem-solving.} Methods with a problem-solving view focus on which actions need to be performed to solve a task and a sequence in which they have to be carried out.
This view can inform the main rationale, meaning the main goal of a knowledge base, specifying what questions a knowledge graph is supposed to answer and how \cite{studerKnowledgeEngineeringPrinciples1998}.

\textbf{Top-down and bottom-up approaches.} A further distinction is made between top-down and bottom-up approaches, the first concerns itself with the graph schema, the ontology, before building the data graph, while the latter starts by examining data that should appear in the final knowledge graph first and focusing on its representation on a schema level later on \cite{li2020research}.

\textbf{Ontology design and construction.} Bravo et al. specify a methodology for the design and construction of ontologies \cite{bravo2019methodology}.
Of the characteristics of their approach we highlight user centric design and evaluation.
An important part of ontology engineering are the requirements of the user group, and involving users \textit{'in the definition of competency questions from the beginning of the methodology to the evaluation'} \cite{bravo2019methodology}.

\textbf{Constructing ontologies from reusable ontologies.} \textit{'Assuming that the world is full of well designed modular ontologies, constructing a new ontology is a matter of assembling existing ones'} \cite{studerKnowledgeEngineeringPrinciples1998}.
Ontologies can be reused by including classes, relations, and axioms of other ontologies into a starting ontologies, and resolving naming conflicts. 
This requires mappings between ontologies \cite{uschold1998ontology, kalfoglou2003ontology, choi2006survey}. 

However, since every ontology has as its premise ontological commitments, this process is not straightforward.
If two ontologies are committed to contradictory ontological commitments, creating mappings between them will be challenging, or even impossible.
Here, upper-level ontologies play another important role: Combining modular ontologies that extend the same upper-level ontology, such as BFO, is easier than mapping between a BFO module and an UFO module.

\textbf{Users and user groups.} Li et al. distinguish three knowledge graph practitioner personas from a user study: KG builders, analysts, and consumers \cite{li2023knowledge}. We quote here the description of KG consumers: \textit{'The KG Consumer is generally an expert in the data domain, business, overarching use case, or the KG’s sociocultural context. While Consumers typically do not interact directly with a KG database or its querying language, they are still a stakeholder or end user of the KG, and know what “types” of insights would be valuable to extract. Consumers tend to rely on KG Analysts, query building GUIs, or automated reporting systems to generate those insights'} \cite{li2023knowledge}.
We mention this especially to highlight the relevance of constructing systems with user's needs in mind and we see it as our responsibility to facilitate interaction and usage of a KG for end users \cite{al2025semantic}, further motivating this thesis. 

With this short overview of knowledge graph and ontology engineering, we conclude the first of three background and related work chapters.
In the following chapter, we present common modelling challenges faced in OWL/RDF knowledge graphs, present background information on semantic units, and motivate their use for this work by showing how they may be used to solve selected modelling challenges.

\chapter{Background and Semantic Units} \label{chapter3}
In the previous chapter we presented background information on the semantic web and concepts related to it.
Our goal in this chapter is to give insights into the knowledge graph user perspective, to list representation and modelling challenges relevant to this work, and introduce semantic units and the purposes we aim to implement them for in our method.

\section{Literature}
In the following sections we will introduce the concept of semantic units and present an overview of common challenges in knowledge graph representation.
We reference the ideas and prior work of Lars Vogt and reference a series of relevant publications in which these concepts were discussed first: 

\begin{itemize}
\item FAIR Knowledge Graphs with Semantic Units: a Prototype \cite{vogt2023fair}
\item Semantic Units: Organizing knowledge graphs into semantically meaningful units of representation \cite{vogt2024asemantic}
\item Suggestions for extending the FAIR Principles based on a linguistic perspective on semantic interoperability \cite{vogt2025suggestions} 
\item Rethinking OWL Expressivity: Semantic Units for FAIR and Cognitively Interoperable Knowledge Graphs -- Why OWLs don't have to understand everything they say \cite{vogt2025rethinkingowlexpressivitysemantic}

\end{itemize}

\section{User Perspective}
The central idea around many challenges in this domain concerns the the loss of human-interpretability of statements modelled in a knowledge graph, as the priority of modelling are reasoning-capabilities, query-functionality, and machine-interpretability, it becomes necessary to make statements about statements in a knowledge graph \cite{vogt2023fair}.

\begin{figure}[h]
   \centering
   \includegraphics[width=1\textwidth]{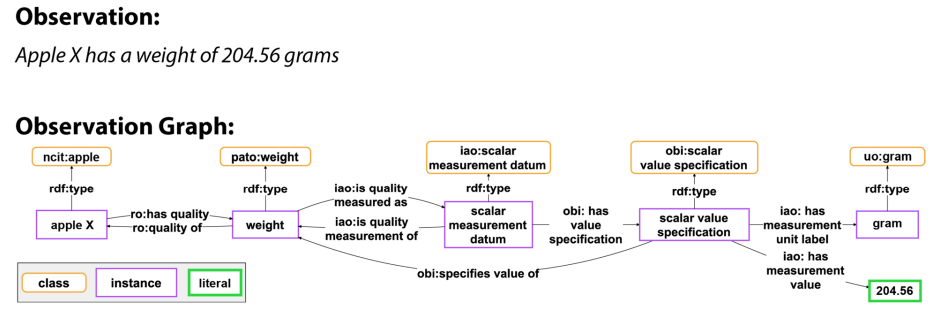}
   \caption{'Comparison of a human-readable statement with its machine-actionable representation as a semantic graph following the RDF syntax. Top: A human-readable statement concerning the observation that a specifc apple (X) weighs 204.56 grams. Bottom: The corresponding representation of the same statement as a semantic graph, adhering to RDF syntax and following the established pattern for measurement data from the Ontology for Biomedical Investigations (OBI) \cite{bandrowski2016ontology} of the Open Biological and Biomedical Ontology Foundry (OBO) \cite{smith2007obo}'. Adapted from \cite{vogt2024asemantic}, Figure 2.}
   \label{fig:Apple}
 \end{figure}

Users, for example experts in ecology and biodiversity research, that want to extract semantically meaningful content from a knowledge graph about a specific domain, are not concerned with how this knowledge graph is structured and why.
Their main concern is the information present in the graph and how to obtain it.
The key point lies in what users deem as relevant. 
A user might want to obtain information about a specific instance of a class, say a specific apple.
One statement of possible interest to users might therefore be its weight. Fig. \ref{fig:Apple} shows the fundamental disconnect: Users care about the observation itself -- '\textit{Apple X has a weight of 204.56 grams}' -- the observational graph however, i.e. the way this observation is modelled in the knowledge graph, must convey considerably more information for precise semantic modelling and to enable machine interoperability. 
For example, the triple \textit{<scalar measurement datum> <obi:hasValueSpecification> <scalar value specification>} contains no semantic significance for domain experts as they are likely not able to infer its meaning.

\section{Knowledge Graph Representation and Modelling Challenges}

Vogt distinguishes twelve challenges in knowledge graph representation and modelling \cite{vogt2025rethinkingowlexpressivitysemantic}. 
From these, we will restate those that are relevant to this work and the publication and dataset metadata of the BE:
\subsection{Modelling Class Axioms that involve triangular Relationships}\label{subs:challenge3}

\begin{figure}[h!]
   \centering
   \includegraphics[width=1\textwidth]{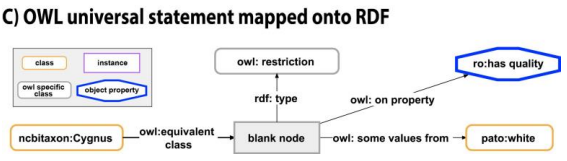}
   \caption{. 'C) All swans have quality some white',  expressed as a class axiom within an ontology. Adapted from \cite{vogt2024asemantic}, Figure 1 C).}
   \label{fig:universal}
 \end{figure}

Fig. \ref{fig:universal} shows the universal statement (class axiom) \emph{'\textbf{All} swans have quality \textbf{some} white'} and how it can be expressed in RDF.
This expression relies on the use of a blank node.

\begin{definition}(Blank Node)
\label{def:bnode}
A blank node is a resource in a knowledge graph that is \textit{'treated as simply indicating the existence of a thing, without using, or saying anything about, the name of that thing'}\footnote{\url{https://www.w3.org/TR/rdf-mt/\#unlabel}}.
\end{definition}

Because of this definition, blank nodes are also referred to as \textit{'existential variables'} \cite{mallea2011blank, hogan2014everything}, meaning variables that do not posses their own \textit{IRI} (Internationalized Resource Identifier) and represent a resource for which no IRI exists.

The aspect of anonymity in blank nodes leads to significant issues because they cannot be referenced or identified, especially when they are used in class axioms as in Fig. \ref{fig:universal}.
Take as example a class axiom in Manchester Syntax\footnote{\url{https://www.w3.org/TR/owl2-manchester-syntax/}} that contains multiple triples referencing a blank node, like this example from \cite{vogt2025rethinkingowlexpressivitysemantic}, in which an antenna is described that is longer than the eyes of the organism both are part of:
\begin{itemize}
    \item 'has part some ((antenna part of \textbf{some multicellular organism}) and has quality some (length and increased in magnitude relative to some (length and inheres in some (eye part of \textbf{some multicellular organism}))))'
\end{itemize}
The statements marked in bold contain references to blank nodes. 
While it seems obvious to readers that '\textbf{some multicellular organism}' must be the same organism for the class axiom to make sense, this connection does not follow from Def. \ref{def:bnode}.
Therefore, triangular relationships in class axioms can not be modelled in OWL in a satisfactory fashion.

\begin{sloppypar}
\subsection{Modelling negations and cardinality restrictions in Knowledge Graphs} \label{subs:challenge4}
\end{sloppypar}
A further challenge is posed by the \emph{Open World Assumption} from description logic \cite{drummond2006open}.
From this assumption it follows that in knowledge graphs, we only have information about what is stated, and we have no information about what is not stated, as it could both be true or false.
Therefore, from the absence of a statement it does not follow that that statement
If we wanted to assert that a particular swan is not white, it does not suffice to exclude triples \emph{swan hasColor white} from the knowledge graph, because from their absence and the open world assumption, we would have to reason that the swan might still be white.

Tools to express that a swan is not white include class expressions as T-Boxes in which we use \emph{owl:complementOf} to exclude an instance from the class of white swans (this, however, needs the class of white swans, and uses blank nodes in its expression), or by including a class axiom that makes it a necessity for instances of the class swan to have a color, then, if a color other than white is stated, it follows that this swan is not white (this also leads to many issues).

A similar issue arises for expressions of cardinality, in which we state that, for example, a swan has exactly two wings. 
These expressions can be made in OWL TBoxes, but translating the information to RDF results in a cluttered graph with at least one blank node \cite{vogt2025rethinkingowlexpressivitysemantic}.

In conclusion, while it is possible to express negations and cardinalities in OWL/RDF, the cognitive interoperability of these expressions suffer significantly due to the use of blank nodes and the corresponding information would be contained in the T-Box, not in the A-Box, and therefore out of the proposition space of an OWL-based knowledge graph.




\subsection{Having to use SPARQL for querying Knowledge Graphs}\label{sec:challenge6}

The main way of interacting with knowledge graphs is through query languages, the most common one for OWL/RDF graphs being SPARQL\footnote{\url{https://www.w3.org/TR/sparql11-query/}}.
As we have covered previously \cite{al2025semantic}, the necessity of relying on SPARQL to interact with knowledge graphs is a hindrance, not only for researchers with a computer science background when learning how to use SPARQL efficiently and effectively, but also for non domain experts, who are a central part of the intended users in this case.


\subsection{Identifying topic-specific subgraphs to provide standardized views on Knowledge Graphs}

As Vogt et al. \cite{vogt2025rethinkingowlexpressivitysemantic} point out, many domains have standardized formats and conventions that data or datasets with specific information follow. 
For example, we can define a schema that describes how to represent scientific papers, or follow best practices for metadata management and capture provenance data for processes following specified recommendations\footnote{\url{https://www.w3.org/TR/prov-overview/}}.
These specifications can also be called data sheets that provide a structure for information to be entered into.

Collections of these schemata or data sheets in a knowledge graph could be leveraged as subgraphs about topics, and, if each subgraph was assigned a Globally Unique Persistent and Resolvable Identifier (GUPRI), they could be referenced.
This would \textit{'significantly simplify making statements about the corresponding information and increase the overall explorability of the graph' }\cite{vogt2025rethinkingowlexpressivitysemantic} by leveraging topic-specific views.






\section{Semantic Units}
In this section we will present the concept of semantic units proposed in the literature, their different types, and how to apply them in OWL/RDF based knowledge graphs.
\begin{sloppypar}
The core principle behind semantic units is to structure a knowledge graph into semantically meaningful, identifiable sets of triples.
These are subgraphs within the graph \textit{'that represent units of representation possessing semantic significance for human readers'} \cite{vogt2024asemantic}.
Each subgraph is represented by its own resource, a GUPRI and thus its own node in the graph that can be referenced.
\end{sloppypar}

\begin{figure}[ht]
   \centering
   \includegraphics[width=1\textwidth]{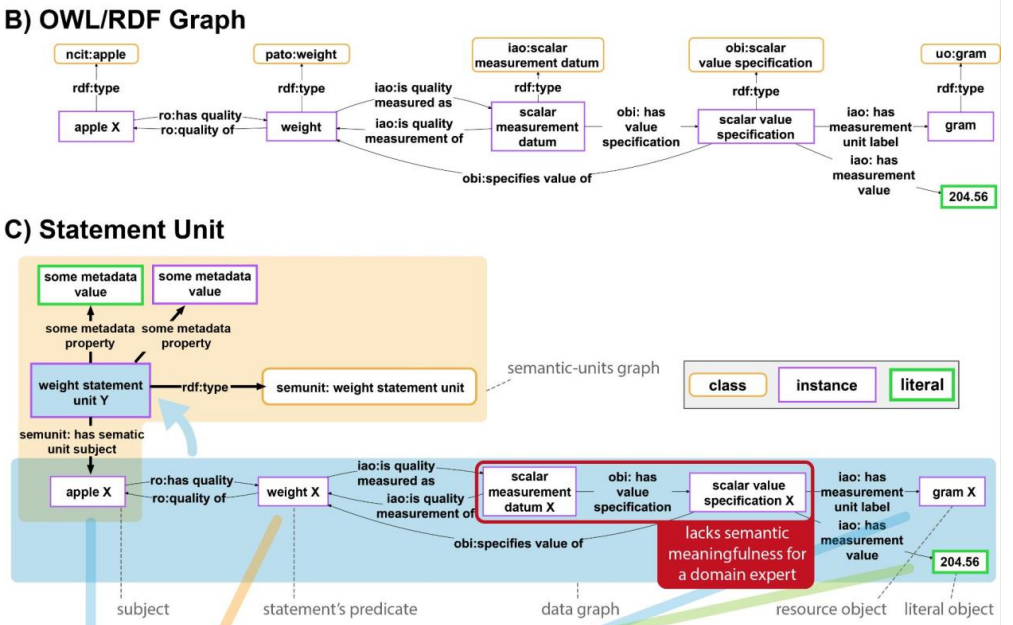}
\caption{'B) The corresponding representation of the same statement [Referring to the weight statement of the apple.] as an instance-based semantic graph, adhering to RDF syntax and following the established pattern for measurement data from the Ontology of Biomedical Investigations (OBI). 
C) The same graph, organized as a statement unit, which is a specific category of semantic unit. 
The data graph, denoted within the blue box at the bottom, articulates the statement with ‘apple X’ as its subject and ‘gram X’ alongside the numerical value 204.56 as its objects. 
Note, how the graph from B) forms the data graph of this statement unit. The peach-colored box encompasses its semantic-units graph.
It explicitly denotes the resource embodying the
statement unit (bordered blue box) as an instance of the *SEMUNIT:weight statement unit* class and the
*SEMUNIT:assertional statement unit* class, with ‘apple X’ identified as the subject.
Notably, the GUPRI of the statement
unit is also the GUPRI of the semantic unit’s data graph (the subgraph in the not bordered blue box). 
The semantic-units graph also contains various metadata triples, here only indicated by some metadata property and some metadata value as their placeholders. 
Highlighted in red within the data graph is an example of a triple that is required for modelling purposes
but lacks semantic meaningfulness for a domain expert.' Adapted from \cite{vogt2025rethinkingowlexpressivitysemantic}, Figure 1 B) and C)
}
   \label{fig:SemUnit}
 \end{figure}
For illustrative purposes, we refer to Fig. \ref{fig:SemUnit} that contains the running example from Fig. \ref{fig:Apple}.
Fig. \ref{fig:SemUnit} B) shows the OWL/RDF graph of the statement -- 'Apple X has a weight of 204.56 grams' -- referred to as the observation graph above.
In Fig. \ref{fig:SemUnit} C), the semantic information of the observation graph is conveyed using a type of semantic unit, a statement unit (to be defined below). 
As demonstrated in the red box, the triple \textit{<scalarMeasurementDatumX> <obi:hasValueSpecification> <scalarValueSpecificationX>} is a vital component of the knowledge graph but it lacks semantic meaningfulness for most users.
The statement unit takes all triples that are relevant and identifies them as a subgraph (blue box), whose subject is 'apple X'.
The peach colored box contains all triples relevant to the statement unit.
It asserts a type definition \textit{semunit:weightStatementUnit} and possible metadata properties and values.
This sets a new layer on top of the data graph called semantic unit layer, containing graphs associated with semantic units and allows for multiple degrees of abstraction.
In this example, it now becomes possible to assert that a particular instance of the class apple, 'apple X' has a statement made about its weight.

This illustrates the main idea behind semantic units: They focus on information that has semantic significance for users.
Their potential lies in being defined for specific purposes, directly linked to a target domain and the users acting within it.

To this end, different types of semantic units exist to enable the expression of different kinds of statements. 
Furthermore, new semantic unit types can be defined depending on the use case.
Two main semantic unit types can be distinguished from the taxonomy of semantic units: Statement units and compound units.
Statement units are concerned with how to express different types of statements (i.e., single propositions), for example, how to identify individuals or instances, how to assert facts or beliefs, or how to make complex statements that reuse other statement units as subjects and/or objects.
Compound units serve \textit{'as a higher-level organizational resource that groups together a semantically meaningful collection of semantic units (either statement or compound units). Each compound unit is represented by its own GUPRI [Globally Unique Persistent and
Resolvable Identifier] and instantiates a corresponding compound unit class'} \cite{vogt2025rethinkingowlexpressivitysemantic}.

With compound units, one can imagine a variety of scenarios that can potentially simplify knowledge graph query.
For example, questions like \textit{'What are all statements made about a specific dataset?'} can be queried and answered much quicker, which enhances graph exploration. 
At the same time, queries like \textit{'List all weight measurement statements made about instances of a specific species'} can be formulated quickly. 
Therefore, in this work, one part of our method concerns itself with what statement and compound units can be useful for publication and dataset metadata of the BE.
Before we present this part of our method however, we first cover related approaches for simplified knowledge graph interaction in the next chapter. 
\chapter{Related Work} \label{chapter4}
In the last chapters we expanded upon background on the semantic web, modelling and representation challenges in OWL/RDF based knowledge graphs, and semantic units.
Below, we give an overview of other approaches concerned with simplifying the interaction between users and knowledge graphs and present how large language models can assist in this challenge, or contribute to knowledge graph and ontology engineering.
\section{Interfaces}
\begin{sloppypar}
One extensively studied approach is that of layering a user interface on top of a knowledge graph.
This allows users to interact with the interface only, and hides querying via SPARQL behind an application.

A popular approach is that of creating faceted search interfaces for knowledge graphs, for example \textit{SemFacet} \cite{arenas2014semfacet, arenas2016faceted} or \textit{Sampo-UI} \cite{ikkala2021sampo, rantala2023create, hyvonen2023digital}.

Further, query builders can be used to forego SPARQL syntax and guide users in assembling queries over the knowledge graph, requiring less prerequisite knowledge.
Examples of these query builder interfaces include \textit{SPARKLIS} \cite{ferre2016sparklis}, \textit{RDF-Explorer} \cite{vargas2019rdf}, the \textit{Wikidata Query Builder}\footnote{\url{https://www.wikidata.org/wiki/Wikidata:Query_Builder}}, and our own prior work \cite{al2025semantic}.  

Visualization approaches can also help users interact with knowledge graphs easier, like \textit{Gruff} from \textit{AllegroGraph}\footnote{\url{https://allegrograph.com/products/gruff/?utm_source=chatgpt.com}}, graph visualization in \textit{Ontotext GraphDB}\footnote{\url{https://graphdb.ontotext.com/documentation/11.0/visualize-and-explore.html}}, the \textit{InK Browser} \cite{christou2025improving}. 

One further application is the \textit{Open Research Knowledge Graph (ORKG)} \cite{jaradeh2019open}, a large scale knowledge graph of scientific publications with an integrated, embedding based, semantic search component, \textit{ORKG Ask} \cite{oelen2024orkg} with which publication contents can be queried with.
\end{sloppypar}

\section{From Natural Language to SPARQL Queries}
\begin{sloppypar}
Another approach next to faceted search and query builders is concerned with the question of how to turn a question formulated in natural language into a SPARQL query.
This idea has evolved over the years. 
Early systems started with rule-based methods \cite{ngonga2013sorry}, query formalization using natural language processing based methods by translating a natural language query into an intermediate \textit{normalized query structure} \cite{dubey2016asknow}, subgraph matching \cite{hu2017answering}, pattern based approaches \cite{steinmetz2019natural}, token-level subgraph generation \cite{jung2020automated}, and incorporating keyword search together with SPARQL translation for better question answering \cite{hu2021natural}.

With the rise of large language models, neural models also became more prevalent \cite{lehmann2023language, yang2023llm, diallo2024comprehensive, emonet2024llm, zahera2024generating, haakansson2025natural}.

Furthermore, a new resource in the form of a dataset containing pairs of natural questions and SPARQL queries has recently been developed \cite{meloni2025exploring}.
\end{sloppypar}

\section{SHACL}
The \textit{Shapes Constraint Language (SHACL)}\footnote{\url{https://www.w3.org/TR/shacl/}} is a language used for schema validation in knowledge graphs.
In SHACL, conditions can be defined to ensure adherence of data that is to be instantiated in a knowledge graph to a schema. 
SHACL shapes can also be defined later for an existing knowledge graph and then a validation check can be run on the KG to ensure integrety of the data.
For example, using SHACL we can express that a publication must have a DOI, it must have a citation string, and it must provide author names.

However, SHACL shapes can not only be used for schema validation, but also present the basis for user interface applications. 
The \textit{Data Shapes Namespace}\footnote{\url{https://datashapes.org/forms.html?utm_source=chatgpt.com, https://www.datashapes.org/}} extends the SHACL vocabulary with terms that pass form information to a user interface, and allows the definition of different views and the display of additional information for users.


\section{LLMs and Knowledge Graphs}
A further point of interest is an examination of the interplay between LLMs and knowledge graphs, as LLMs could deliver assistance in a variety of applications related to knowledge graphs, for example by reducing the workload of KG construction or by extracting latent information from data that might enrich the contents of knowledge graphs.
In recent times this has been a core question in the domain, examining how LLMs could support KG usage in ways mentioned above, but also assist KG engineering, particularly generation and answering of competency questions, automatic generation of ontologies, automatic instantiation, schema and graph enrichment, and data processing \cite{meyer2023llm, kommineni2024human, zhang2024extract}.
One further point of application is the usage of LLMs for text mining that is currently studied extensively \cite{fink2023potential, wan2024tnt, yang2024integrated, mukanova2024llm, d2025mining} and may be applicable for our purposes. 

Although it would certainly be an interesting question whether LLMs could support the schema-level modelling of the KG including creation of semantic units, in the context of this thesis the modelling process is done manually.
One aspect we aim to investigate however, is whether LLMs can assist in two applications: The first being a support for researchers entering their publication or dataset metadata, by investigating whether some metadata categories can be extracted from titles and abstracts automatically, and the second one being the extraction of latent information in titles and abstracts, and whether publications and datasets can be linked through the research goals of the Biodiversity Exploratories utilzing embedding methods.

With this overview of related work we conclude the background chapters of this thesis. 
In the following chapters, we present more background on the Biodiversity Exploratories and the data used as basis for this work, and then proceed by presenting our method in detail. 
\chapter{Biodiversity Exploratories, BExIS, and Source Data} \label{chapter4-5}

In this chapter we provide explanations and details about the Biodiversity Exploratories, their components, and the data collected about related studies and datasets, available on BExIS, the \emph{Biodiversity Exploratories Information System}. 
As this thesis is concerned with transforming this metadata into a knowledge graph, and this process requires deep understanding of how experiments and infrastructures of the Biodiversity Exploratories relate to each other, we collect this information in a distinct chapter so it can be referenced at later points of this thesis.

\section{Biodiversity Exploratories} \label{sec:BE}
The Biodiversity Exploratories (BE) are a DFG-funded Infrastructure Priority Programme (SPP 1374) that present a research platform for functional biodiversity research\textsuperscript{\ref{fn:BEMAIN}} on selected research plots across Germany.
Fisher et al. explain the structure and rationale behind the BE in the publication \emph{'Implementing large-scale and long-term functional biodiversity research: The Biodiversity Exploratories'} \cite{fischer2010implementing}.
We compile information from that source and the main websites of the BE\footnote{\label{fn:BEMAIN}\url{https://www.biodiversity-exploratories.de/en/}} below.
The BE's main research objectives examine:
\begin{itemize}
    \item \textit{'how the form and intensities of land use affect biodiversity and ecosystem processes'\footnote{\label{fn:BEOBJ}\url{https://www.biodiversity-exploratories.de/en/about-us/research-objectives-and-background/}}}, 
    \item \textit{'how different components of biodiversity interact'\textsuperscript} and
    \item \textit{'how different components of biodiversity influence ecosystem processes and ecosystem services.'}\textsuperscript{\ref{fn:BEOBJ}}
\end{itemize}

\begin{figure}[t]
   \centering
   \includegraphics[width=1\textwidth]{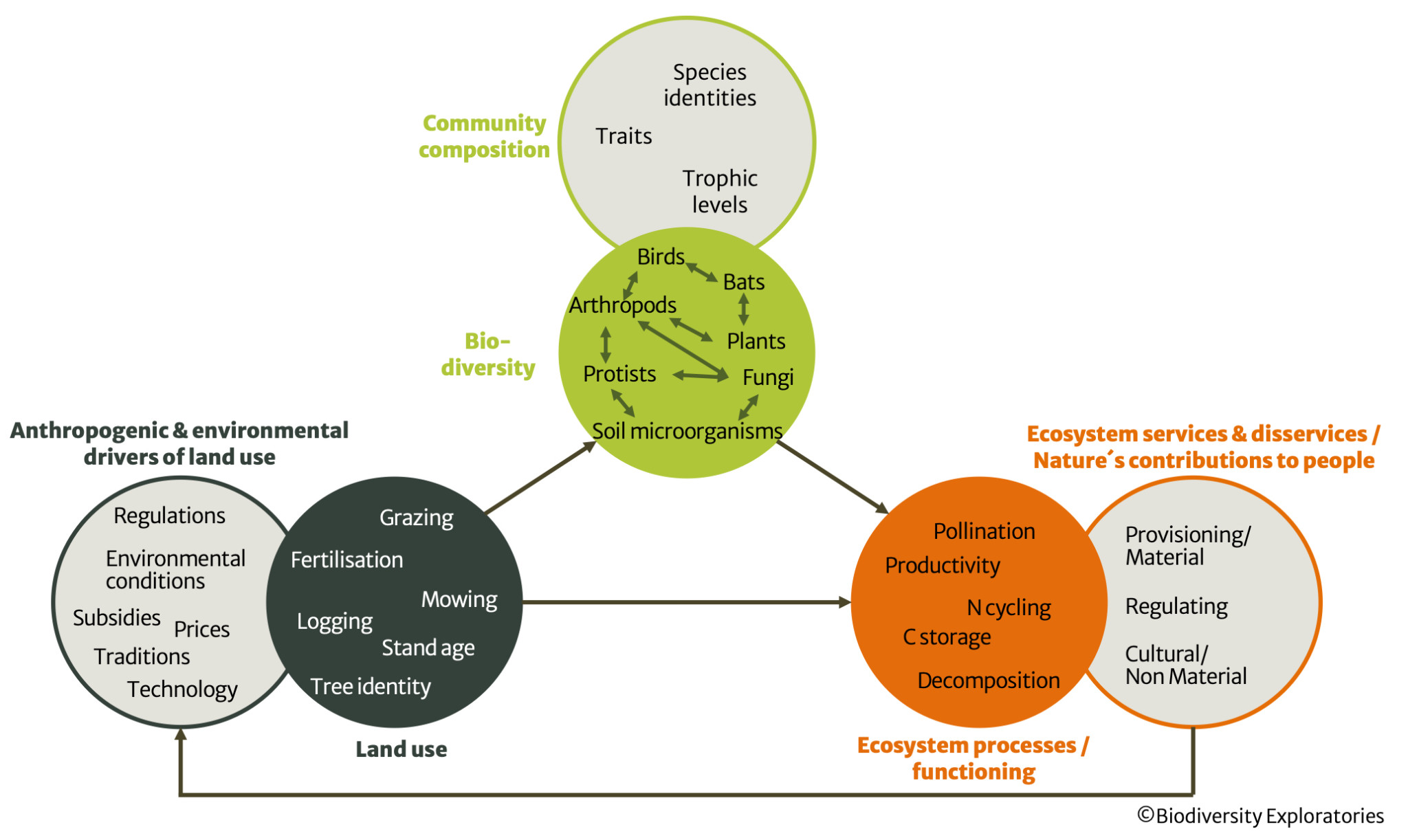}
   \caption{'Conceptual framework and basis of biodiversity exploratories.'\textsuperscript{\ref{fn:BEOBJ}}}
   \label{fig:BE}
 \end{figure}

Figure \ref{fig:BE} presents the conceptual framework and basis of research conducted in the BE\textsuperscript{\ref{fn:BEOBJ}}. 
The coloured circles at the centre of the figure show how experiments aim to gain insight on how land use affects biodiversity and ecosystem processes and functioning, and how biodiversity itself affects ecosystems.
On the outside, three further research interests are distinguished: How are communities composed; what are ecosystem services and disservices, and what are nature's contributions to people; and thirdly, how does that link back to the anthropogenic and environmental drivers of land use.
In the following sections, we present the components of the BE that make this research possible.

\section{Exploratories}
Research is conducted on three big regions across Germany.
In all three regions, a network of field plots is established in grasslands and forests, and of different land use types and intensities \cite{fischer2010implementing}.
The collection of all plots within a specific region is called an exploratory \cite{fischer2010implementing}.
We quote more information about the exploratories below:

\textit{'(1) the UNESCO Biosphere Reserve Schorfheide-Chorin, which is situated in the lowlands of North-eastern Germany, a young glacial landscape with many wetlands, (2) the National Park Hainich and its surrounding areas, situated in the hilly lands of Central Germany, and (3) the UNESCO Biosphere Reserve Schwäbische Alb (Swabian Jura), which is situated in the low mountain ranges of South-western Germany'} \cite{fischer2010implementing}.

We introduce the following abbreviations for the exploratories for further use in this thesis: 
Schorfheide-Chorin (SCH), Hainich-Dün (HAI), and Schwäbische Alb (ALB).

\section{Plots and Plot Types}
There are four types of plots within each exploratory, and all plot types have a standardized number of plots within each type. We quote the description of plot types below:

\textit{'(1) about 1000 grid plots (GPs), 500 in grasslands and 500 in forests, which are mainly used for large-scale analyses of biodiversity data and their relationships to land use and other environmental factors, (2) 100 experimental plots (EPs), 50 in grasslands and 50 in forests, which are a selected subset of the respective grid plots serving as a platform for more thorough biodiversity assessment and environmental monitoring, as well as for several manipulative experiments, and (3) 18 very intensive plots (VIPs), half in grasslands and half in forests, which are a subset of the experimental plots used for studying  biodiversity or ecological processes in extreme detail for requiring very labour-intensive methods, for which the use of the experimental plots is not feasible. In one of the exploratories (Hainich-Dün), where selection forests are an especially important forest type, we established three further very intensive plots in these forests. Altogether, the Biodiversity Exploratories comprise 57 very intensive plots, 300 experimental plots, and some 3000 grid plots'} \cite{fischer2010implementing}.

The fourth plot type not mentioned in the quote above are plots of medium research intensity (MIPs) that are categorized between the experimental plots, and the very intensive research plots.

\section{Joint Experiments and Types}
In addition to the plot categories described above, three new kinds of experiments called joint multi-site experiments were introduced in 2020.
These experiments are conducted on experimental plots (EPs).

\subsection{Reduced Land-use intensity Experiment (REX)}
The first experiment type is concerned with the question of \textit{'whether biodiversity and ecosystem functions of intensively used grasslands can be restored by reducing land use intensity'.}\footnote{\label{fn:JOINTEXP}\url{https://www.biodiversity-exploratories.de/en/about-us/research-design/}}
Subjects of this experiment are a subset of 15 grassland sites per exploratory with a substantial decrease in land use. 
These 45 sites are part of design one of the REX experiments, REX I.

For the second experiment design, REX II, a subgroup of 16 from the 45 sites was taken to test additional parameters related to the sowing of new and/or old species (or no sowing) and the effects this has on productivity. 

\subsection{Land-use Experiment (LUX)}
The second experiment type is concerned with the question of \textit{'how changes in land use (mowing, grazing and fertilisation) affect the diversity and functions of the ecosystem'.}\textsuperscript{\ref{fn:JOINTEXP}}
For this purpose, additional plots were established.
There, fertilization was discontinued while other land use components continued.

\subsection{Forest Gap Experiment (FOX)}
The third experiment type aims to \textit{'analyze the effects of creating gaps with and without deadwood enrichment as targeted experimental interventions in 29 forests with different management, biodiversity and ecosystem functions.'}\textsuperscript{\ref{fn:JOINTEXP}}
Specifically, they investigate changes in abiotic conditions and the availability of biotic resources through a variety of experiment variants and additional experimental plots.

\section{Projects and Infrastructures within the BE}
Research conducted in the Biodiversity Exploratories is categorized into few core and a multitude of contributing projects and cooperation projects. 
The six original core projects are described as follows:

\textit{'Starting in 2006, six core projects carried out the environmental, soil and land use inventories described above, selected experimental and very intensive plots, and implemented the necessary technical and administrative infrastructure for the project. Moreover, the core projects are ensuring consistent long-term monitoring of the species diversity of selected taxa of plants, fungi, birds, bats and other mammals, selected insect taxa and microorganisms, and they assess genetic diversity of selected taxa of plants and microorganisms. The core projects address selected ecosystem processes and services including pools of biomass and carbon, pollination services, predation, seed dispersal and community stability and are also carrying out experimental studies. These include experiments manipulating deadwood availability in the very intensive forest plots, fencing off subplots of all forest experimental plots to exclude large herbivores, and increasing plant species diversity by seeding in subplots of the grassland experimental plots'} \cite{fischer2010implementing}.

In the publication and dataset metadata used as a starting point for this work, these core projects are also called \textit{infrastructures}, and six types are distinguished (Note that the core projects have since been expanded to ten projects\footnote{\url{https://www.biodiversity-exploratories.de/en/about-us/infrastructure/further-core-projects/}}, but only six are distinguished in the metadata:
\begin{itemize}
    \item Central Coordination / Local Management
    \item Data Management
    \item Inventory / Campaign
    \item Instrumentation / Remote Sensing
    \item Synthesis
    \item Theory / Modeling / Methodological approaches
\end{itemize}

The core projects of the BE deliver the foundation, such that contributing and cooperation projects can achieve their goals to \textit{'investigate the cause-and-effect relations in the enlivened nature, both, by monitoring as well as experiments'}\footnote{\url{https://www.biodiversity-exploratories.de/en/projects/}}, and address \textit{the
genetic and species diversity of many taxa, including soil
organisms, ecosystem processes related to nutrients, food
webs and pollination webs, and modelling of species interactions and ecosystem processes in relation to land use} \cite{fischer2010implementing}.

\section{BExIS - The Biodiversity Exploratories Information System}\label{sec:bexis}
To appropriately manage the research data associated with the BE, the core project related to data management conceived the research data management system BExIS, \emph{The Biodiversity Exploratories Information System}\footnote{\url{https://www.bexis.uni-jena.de/}}.
The technical implementation of BExIS2 (an later version of the system) is detailed in \cite{chamanara2021bexis2}.

BExIS not only makes metadata and data for publications and datasets of the BE available and delivers a search interface over its various data sources, but it also delivers functionalities for researchers to upload their data, an API interface, and various helpful tools for researchers.

We pull all publication and dataset metadata from the BExIS API for further processing. In the following sections, we will present the metadata variables and structure in detail.

\section{Abstract Volumes}
The Biodiversity Exploratories provide a variety of media to communicate work created within them for outreach and research communication purposes.
One part of this material relevant for this thesis are so called abstract volumes:

\textit{'The abstract volume comprises a collection of summaries (‘abstracts’) of all scientific publications of the Biodiversity Exploratories in German. The texts are subdivided into individual topics. This is intended to provide an overview of the individual projects and their results'}.\footnote{\url{https://www.biodiversity-exploratories.de/en/public-releases/public-relations-material/}}
Three of these volumes exist, ranging from the years of 2013-2016, 2017-2019, and 2020-2022.
For each publication contained in a volume, a double page is reserved on which authors are able to communicate their research in simplified german language, and supply relevant photographs or figures.

Originally, the core idea of this thesis was to extract information from the PDF files of these abstract volumes, and create a knowledge graph from these sources, however, upon further inspection, BExIS already provides this data in a convenient fashion, which is why we decided to forego this step and focus on data obtainable via the BExIS API directly.

\section{Metadata Presentation}
In the final section of this chapter, we present the data used as basis for the knowledge graph in detail. Data and metadata stored in BExIS follow schemata with complete descriptions and definitions for all fields. These metadata schemata were provided by the BExIS developers\footnote{\url{https://github.com/BEXIS2}} and the information presented below is quoted from those files.

Metadata for both sources has a hierarchical structure. 
For Publications, there are three main categories, \textit{Publication Details}, \textit{Exploratory Objects}, and \textit{Domain Objects}.
Datasets on the other hand, have the main categories \textit{General}, \textit{Contacts}, \textit{Coverage}, \textit{Keywords}, and \textit{Data Types}.

\subsection{Publications}

\textbf{Publication Details:}

\begin{description}[style=nextline, labelwidth=4cm, labelsep=1em, leftmargin=5.5cm]
  \item[\textbf{Title}] 
    \tagbadge{stringNotEmptyLong} \\
    Title of the publication (in English).

  \item[\textbf{Citation}] 
    \tagbadge{stringNotEmptyLong} \\
    Full citation string of the publication.

  \item[\textbf{First Author}] 
    \tagbadge{stringNotEmptyShort} \\
    Main Author

  \item[\textbf{Corresponding E-Mail}] 
    \tagbadge{stringNotEmptyShort} \\
    The email address of the author.

  \item[\textbf{Year}] 
    \tagbadge{stringNotEmptyShort} \\
    Year of publication.

  \item[\textbf{Co-Authors}] 
    \tagbadge{stringNotEmptyLong} \\
    Co-author list.

  \item[\textbf{Institutions}] 
    \tagbadge{stringNotEmptyLong} \\
    Involved institutions.  
    
  \item[\textbf{Project}] 
    \tagbadge{stringNotEmptyShort} \tagbadge{minOccurs="1"}
    \tagbadge{maxOccurs="unbounded"}\\
    Associated project within the DFG-funded joint project Biodiversity Exploratories.
    
  \item[\textbf{Status}] 
    \tagbadge{Submitted} \tagbadge{In press} \tagbadge{Published}\\
    Status of the publication. 

  \item[\textbf{Open Access}] 
    \tagbadge{booleanYesNoUnknown} \\
    Information on open access. 

  \item[\textbf{Type of Publication}] 
    \tagbadge{Technical report} \tagbadge{Paper} \tagbadge{Book} \tagbadge{Book section} \tagbadge{Thesis} \tagbadge{Conference Proceedings} \tagbadge{German article} \\
    Type of Publication. 

  \item[\textbf{Published In}] 
    \tagbadge{stringNotEmptyShort} \\
    Information on where it was published (e.g. journal name). 

  \item[\textbf{DOI}] 
    \tagbadge{stringNotEmptyShort}\\
    DOI link of the publication. 

  \item[\textbf{URL}] 
    \tagbadge{stringEmptyLong}\\
    URL link to the publication.

  \item[\textbf{Keywords}] 
    \tagbadge{stringEmptyLong} \tagbadge{minOccurs="1"} \\
    Title of the publication (in English).

  \item[\textbf{Summary}] 
    \tagbadge{stringEmptyLong} \\
    Summary of the publication (in English).

  \item[\textbf{Metadata Schema}] 
    \tagbadge{stringEmptyShort} \\
    Name and version of metadata schema.

  \item[\textbf{German Title}] 
    \tagbadge{stringEmptyLong} \\
    Title of the publication (in German).

  \item[\textbf{German Summary}] 
    \tagbadge{stringEmptyLong} \\
    Summary of the publication (in German).  
\end{description}

\textbf{Exploratory Objects:}

\begin{description}[style=nextline, labelwidth=4cm, labelsep=1em, leftmargin=5.5cm]
  \item[\textbf{Exploratory}] 
    \tagbadge{ALB} \tagbadge{booleanYesNo}
    \tagbadge{SCH} \tagbadge{booleanYesNo}\\
    \tagbadge{HAI} \tagbadge{booleanYesNo}\\
    Exploratory covered by the described study.

  \item[\textbf{Plot Level}] 
    \tagbadge{GP (all)} \tagbadge{GP (few)} \tagbadge{EP (all)} \tagbadge{EP (few)} \\ \tagbadge{MIP (all)} \tagbadge{MIP (few)} \tagbadge{VIP (all)} \tagbadge{VIP (few)} \tagbadge{none}\\
    Plot level (none = the study did not take place in the field).

  \item[\textbf{Joint Experiment 2020}] 
    \tagbadge{REX 1} \tagbadge{REX 2} \tagbadge{LUX} \tagbadge{FOX}\\
    Joint Experiment 2020 includes all experimental plots installed in 2020 in addition to the 300 EPs. Three grassland (REX 1 and 2, and LUX) and one forest experiment (FOX) are distinguished.

  \item[\textbf{Habitats}] 
    \tagbadge{Grassland} \tagbadge{booleanYesNo}
    \tagbadge{Forest} \tagbadge{booleanYesNo}\\
    Observed habitat (grassland or forest).

  \item[\textbf{Relative position to ground}] 
    \tagbadge{Aboveground} \tagbadge{booleanYesNo}
    \tagbadge{Belowground} \tagbadge{booleanYesNo}\\
    Location of the observations - below or/and above ground.

  \item[\textbf{Data Source}] 
    \tagbadge{Field} \tagbadge{booleanYesNo}
    \tagbadge{Lab} \tagbadge{booleanYesNo}\\
    \tagbadge{Review} \tagbadge{booleanYesNo}\\
    Source of the used data (field, laboratory, review).

  \item[\textbf{Infrastructures}] 
    \tagbadge{Central coordination/LMT} \tagbadge{Data management}\\ \tagbadge{Instrumentation/Remote sensing} \tagbadge{Inventory/Campaign} \tagbadge{Synthesis} \tagbadge{Theory/Modeling/Methodological approaches} \tagbadge{none}\\
    Infrastructure type the publication belongs to.  
\end{description}

\textbf{Domain Objects:}

\begin{description}[style=nextline, labelwidth=4cm, labelsep=1em, leftmargin=5.5cm]
  \item[\textbf{Biotic Data Taxa}] 
    \tagbadge{Animals - Invertebrates} \tagbadge{Animals - Vertebrates} \\
    \tagbadge{Fungi} \tagbadge{Microbes}
    \tagbadge{Plants} \tagbadge{none}\\
    The taxa covered by the publication.

    \item[\textbf{Biotic Data Types}] 
    \tagbadge{Single Species} \tagbadge{Multiple Species}\\
    \tagbadge{Genetic} \tagbadge{Total Abundance}
    \tagbadge{None}\\
    Selection of biotic data types (none = biotic data type is not listed or no biotic data described in the publication).

    \item[\textbf{Environmental Descriptors}] 
    \tagbadge{Land use (managed)} \tagbadge{Land use (unmanaged)}\\
    \tagbadge{Soil} \tagbadge{Climate}\tagbadge{none}\\
    List of environmental descriptions to characterize the observations in the publication (none = no appropriate descriptor listed or publication can not be assigned to an environmental descriptor).

    \item[\textbf{Processes and Services}] 
    \tagbadge{Biogeochemical cycle} \tagbadge{Decomposition}
    \tagbadge{Productivity} \tagbadge{Herbivory}
    \tagbadge{Pollination} \tagbadge{Predation/Parasitism}
    \tagbadge{Dispersal} \tagbadge{none}\\
    Selection of ecosystem processes and service classes covered by the publication (none = appropriate process or service is not predefined or publication describes no ecosystem processes or services).
\end{description}

\subsection{Datasets}
\textbf{General:}

\begin{description}[style=nextline, labelwidth=4cm, labelsep=1em, leftmargin=5.5cm]
    \item[\textbf{Title}] 
    \tagbadge{stringNotEmptyShort} \\
    Title of the dataset well describing the data (what, where, when).

    \item[\textbf{Abstract}] 
    \tagbadge{stringNotEmptyLong} \\
    General description of the dataset including:(i) motivation and background, description of the study or analysis (ii) study site and timeframe (e.g. The 35 study sites are 50 by 50m plots located in the Hainich, Germany. The data were collected monthly from April 2014 to October 2019.) (iii) kind of data (e.g. species occurrence, GIS, OTUs, trait data) (iv) funding resources (e.g. The data collection/analysis was part of the project BIODIV within the framework of the DFG-funded joint project Biodiversity Exploratories (https://www.biodiversity-exploratories.de). Do not use abbreviations and do not include references to publications or methods.

    \item[\textbf{Project Name}] 
    \tagbadge{stringNotEmptyShort} \\
    https://www.biodiversity-exploratories.de/en/projects/ Associated project within the DFG-funded joint project Biodiversity Exploratories.

    \item[\textbf{Consortium}] 
    \tagbadge{stringEmptyShort} \\
    Name of a group, either consisting of: BE projects collaborating on data collection and analysis or topic-specific datasets. Please contact the BExIS team if you miss a consortium.

    \item[\textbf{Metadata Creation Date}] 
    \tagbadge{xsd:string} \\
    Entry date of the metadata.

    \item[\textbf{Metadata Last Modification Date}] 
    \tagbadge{xsd:string} \\
    Date on which the metadata was last modified.

    \item[\textbf{Last Modification Date}] 
    \tagbadge{xsd:string} \\
    Date on which the data was last modified.

    \item[\textbf{ID}] 
    \tagbadge{positiveInteger} \\
    Identifier of the dataset.

    \item[\textbf{Version}] 
    \tagbadge{positiveInteger} \\
    Version of the dataset.

    \item[\textbf{DOI}] 
    \tagbadge{stringEmptyShort} \\
    DOI (Digital Object Identifier) of the dataset.

    \item[\textbf{License}] 
    \tagbadge{stringEmptyShort} \\
    License short name, e.g. CC BY 4.0 (Creative Commons Attribution 4.0 International).

    \item[\textbf{Funder}] 
    \tagbadge{minOccurs="1"} \\
    Information about the funder provider.

    \item[\textbf{Funder Name}] 
    \tagbadge{stringEmptyShort} \\
    Name of the funder provider, e.g. German Research Foundation.

    \item[\textbf{Funder Identifier}] 
    \tagbadge{stringEmptyShort} \\
    Funder identifier, e.g.https://ror.org/018mejw64 for the German Research Foundation issued by the Research Organization Registry (ROR).

\end{description}

\textbf{Contacts:}

\begin{description}[style=nextline, labelwidth=4cm, labelsep=1em, leftmargin=5.5cm]

    \item[\textbf{Metadata Creator}] 
    \tagbadge{stringNotEmptyShort} \\
    The person filling out the metadata. They have write permissions.

    \item[\textbf{Data Creator}] 
    \tagbadge{stringNotEmptyShort} \\
    Main person(s) involved in producing the data. They have write permissions. Only they are taken for the citation string.

    \item[\textbf{Data Collector}] 
    \tagbadge{stringNotEmptyShort} \\
    Person(s) responsible for gathering/collecting data under the guidelines of the data creator or PI. The collector and creator can be the same person. In assembled datasets, data collector(s) are the contact person(s) of the dataset(s) contributing data to the new dataset.

    \item[\textbf{Project Leader}] 
    \tagbadge{stringNotEmptyShort} \\
    The PI(s) of the project. They have write permissions. PIs are not part of the citation string.    

    \item[\textbf{Contact Person}] 
    \tagbadge{stringNotEmptyShort} \\
    The person managing and granting access to the dataset. In general, the contact person is the PI of the project or a person designated by the PI. They have write permissions.

    \item[\textbf{Contact E-mail}] 
    \tagbadge{xsd:pattern} \\
    The email address of the contact person.

    \item[\textbf{Institute}] 
    \tagbadge{stringNotEmptyShort} \\
    The institute of the contact person.

    \item[\textbf{Street}] 
    \tagbadge{stringNotEmptyShort} \\
    The street of the institute.

    \item[\textbf{Postal Code}] 
    \tagbadge{stringNotEmptyShort} \\
    The postal code of the institute.

    \item[\textbf{City}] 
    \tagbadge{stringNotEmptyShort} \\
    The city of the institute.

    \item[\textbf{Phone}] 
    \tagbadge{stringEmptyShort} \\
    The phone number of the contact person.

    \item[\textbf{Mobile}] 
    \tagbadge{stringEmptyShort} \\
    The mobile phone number of the contact person.

    \item[\textbf{URL}] 
    \tagbadge{stringNotEmptyLong} \\
    The web page of the contact person, the project or the institute.    
\end{description}

\textbf{Methods:}

\begin{description}[style=nextline, labelwidth=4cm, labelsep=1em, leftmargin=5.5cm]
\item[\textbf{Study Design}] Description of the study design.

    \item[\textbf{Experimental Manipulation}] 
    \tagbadge{booleanYesNo} \\
    Indicates experiments with manipulations on the study area or site, e.g. fertilization or fencing. Further information should be provided in the section 'Detailed description of study design'.

    \item[\textbf{Temporal Repetitions per Study Site}] 
    \tagbadge{booleanYesNo} \\
    Indicates if measurements were repeated over time, e.g. every day, weekly, or yearly. Further information should be provided in the section 'Detailed description of study design'.

    \item[\textbf{Plot Based Repetitions of Measurements}] 
    \tagbadge{booleanYesNo} \\
    Indicates if measurements were repeated on subplots (within a plot). Further information should be provided in the section 'Detailed description of study design'.

    \item[\textbf{Detailed Description of Study Design}] 
    \tagbadge{stringEmptyLong} \\
    Detailed description of the study design (e.g. number and size of plots or transects) together with details on temporal and spatial repetitions.

    \item[\textbf{Measurements}] 

    \item[\textbf{Sampling Design}] 
    \tagbadge{stringEmptyLong} \\
    Detailed description of the sampling design at EP level (e.g. location of transects, study plots, soil cores, or devices) and the number of samples taken at an EP. Furthermore, a detailed description of the sampling methods (e.g. size of the soil core) is provided.

    \item[\textbf{Sample Preparation}] 
    \tagbadge{stringEmptyLong} \\
    Description of the further preparation of samples before analysis (e.g. dry biomass, ground soils, freeze samples).

    \item[\textbf{Sample Analysis}] 
    \tagbadge{stringEmptyLong} \\
    Description of all measurements in the field or lab (e.g. pH, sequencing, trait measurements, soil analysis).

    \item[\textbf{Equipment}] 
    \tagbadge{stringEmptyLong} \\
    Equipment used for sample collection, measurements, and analysis (including calibration method of equipment).

    \item[\textbf{Data Processing}] 
    Description of the analytical and statistical analysis of the data.

    \item[\textbf{Data Preparation}] 
    \tagbadge{stringEmptyLong} \\
    Steps of pre-processing of raw data before data analysis (e.g. data fusion, data cleaning).

    \item[\textbf{Data Analysis}] 
    \tagbadge{stringEmptyLong} \\
    Description of the data analyis (e.g. statistical methods, indices, programs used).

    \item[\textbf{References}] 

    \item[\textbf{Literature References}] 
    \tagbadge{stringEmptyLong} \\
    List of references mentioned in the method section. Citation sample: First author et al. (year). Title. Journal. DOI.

    \item[\textbf{Data References}] 
    \tagbadge{stringEmptyLong} \\
    List of datasets referred to in the method section.

    \item[\textbf{Acronyms}] Acronyms used in the method section and their meaning.

    \item[\textbf{Important Notes}] 
    \tagbadge{stringEmptyLong} \\
    Very important information concerning the use of the primary data.   
\end{description}

\textbf{Keywords:}

\begin{description}[style=nextline, labelwidth=4cm, labelsep=1em, leftmargin=5.5cm]
\item[\textbf{Environmental Descriptors}] 
    \tagbadge{Land Use} \tagbadge{Soil} \tagbadge{Climate} \tagbadge{Habitat structure} \tagbadge{Spatial} \tagbadge{Other} \tagbadge{None}\\
    List of environmental descriptions to characterize the observations in the dataset (other =  no appropriate descriptor listed, none = dataset can not be assigned to an environmental descriptor, e.g. R script).

    \item[\textbf{Processes and Services}] 
    \tagbadge{Productivity} \tagbadge{Decomposition}\\ \tagbadge{Biogeochemical cycle - Water}\\ \tagbadge{Biogeochemical cycle - Carbon}\\ \tagbadge{Biogeochemical cycle - Nitrogen}\\ \tagbadge{Biogeochemical cycle - Other nutrients} \\
    \tagbadge{Ecological process - Herbivory} \\ \tagbadge{Ecological process - Soil aggregation}\\ \tagbadge{Ecological process - Mycorrhization}\\ \tagbadge{Ecological process - Pollination}\\ \tagbadge{Ecological process - Predation/Parasitism}\\ \tagbadge{Ecological process - Dispersal}\\ \tagbadge{Function/Service - Pest control}\\ \tagbadge{Function/Service - Aesthetic} \tagbadge{Other} \tagbadge{None}\\
    Selection of ecosystem process and service classes covered by the dataset (other = appropriate process or service is not predefined, none = dataset describes no ecosystem process or service).

    \item[\textbf{Biotic Data Types}] 
    \tagbadge{Single species} \tagbadge{Multiple species} \tagbadge{Abundance}\\ \tagbadge{Total abundance} \tagbadge{Presence/Absence} \tagbadge{Aggregated diversity} \tagbadge{Demographic} \tagbadge{Morphological} \tagbadge{Chemical} \tagbadge{Genetic} \tagbadge{Trait} \tagbadge{Other} \tagbadge{None}\\
    Selection of biotic data types (other = biotic data type is not listed, none = no biotic data in the dataset).

    \item[\textbf{Habitat}]
    Observed habitat (grassland or forest).

    \item[\textbf{Grassland}] 
    \tagbadge{booleanYesNo} \\
    .

    \item[\textbf{Forest}] 
    \tagbadge{booleanYesNo} \\
    .

    \item[\textbf{relativePositionToGround}]
    Location of the observations - below or/and above ground.

    \item[\textbf{Aboveground}] 
    \tagbadge{booleanYesNo} \\

    \item[\textbf{Belowground}] 
    \tagbadge{booleanYesNo} \\

    \item[\textbf{Additional Keywords}] 
    \tagbadge{stringNotEmptyShort} \\
    Further keywords describing the dataset.    
\end{description}

\textbf{Coverage:}

\begin{description}[style=nextline, labelwidth=4cm, labelsep=1em, leftmargin=5.5cm]
\item[\textbf{Temporal Coverage}] Time period covered by the dataset.

    \item[\textbf{Start Date}] 
    \tagbadge{xsd:date} \\
    Start date of the time period covered by the dataset.

    \item[\textbf{End Date}] 
    \tagbadge{xsd:date} \\
    End date of the time period covered by the dataset. 

    \item[\textbf{Geographic Coverage}] Spatial extent of the whole dataset and/or number of plots included in the dataset. 

    \item[\textbf{Number of Plots}] Number of investigated plots classified by plot level (GP, EP, MIP, VIP). 

    \item[\textbf{Number of GP}] 
    \tagbadge{positiveInteger} \\
    Number of investigated GPs. Usually, it is equal to the number of EPs. If the entire exploratory is covered, the following numbers apply (all: 3967, ALB: 1150, HAI: 1656, SCH: 1161). 

    \item[\textbf{Number of EP}] 
    \tagbadge{positiveInteger} \\
    Number of investigated EPs (all: 300). Each exploratory has 100 EPs (50 grassland, 50 forest). 

    \item[\textbf{Number of MIP}] 
    \tagbadge{positiveInteger} \\
    Number of investigated MIPs (all: 150). Each exploratory has 50 MIPs (25 grassland, 25 forest). 

    \item[\textbf{Number of VIP}] 
    \tagbadge{positiveInteger} \\
    Number of investigated VIPs (all: 57). Each exploratory has 18 VIPs (9 grassland, 9 forest, but in the Hainich there are 3 additional forest plots). 

    \item[\textbf{Joint Experiment 2020}] Joint Experiment 2020 includes all experimental plots installed in 2020 in addition to the 300 EP. Three grassland (REX 1 and 2, and LUX) and one forest experiment (FOX) are distinguished. REX 1: Plots of the restoration experiment I in grassland. REX 2: Plots of the restoration experiment II in grassland. LUX: Plots of the land-use experiment in grassland. FOX: Plots of the gap experiment in the forest. Datasets can include different joint experiments. 

    \item[\textbf{Experiment Type}] 
    \tagbadge{REX1} \tagbadge{REX2} \tagbadge{LUX} \tagbadge{FOX}\\
    . 

    \item[\textbf{Coordinates WGS84}] Point locations are provided as longitude and latitude (geographical coordinates) in degree (WGS84). 

    \item[\textbf{Longitude}] 
    \tagbadge{xsd:decimal} \\

    \item[\textbf{Latitude}] 
    \tagbadge{xsd:decimal} \\

    \item[\textbf{Bounding Box}] Covered area described by the four edge coordinates in degree (WGS84). 

    \item[\textbf{Westernmost Longitude}] 
    \tagbadge{xsd:decimal} \\

    \item[\textbf{Northernmost Latitude}] 
    \tagbadge{xsd:decimal} \\

    \item[\textbf{Easternmost Longitude}] 
    \tagbadge{xsd:decimal} \\

    \item[\textbf{Southernmost Latitude}] 
    \tagbadge{xsd:decimal} \\

    \item[\textbf{Taxonomic Coverage}] Taxonomic groups investigated in this dataset.

    \item[\textbf{Taxon}] 
    \tagbadge{Microbes} \tagbadge{Fungi} \tagbadge{Plants} \tagbadge{Animals - Vertebrates}\\ \tagbadge{Animals - Invertebrates} \tagbadge{None}\\
    The taxa covered by the dataset.   
\end{description}

\textbf{Data Type and Status:}

\begin{description}[style=nextline, labelwidth=4cm, labelsep=1em, leftmargin=5.5cm]
    \item[\textbf{Data Processing Status}] 
    \tagbadge{Raw} \tagbadge{Processed} \tagbadge{Assembled}\\
    Status of data processing. Raw = original data without any processing. Processed = data was processed (e.g., validated, aggregated, classified). Assembled data = data is originally part of several datasets.

    \item[\textbf{Data Upload Status}] 
    \tagbadge{Incomplete} \tagbadge{Ongoing} \tagbadge{Complete} \\
    Status of data uploading. Complete = all data is uploaded. Incomplete = more data will be uploaded in the next weeks. Ongoing = dataset contains a time series; data will be continuously uploaded, e.g. yearly. 

    \item[\textbf{Content Type}] 
    \tagbadge{Structured Data} \tagbadge{Unstructured Data} \tagbadge{GIS} \tagbadge{Model/Scripts} \tagbadge{External Link}\\
    The content type of the data. Structured data= tables. Unstructured data = further data types, e.g. documents, images. GIS = GIS file formats encoding geographical information. Models/Scripts = code of models or scripts (e.g., R scripts). External link = data is stored in an external repository.

    \item[\textbf{Repository}] Information about the repository hosting the primary data.

    \item[\textbf{Name of Repository}] 
    \tagbadge{stringEmptyShort} \\
    Name of the repository hosting the primary data (For example, NCBI).

    \item[\textbf{URL}] 
    \tagbadge{stringNotEmptyLong} \\
    The URL of the repository. 

    \item[\textbf{Identifier}] 
    \tagbadge{stringEmptyLong} \\
    Identifier or DOI link of the described datasets.

    \item[\textbf{Additional Information}] 
    \tagbadge{stringEmptyLong} \\
    Additional information about finding, accessing, and downloading the data. 
\end{description} 
\chapter{Method - Problem Statement, Research Question, and Method Overview} \label{chapter5}
The method section of this work is split into three parts to provide better structure to the different aspects of our approach.
In the previous chapter we detailed different aspects of the Biodiversity Exploratories, its associated experiments, projects, infrastructures, BExIS, and the publication and dataset metadata used as a starting point for our methodology.
In this chapter, we restate the main research problem, formulate the main research questions, and give an overview of the chapters and sections to follow.

\section{Problem Statement}
\begin{itemize}
    \item \textit{There is a fundamental mismatch between the requirements that ontology and knowledge graph engineers must satisfy to comply with best practices regarding data governance, technical requirements, and enabling machine interoperability, and the requirements that users of these technologies impose on their ability to provide semantically meaningful statements to convey information.}
\end{itemize}

To address this problem, we first model and construct a knowledge graph of publication and dataset metadata of the Biodiversity Exploratories.
Afterwards, we implement semantic units for this knowledge graph that may provide solutions to bridge the gap between requirements, allowing users to query for semantically meaningful information easier.

We also investigate an LLM approach to extract structured metadata categories from publication and dataset titles and abstracts to possibly support KG creation workflows, and aid researchers in creating structured metadata.

Finally, in a second LLM application, we aim to leverage embedding models to extract latent topics from publication and dataset titles and abstracts, categorize documents by providing topic hierarchies as anchors in an embedding space, and link documents to the overarching research goals of the Biodiversity Exploratories to further enrich BE metadata.

\section{Research Question}
To operationalize our investigation into solutions of these problems, we formulate three research questions below:
\begin{itemize}
    \item \textbf{RQ1:} Does semantic unit modelling reduce SPARQL query complexity for representative tasks compared to querying the data graph?
    \item \textbf{RQ2:} Can large language models be used to extract metadata categories from publication and dataset titles and abstracts?
    \item \textbf{RQ3:} Can embedding models be used to extract latent information from publication and dataset titles and abstracts to enrich BE metadata?
\end{itemize}
We further split the research questions into sub-questions:
\begin{sloppypar}
\begin{itemize}
    \item \textbf{SQ1:} Can SPARQL queries on the constructed knowledge graph answer sets of competency questions?
    \item \textbf{SQ2:} Can the same SPARQL queries be formulated over semantic units to answer the same sets of competency questions?
    \item \textbf{SQ3:} Is SPARQL query complexity lower for queries that answer competency question when querying over semantic units compared to queries on the data graph only?
    \item \textbf{SQ4:} Which metadata categories can the LLM extraction approach predict reliably (F1-score of 0.8 or higher)?
    \item \textbf{SQ5:} Can an embedding approach assign publication and dataset titles and abstracts to the research goals of the BE?
\end{itemize}
\end{sloppypar}
The competency questions mentioned above will be listed in Sec. \ref{sec:compq}.

\section{Method Overview}
The methodology of this thesis consists of two main parts: 
\begin{itemize}
    \item \textbf{(1)} The modelling and construction of a knowledge graph for publication and dataset metadata associated with the Biodiversity Exploratories and an implementation of semantic units for better cognitive interoperability between users and the knowledge graph.
    \item \textbf{(2)} An investigation of LLM approaches for information extraction on two tasks:
    First, the extraction of structured metadata categories from publication and dataset titles and abstracts and second, extraction of latent information on the same titles and abstracts using an embedding model.
\end{itemize}

In the following sections we will give a short presentation of what contents are included in both parts of the methodology.

\subsection{Knowledge Graph Engineering}
The first part of our methodology, which also presents the main contribution of this work, consists of the modelling and construction of a knowledge graph for publication and dataset metadata of the Biodiversity Exploratories.
At first, we derive competency questions that will allow evaluation of the resulting knowledge graph.
Since we want to explore the potential of semantic units for easier interaction with the knowledge graph, we will show how they can be constructed to answer competency questions directly.

In the next section, we present the process of modelling in detail and show the results and intricacies for publications and datasets, presenting the terms chosen as mappings from BExIS metadata to ontology vocabulary.
Thereafter, a discussion of the substantial modelling challenges will follow, which contains challenges for negations, the open world assumption, negations, and challenges derived from the source data.

Following the discussion of modelling challenges, we describe the implementation in detail, covering the mapping language used, how semantic units can be implemented as named subgraphs, and present possible extensions for schema validation.

\subsection{Information Extraction using Large Language Models}

The second part of the methodology section covers information extraction using large language models on publication and dataset titles and abstracts for two tasks.

In the first task we investigate if an LLM can predict a list of structured metadata categories if given an extraction schema consisting of the category name, definition, and a list of allowed answers.
We present the instructions sent to the LLM, and show the results for multiple sets of categories, those that allow only yes or no answers, and those that allow a selection from multiple keywords.
We evaluate the model's predictions using precision, recall, F1-scores, accuracy, and, if applicable, the average jaccard distance.

For the second task, we investigate if an embedding model can be used to extract latent information from the input data for three subtasks.
First, we embed all titles and abstracts and attempt to identify clusters of documents in the embedding space, and list the top keywords per cluster.
Second, we provide a controlled vocabulary of concepts from outside our input data as anchors in the embedding space, and investigate which clusters emerge around anchor concepts in the embedding space.
Third, we attempt to link documents to the overarching research goals of the Biodiversity Exploratories by embedding goals and their descriptions, and applying the same clustering approach as in earlier tasks.
Finally, we conclude the chapter by visualizing the results of the final embedding task. 
\chapter{Method - Knowledge Graph and Semantic Unit Modelling} \label{chapter6}
In this chapter, we will present the methodology behind our approach.
With the goal of constructing a knowledge graph for publication and dataset metadata of the Biodiversity Exploratories in mind, we first present our approach for knowledge graph modelling and its application on our use case.
We then present the modelling process and its results, which leads into a discussion of the substantial modelling challenges faced due to the source data used, and general KG engineering challenges regarding negations and the open world assumption.
Afterwards, we present how semantic units can be developed on top of the constructed knowledge graph.
Finally, we discuss how SHACL shapes can be used for schema validation.

\section{Modelling Approach}
Our modelling approach for the knowledge graph and its schemata developed in this thesis aligns with the knowledge graph and ontology engineering approaches detailed in Sec. \ref{sec:KGENG}. 
With knowledge about the Biodiversity Exploratories, their components, and data (Ch. \ref{chapter4-5}) in mind, we approach graph modelling from multiple perspectives, combining insights from a top-down and bottom-up view, the domain the data originates from, and the end users of the graph and their requirements. 

\subsection{Top-down versus Bottom-up}\label{sec:topbottom}
In the top-down approach we focus on the schema level of the knowledge graph first, and consider existing ontologies and their specifications on how to represent data points and processes.
On closer inspection, our source data is modular in nature,
for example, publication metadata is split onto two main parts: (1) The metadata of the publication itself (who authored it etc.), and (2) Metadata in a schema specific to the Biodiversity Exploratories about the contents of a publication (e.g., is the publication about a grassland, forest, or both?).

Considering these modules, there exist a variety of domain specific and domain agnostic ontologies for the data we want to represent.
For example, \textit{The OpenCitations Data Model} \cite{garcia2013biotea, daquino2020opencitations} is a domain agnostic data model to represent publications, authors, publication parts and institutions, venues, and more. 
Therefore, it is a prime candidate for reuse for our purposes.
However, when reusing an existing data model, the top-down approach must contend with a question: Is it possible to fit source data to the model, and how can that fitting be achieved?

As for the bottom-up approach, our perspective changes.
We first look at the data we want to represent in the knowledge graph, and then consider the graph schema in a later step.
Questions that arise when looking at the data first are: What are the Biodiversity Exploratories and their components?
What is a plot and what are different plot categories?
Why do the metadata fields and categories exist in the metadata schemata for publications and datasets, and what information and purpose do these fields represent, and finally, how do we represent this in a knowledge graph?
This second step of representing data in a knowledge graph leads to substantial challenges that we discuss in Sec. \ref{sec:modellingChallenges}.

Both of these modelling perspectives do not take place in isolation, rather, they inform each other and lead to multiple iterations of the graph schema.

\subsection{Competency Questions}\label{sec:compq}
As discussed in Sec. \ref{sec:KGENG}, a knowledge graph is modelled using a problem-solving approach, meaning that it is directed towards an application and solves problems that arise inside this application.
For this specific use case, we develop a knowledge graph in the domain of biodiversity research for dataset and publication metadata.
To properly evaluate the knowledge graph later, we formulate competency questions in accordance with users that aim to cover a multitude of scenarios of questions users may have for the knowledge graph.
We also include competency questions for information that could be extracted using the embedding approach presented in the next chapter to outline application scenarios this extracted information were to be included in the KG:

\textbf{Competency Questions}

\begin{itemize}
    \item \textbf{CQ P1:} Who authored the publication \textbf{\emph{P}}?
    \item \textbf{CQ D1:} Who authored dataset \textbf{\emph{D}}?
    
    \item \textbf{CQ P2:} What datasets is publication \textbf{\emph{P}} linked to and what are the types of connections? 
    \item \textbf{CQ D2:} What publications is dataset \textbf{\emph{D}} linked to and what are the types of connections?
    
    \item \textbf{CQ P3:} For publication \textbf{\emph{P}}, what are the plot levels of the plots it investigated, and what exploratory or multiple exploratories do those plots cover?
    \item \textbf{CQ D3:} For dataset \textbf{\emph{D}}, what are the plot levels of the plots it investigated, what exploratory or multiple exploratories do those plots cover, and how many plots per plot level are investigated?
    
    \item \textbf{CQ P4:} What are all semantic units that have publication \textbf{\emph{P}} as subject, or that are associated with it?  
    \item \textbf{CQ D4:} What are all semantic units that have dataset \textbf{\emph{D}} as subject, or that are associated with it? 
    \item \textbf{CQ 5:} List all publications that belong to the infrastructure \textit{Instrumentation/Remote sensing} that investigate pollination of plants on grassland plots. What are the datasets they reference and what projects do those belong to?
\end{itemize}

\subsection{Connection between competency questions and semantic units}
As the central research question of this work aims to investigate how semantic units can be used for simpler interactions between users and a knowledge graph, we observe a link that can be made between competency questions and semantic units:
Many of the competency question listed above can be answered simpler and directly with compound units. 
Once the knowledge graph is able to answer these questions, compound units can be defined to bundle relevant statements.


\section{The process of modelling BE Publication and Dataset Metadata}
In this section, we present the process of modelling a knowledge graph for publication and dataset metadata of the Biodiversity Exploratories in detail and show the modelled graph schemata for different modules of the graph.
We present not only finished schemata, but also aim to display the process of modelling by showing ideas that were discontinued or changed substantially over the course of this work.

The majority of modelling work took place on a Miro\footnote{\url{https://miro.com/}} board that we also provide on our GitHub repository\footnote{\url{https://github.com/fusion-jena/Biodiversity-Exploratories-Knowledge-Graph}}.

\subsection{Ontology Reuse}
We prioritize reusing existing ontology terms whenever possible.
For this we mainly use two search portals to lookup ontology vocabulary: \textit{BioPortal}\footnote{\url{https://bioportal.bioontology.org/}} \cite{noy2009bioportal}, a service to search for biomedical ontologies and terms, and the \textit{Ontology Lookup service}\footnote{\url{https://www.ebi.ac.uk/ols4/}} \cite{jupp2015new}.
By searching for terms or browsing specific ontologies, we are able to decide on ontology vocabulary depending on how fitting a term's definition, description, or usage instructions are for our modelling purposes.


\subsection{First Steps}
In the first stages of modelling we focused on three core parts of what the knowledge graph should be able to represent: A publication and metadata associated with it, a dataset and metadata associated with it, and the process of extracting information from a free text field via LLM and the products of that extraction.

\begin{landscape}
\begin{figure}[p]
  \centering
  \includegraphics[width=\linewidth,height=\linewidth,keepaspectratio]{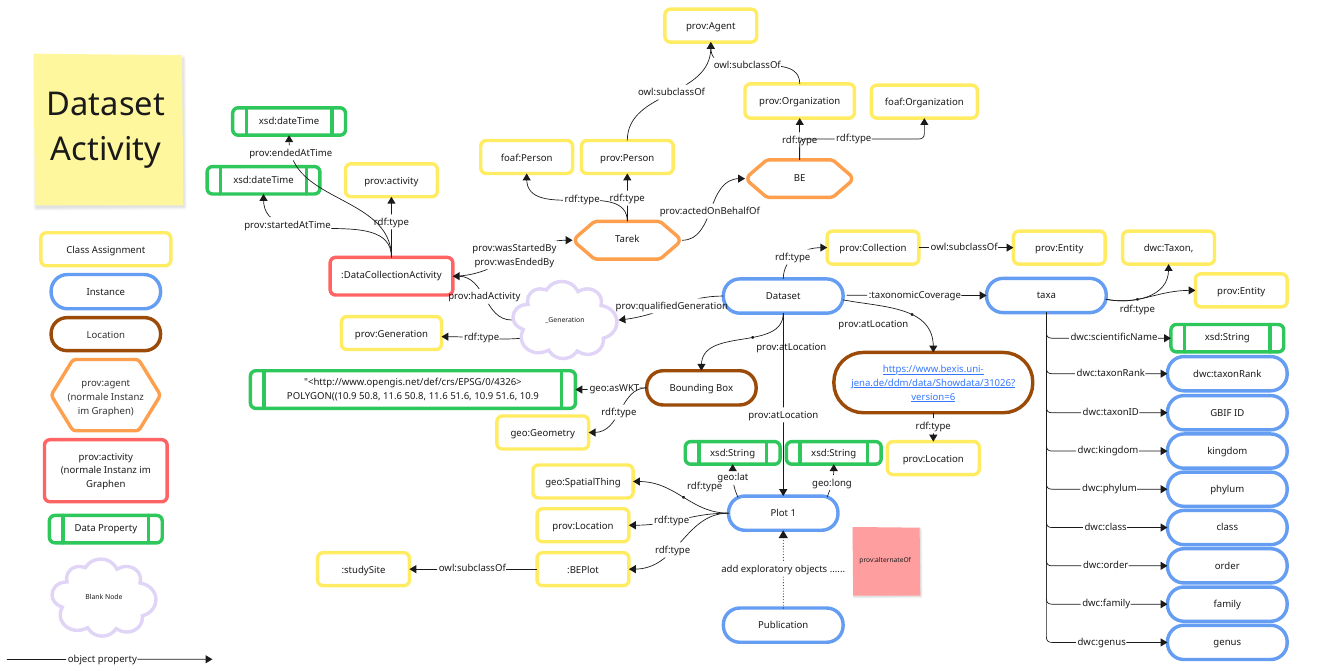}
  \caption{Figure of a first attempt at modelling an activity that leads to the generation of a dataset using the prov vocabulary. The left hand side contains a legend explaining the meaning of the different shapes and colors used.}
  \label{fig:provdataset}
\end{figure}
\end{landscape}

\begin{landscape}
\begin{figure}[p]
  \centering
  \includegraphics[width=\linewidth,height=\linewidth,keepaspectratio]{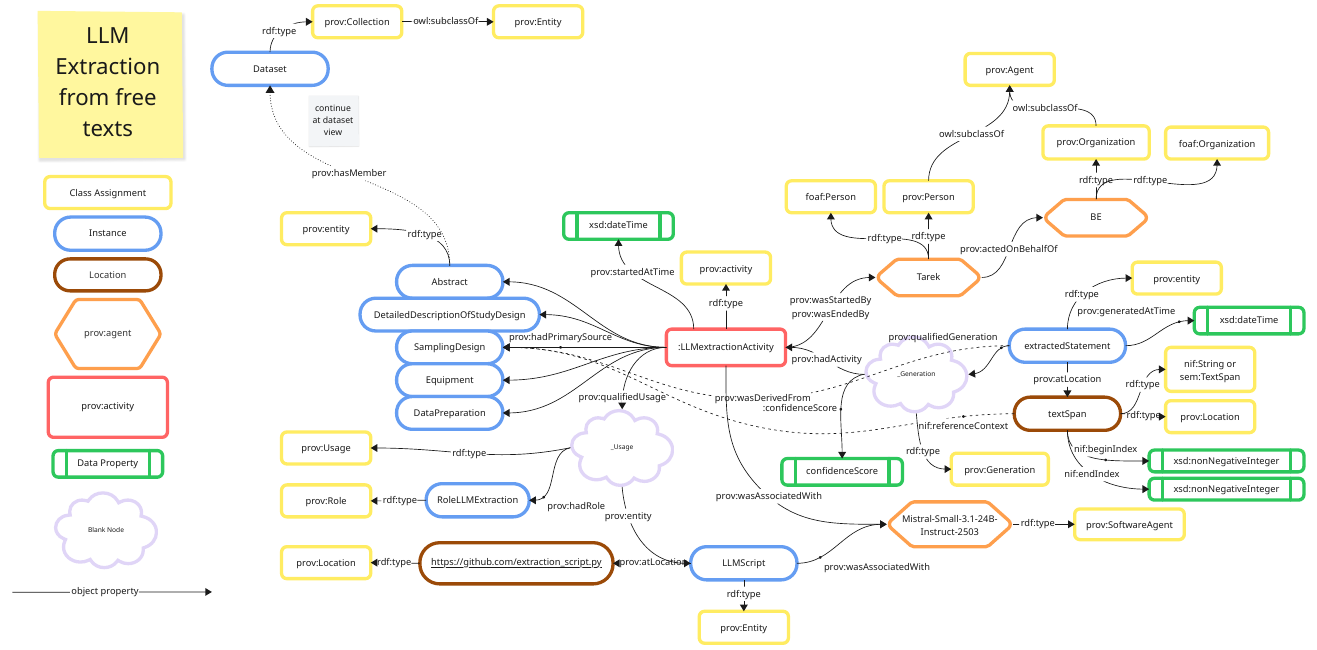}
  \caption{Figure of a schema that may be used for an activity that leads to the generation of a statement with a large language model using the prov vocabulary. The left hand side contains a legend explaining the meaning of the different shapes and colors used.}
  \label{fig:provllm}
\end{figure}
\end{landscape}

We present early results for datasets and LLM extracting of the modelling process in Fig. \ref{fig:provdataset} and Fig. \ref{fig:provllm} (we leave out publications because of space limitations for figures).
For both figures, we refer readers to the left hand side that contains a legend of the colored shapes and what they represent.
Yellow rectangles represent OWL class assignments, meaning that we not only model a preliminary version of the RDF graph, but also represent schema information.
Green rectrangles with crossing lines at the side represent OWL data properties, such as names or timestamps. 
The rest of the shapes represent OWL instances, meaning actual instances of the knowledge graph. 
Those colored in blue are instances without any special purpose, brown ovals are locations, orange hexagons are agents from the prov ontology, red rectangles are activities from the prov ontology, and pale cloud shaped objects are blank nodes. 
Finally, black arrows connecting the shapes are OWL object properties and the text tags on or next to those arrows are specific relations from ontologies, such as \textit{RDF:type} for type assignments between instances and ontology classes.

We chose the PROV\footnote{\url{https://www.w3.org/TR/prov-overview/}} vocabulary at the beginning of the modelling process, as it provides a simple but powerful framework to describe our source data with.
PROV defines three main types of instances that can exist in the knowledge graph: Entities, activities, and agents.
In simple terms, the framework provides us with the tools to represent activities and their inputs and outputs, who started them, and which agents were associated with them.
Considering the example from Fig. \ref{fig:provdataset}, the activity of collecting data (red rectangle) can be started by and attributed to the scientists that conducted some research on behalf of the Biodiversity Exploratories (orange hexagons), and the output of that activity can be the dataset itself (blue rectangle), that has a variety of other data points connected to it like plot information, taxa, or locations.

Similarly, the example Fig. \ref{fig:provllm} presents, shows the activity of extracting information from a free text field from the BE metadata (abstract, detailed description of study design, sampling design, equipment, data preparation), describes the person the activity was started by and that they acted on behalf of the BE, and that the outcome of the activity is an entity called \textit{extractedStatement} with multiple types, data, and metadata. 
We can also show that the activity was associated with a software agent (a subclass of \textit{prov:agent} named \textit{Mistral-Small-3.1-24B-Instruct-2503}\footnote{\url{https://huggingface.co/mistralai/Mistral-Small-3.1-24B-Instruct-2503}} and attach further provenance data to its usage.

Unfortunately, the use of PROV vocabulary has three limitations: 
\begin{itemize}
    
\item \textbf{(1)} As shown in Fig. \ref{fig:provdataset} and Fig. \ref{fig:provllm}, blank nodes are used at central points for the representations of publications, datasets, and LLM extractions. 
These are necessary to make a \textit{prov:qualifiedGeneration}\footnote{\url{https://www.w3.org/TR/prov-o/\#qualifiedGeneration}} or a \textit{prov:qualifiedUsage}\footnote{\url{https://www.w3.org/TR/prov-o/\#qualifiedUsage}}, which are used to attach additional information to a generic generation or usage.
For possible disadvantage of using blank nodes, we refer to Subsec. \ref{subs:challenge3} and \ref{subs:challenge4}. 
\item \textbf{(2)} PROV in itself lacks some expressions necessary to convey nuanced information. For example, the only object properties in PROV-O between two activities are \textit{prov:qualifiedUsage}\footnote{\url{https://www.w3.org/TR/prov-o/\#wasInformedBy}} or \textit{prov:Communication} / \textit{prov:}\textit{qualifiedCommu-nication}.\footnote{\url{https://www.w3.org/TR/prov-o/\#Communication}}
This means that native PROV lacks some expressiveness, which could be circumvented by using other vocabularies, but, 
\item \textbf{(3)} native PROV-O is not BFO compliant. This is not a substantial limitation as PROV activities can be mapped to BFO occurrents, entities to continuants, and agents can be decided on a case by case basis. 
However, as will be presented in the next section, we rely heavily on BFO compliant ontologies and are able to find expressions outside of the PROV framework.

\end{itemize}

For these reasons, we decided against using the PROV framework in the final knowledge graph.
However, with this first modelling approach we still gained valuable insights for the end result of the graph schema that we present in the next sections.

\subsection{Knowledge Graph Schema}
\begin{landscape}
\begin{figure}[p]
  \centering
  \includegraphics[width=\linewidth,height=\linewidth,keepaspectratio]{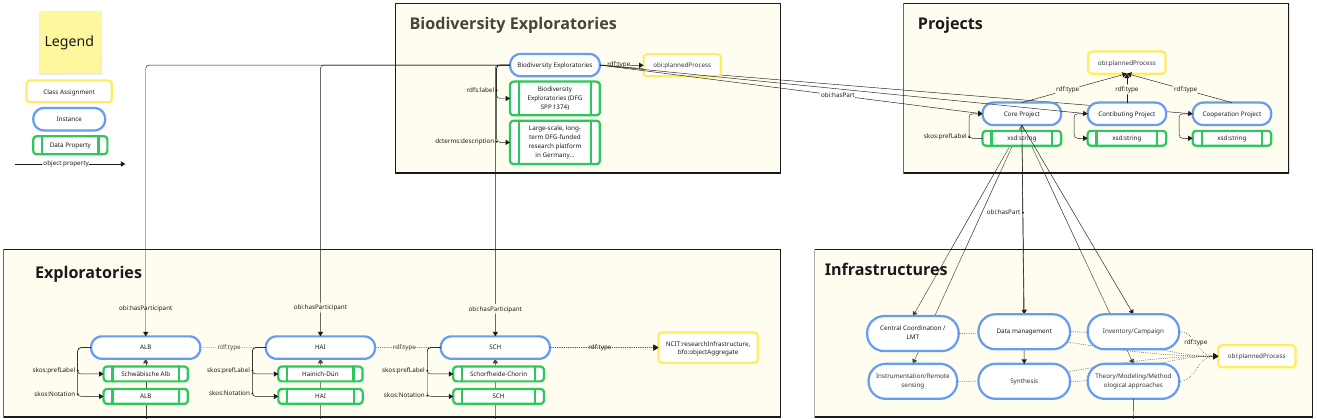}
  \caption{Schema for the Biodiversity Exploratories, its projects and infrastructures, and the exploratories. As the legend on the left hand side suggests, yellow rectangles contain class assignments, blue ovals denote instances, and green rectangles denote data properties.}
  \label{fig:schemaBE}
\end{figure}
\end{landscape}

In this section we present the schema of the knowledge graph modelled for this thesis. 
Below, we explain our modelling choices and provide figures to support visual understanding.
Starting from a top-down view, Fig. \ref{fig:schemaBE} shows our modelling choices for the Biodiversity Exploratories, its projects, infrastructures, and the three exploratories themselves.
In the top left corner of the figure, we provide a legend similar to the previous section.
A yellow rectangle stands for OWL classes, a blue oval for instances in the knowledge graph, a green rectangle for data properties, and a black arrow between any two objects represent object properties.
We will use these shapes and their interpretations for the remainder of the thesis.
To further enhance visual interpretability of the schema, we draw bigger boxes around groups of shapes and provide a title in the boxes' top left corner to categorize its contents.

In the top-middle of Fig. \ref{fig:schemaBE}, we instantiate the Biodiversity Exploratories of the type \textit{obi:plannedProcess}\footnote{\url{http://purl.obolibrary.org/obo/OBI_0000011}} and attach a label and description.
Below, we instantiate the three exploratories \textit{ALB, HAI}, and \textit{SCH} with two types, declaring them as an \textit{ncit:researchInfrastructure}, a class for \textit{'physical structures needed to conduct research.'}\footnote{\url{http://purl.obolibrary.org/obo/NCIT_C19158}}, and as a \textit{bfo:objectAggregate}\footnote{\url{http://purl.obolibrary.org/obo/BFO_0000027}}, as the exploratories are aggregates of plots.

We also attach labels and notations, and connect the BE to the exploratories via \textit{obi:hasParticipant}\footnote{\url{http://purl.obolibrary.org/obo/RO_0000057}}, representing that the exploratories as continuants are participating in the planned process of the BE.
At the right hand side, we model the BE core, contributing, and cooperation projects as instances of \textit{obi:plannedProcess}.
Below, we instantiate the infrastructures of the BE in a similar fashion, as planned processes that are part of the core projects.
We connect the BE, its projects, and infrastructures via \textit{obi:hasPart}\footnote{\url{http://purl.obolibrary.org/obo/BFO_0000051}}. 
Note that this is possible as continuants are allowed to have other continuants as part and also note that whenever object properties are mentioned in this textual description, readers can assume that the inverse relationship also holds even if not mentioned explicitly.

\begin{figure}[h]
  \centering
\includegraphics[width=\linewidth,height=\linewidth,keepaspectratio]{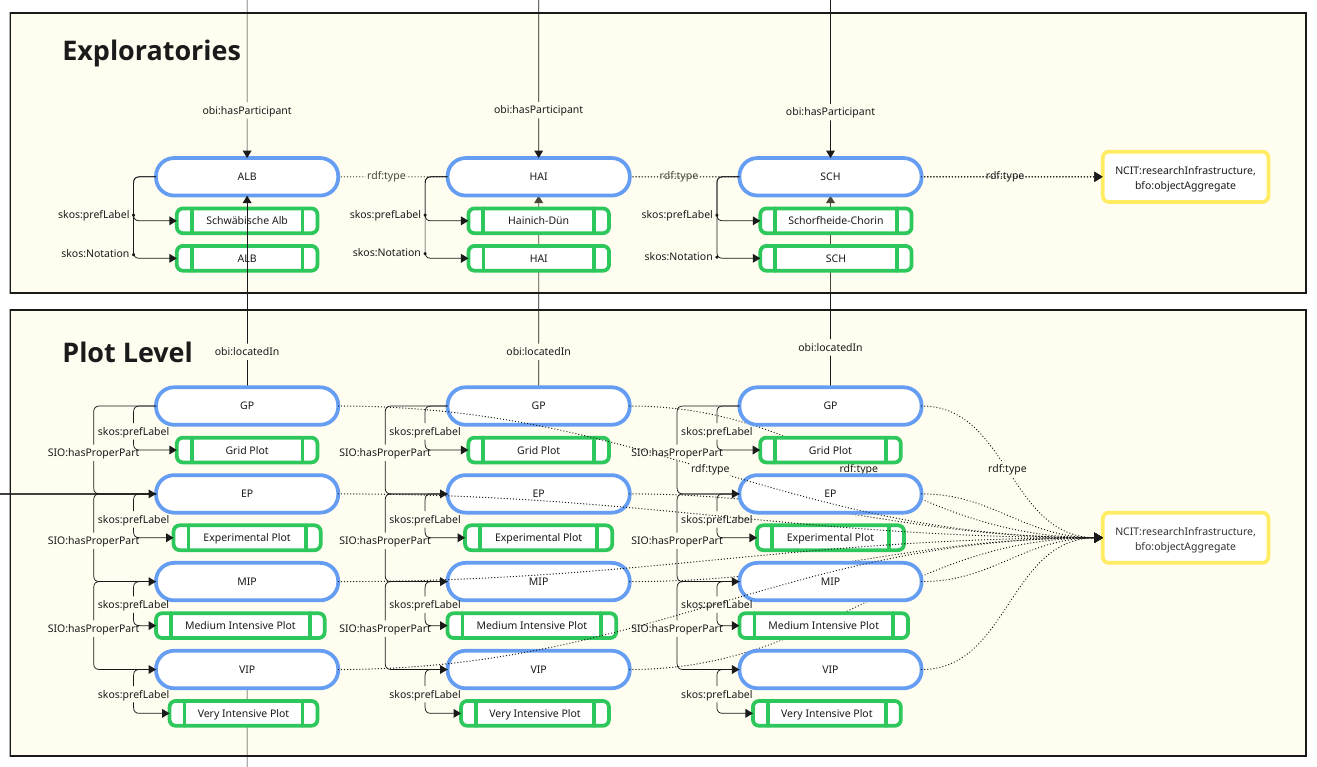}
  \caption{Schema of how different plot levels within the BE are modelled and attached to the exploratories.}
  \label{fig:schemaPlots}
\end{figure}

Fig. \ref{fig:schemaPlots} shows the research plots of the BE and different plot levels.
The most appropriate solution to modelling plot level in our view is to consider them object aggregates, and to create one aggregate per exploratory, as, for example, the grid plots located in \textit{ALB} can be viewed as a collection of plots that is characterized by their location.
As we will cover in Sec. \ref{sec:modellingChallenges}, the source data keeps singular plots as anonymous, and does not present us with the means to disseminate what study took place on which exact plots, thus we decide to represent plot levels as collections of plots characterized by being of a specific plot type.
Next to being object aggregates, we also assign the type \textit{ncit:researchInfrastructure} to all plot collections.
As for the collections and their relationship to each other, grid plots, experimental plots, medium intensive plots, and very intensive plots, form a partonomy, meaning that each lower level is a subset of the one above it, and, as per definition of the plot types, they are proper subsets, allowing us to relate them with \textit{sio:properPart}\footnote{\url{http://semanticscience.org/resource/SIO_000053}}.

\begin{figure}[h]
  \centering
\includegraphics[width=\linewidth,height=\linewidth,keepaspectratio]{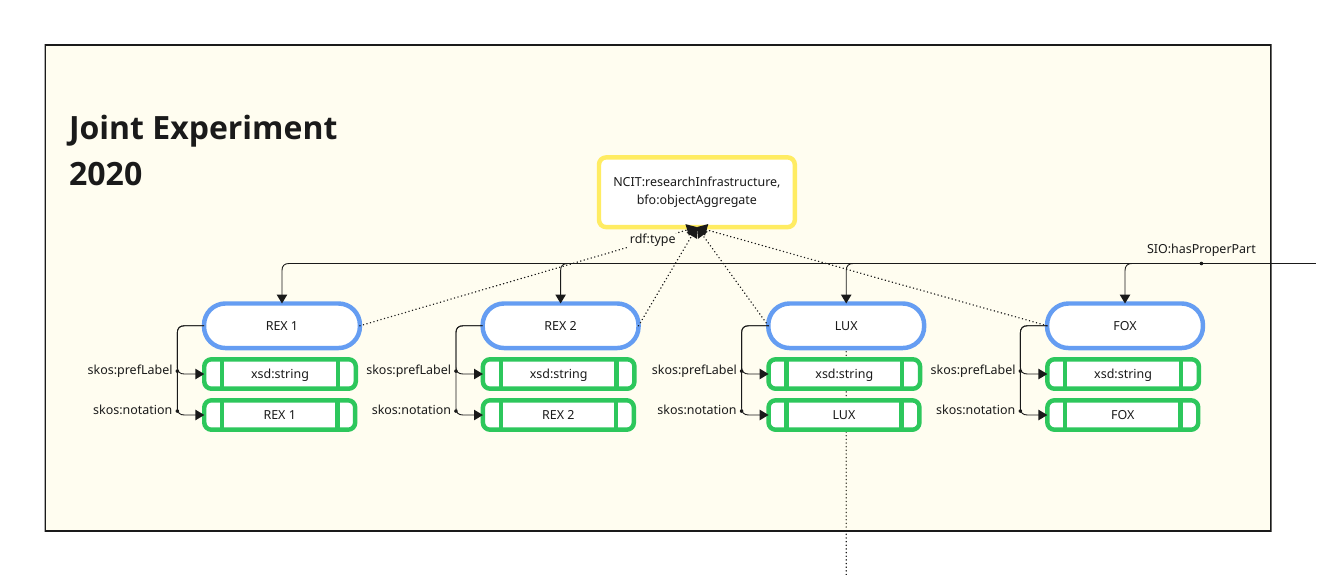}
  \caption{Model of the REX1, REX2, LUX, and FOX experiments and how they connect to the exploratories (analogously to the different plot levels).}
  \label{fig:schemaEXP}
\end{figure}
Analogously, in Fig. \ref{fig:schemaEXP}, we model experiments as parts of the experimental plots.
\begin{figure}[h]
  \centering
\includegraphics[width=\linewidth,height=\linewidth,keepaspectratio]{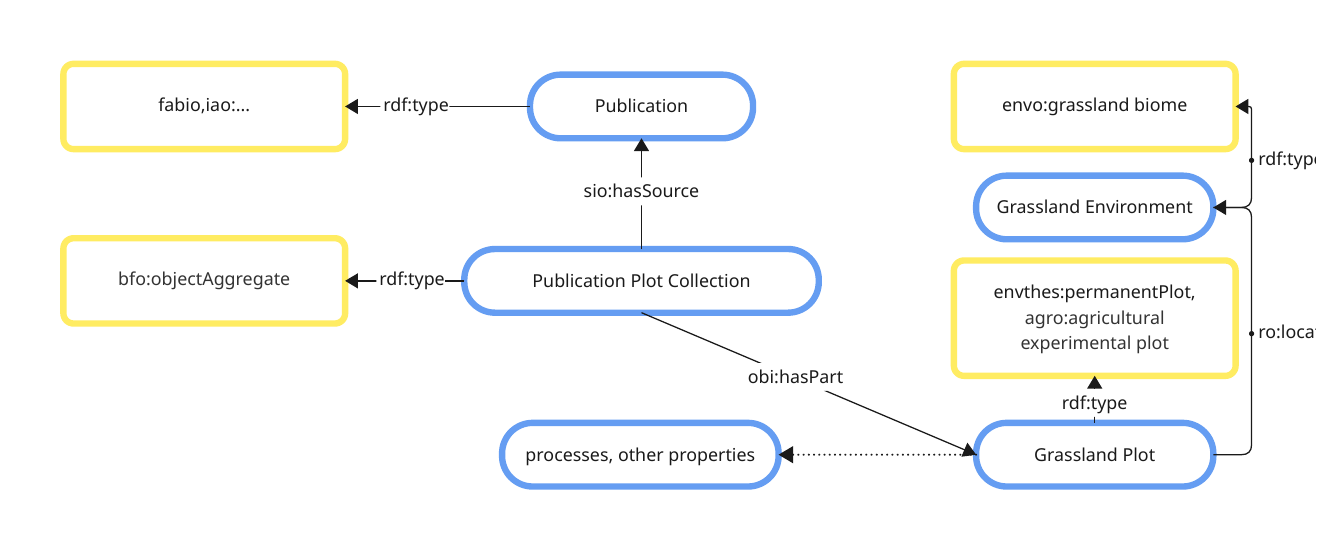}
  \caption{Simplified schema of instantiated plot collections for each publication, how they connect to their source publications, what plots they contain, and how processes and other properties are connected to those plots.}
  \label{fig:schemaSimple}
\end{figure}

Instead of presenting the schema for the relationship between publication metadata and the research plots in detail, we instead show a simplified version in Fig. \ref{fig:schemaSimple} and list relevant vocabulary below.
The core idea is to create a collection of plots (via bfo:objectAggregate) for any publication and state that the publication is the source (in a bibliographic sense, therefore \textit{sio:hasSource}\footnote{\url{http://semanticscience.org/resource/SIO_000253}} is used).
This collection has as parts different kinds of plots that are characterized by either being located in a grassland or forest environment.
Any taxa, processes, properties, or important keyword mentioned in a publication's metadata can be appended to a plot using appropriate object properties.
Instantiated plots are of types \textit{envthes:permanentPlot}\footnote{\url{http://vocabs.lter-europe.net/EnvThes/20307}} and \textit{agro:agriculturalExperimentPlot}\footnote{\url{http://purl.obolibrary.org/obo/AGRO_00000301}}.
For forest and grassland environments we use \textit{envo:grasslandBiome}\footnote{\url{http://purl.obolibrary.org/obo/ENVO_01000177}} and \textit{envo:forestBiome}\footnote{\url{http://purl.obolibrary.org/obo/ENVO_01000174}} respectively.
\clearpage

\begin{landscape}
\begin{figure}[p]
  \centering
  \includegraphics[width=\linewidth,height=\linewidth,keepaspectratio]{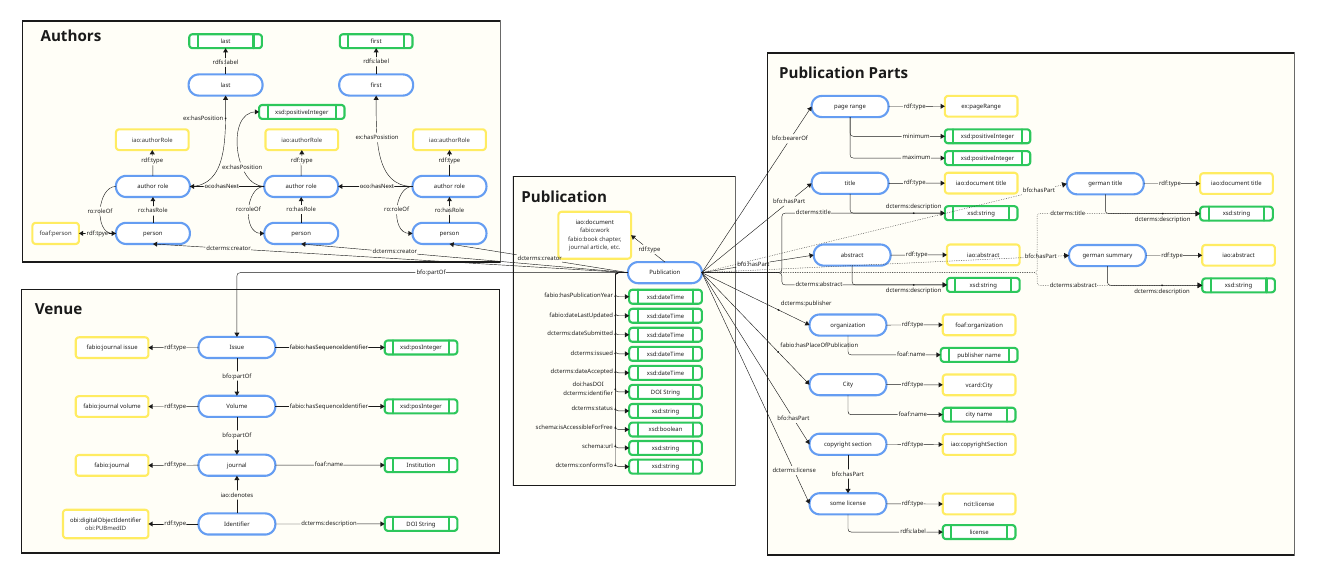}
  \caption{Representation of the Open Citations Data Model, adapted to the publications of the BE. Publications are depicted in the center and are connected to its authors, the publishing venue, and the parts of the publication itself.}
  \label{fig:schemaPub}
\end{figure}
\end{landscape}

As for publication metadata relevant to the investigated plots, we list ontology classes below in Table \ref{tab:plotterms}, which leaves only a description of publications and datasets, and how they are connected to the plots they study.
As stated in Sec. \ref{sec:topbottom}, we reuse the \textit{Open Citations Data Model}, shown in Fig. \ref{fig:schemaPub}.

\begin{table}[h]
\begin{center}
\small
\begin{tabular}{|c c c|} \hline
   Metadata Slot  & Ontology Classes & Object Property\\ \hline
   \textbf{processes/services:} & & \\
    
    herbivory & envthes:herbivory\tablefootnote{http://vocabs.lter-europe.net/EnvThes/21104}, obi:process\tablefootnote{\url{http://purl.obolibrary.org/obo/BFO_0000015}} & obi:hasParticipant\\
    predation/parasitism & envthes:predation\tablefootnote{\url{http://vocabs.lter-europe.net/EnvThes/21113}}/parasitism\tablefootnote{\url{http://vocabs.lter-europe.net/EnvThes/21106}}, obi:process &  obi:hasParticipant\\
    biogeochemical cycling & envo:biogeochemicalCycling\tablefootnote{\url{http://purl.obolibrary.org/obo/ENVO_02500009}}, obi:process & obi:hasParticipant\\
    productivity & envthes:productivity\tablefootnote{\url{http://vocabs.lter-europe.net/EnvThes/21417}}, obi:process & obi:hasParticipant\\
    pollination & mesh:pollination\tablefootnote{\url{http://id.nlm.nih.gov/mesh/D054817}}, obi:process & obi:hasParticipant\\
    dispersal & envthes:dispersal\tablefootnote{\url{http://vocabs.lter-europe.net/EnvThes/21038}}, obi:process & obi:hasParticipant\\
    decomposition & envthes:decomposition\tablefootnote{\url{http://vocabs.lter-europe.net/EnvThes/20952}}, obi:process & obi:hasParticipant\\
    aboveground layer & envo:layer\tablefootnote{\url{http://purl.obolibrary.org/obo/ENVO_01000281}} & obi:partOf\tablefootnote{\url{http://purl.obolibrary.org/obo/BFO_0000050}}\\
    belowground layer & envo:layer & obi:partOf\\
    \hline
    
    \textbf{taxa:}& & \\
    invertebrates & ex:wikidataID\tablefootnote{\url{http://www.wikidata.org/wiki/Q43806}} &obi:partOf\\
    vertebrates & ex:wikidataID\tablefootnote{\url{http://www.wikidata.org/wiki/Q25241}} &obi:partOf\\
    fungi & ex:gbifID\tablefootnote{\url{http://www.gbif.org/species/5}} &obi:partOf\\
    microbes & ex:gbifID\tablefootnote{\url{http://www.gbif.org/species/3}} &obi:partOf\\
    plants & ex:gbifID\tablefootnote{\url{http://www.gbif.org/species/6}} &obi:partOf\\
    \hline
    
    \textbf{environmental:}& & \\
    climate & envo:climate\tablefootnote{\url{http://purl.obolibrary.org/obo/ENVO_01001082}} & ro:dispositionOf\tablefootnote{\url{http://purl.obolibrary.org/obo/RO_0000092}}\\
    soil & envo:soil\tablefootnote{\url{http://purl.obolibrary.org/obo/ENVO_00001998}} & obi:partOf\\
    managed land use & envo:landUseProcess\tablefootnote{\url{http://purl.obolibrary.org/obo/ENVO_01001431}}, obi:\textbf{planned}Process & obi:hasParticipant\\
    unmanaged land use & envo:landUseProcess, obi:process & obi:hasParticipant\\
    \hline
\end{tabular} 
\end{center} 
\caption{Table of publication metadata slots, ontology classes assigned to them, and the object property they are connected to the plot with.}
\label{tab:plotterms}
\end{table}

The publication and its data properties are shown in the center, and around it are different parts concerning authors, venue, publication parts, and, added to the model in our case, the process of how the publication came to be.
Attached directly to the publication are several literals regarding publication year, submission date, open access status, and other information that does not require its own resource.
Additionally, we attach information about what project a publication belongs to via \textit{obi:partOf} by dynamically instantiating projects as resources, and distinguish those as belonging to either the core or contributing projects.
Further, we add the source of data that is discussed in the publication as originating from the field (envthes:fieldStudy\footnote{\url{http://vocabs.lter-europe.net/EnvThes/20224}}), lab (ncit:laboratoryStudy\footnote{\url{http://ncicb.nci.nih.gov/xml/owl/EVS/Thesaurus.owl\#C28278}}), or a review(vivo:review\footnote{\url{http://vivoweb.org/ontology/core\#Review}}) and attach this information by specifying that the publication \textit{iao:isAbout}\footnote{\url{http://purl.obolibrary.org/obo/IAO_0000136}} particular study types.
We also adopt the category \textit{biotic data types} from the metadata that contains the options listed in Table \ref{tab:bioticdatatypes} below:

\begin{table}[h]
\begin{center}
\small
\begin{tabular}{|c c c|} \hline
   Metadata Slot  & Ontology Classes & Object Property\\ \hline    
   \textbf{Publications:}& & \\
   single species &pco:singleSpeciesCollectionOfOrganisms\tablefootnote{\url{http://purl.obolibrary.org/obo/PCO_0000052}}&iao:isAbout\\
   multiple species &pco:multiSpeciesCollectionOfOrganisms\tablefootnote{\url{http://purl.obolibrary.org/obo/PCO_0000029}}&iao:isAbout\\
   total abundance &envthes:abundance\tablefootnote{\url{http://vocabs.lter-europe.net/EnvThes/21541}}&iao:isAbout\\
   genetic &sio:geneticData\tablefootnote{\url{http://semanticscience.org/resource/SIO_010028}}&iao:isAbout\\
   \hline
   
    \textbf{Additional for Datasets:}& & \\
    aggregated diversity & envthes:diversity\tablefootnote{\url{http://vocabs.lter-europe.net/EnvThes/10087}} & iao:isAbout\\
    chemical & envthes:chemical\tablefootnote{\url{http://vocabs.lter-europe.net/EnvThes/21964}} & iao:isAbout\\
    demographic & envthes:demography\tablefootnote{\url{http://vocabs.lter-europe.net/EnvThes/21670}} & iao:isAbout\\
    morphological & envthes:morphological\tablefootnote{\url{http://vocabs.lter-europe.net/EnvThes/21995}} & iao:isAbout\\
    presence/absence & envthes:presence\tablefootnote{\url{http://vocabs.lter-europe.net/EnvThes/10080}} & iao:isAbout\\
    trait & envthes:trait\tablefootnote{\url{http://vocabs.lter-europe.net/EnvThes/10023}} & iao:isAbout \\
    \hline
\end{tabular} 
\end{center} 
\caption{Table of metadata slots from the category \textit{'biotic data types'}, their ontology class mappings, and how a publication or dataset connects to them. Note that for datasets the same slots exist as for publications, but that also six more choices are allowed.}
\label{tab:bioticdatatypes}
\end{table}

For datasets, modelling follows a similar structure in which a collection of plots is instantiated that has the dataset as its \textit{sio:hasSource}.
In the BExIS metadata schemata, datasets have more selection options for various metadata categories than publications.
Therefore, we list additional ontology terms used to represent those concepts in Table \ref{tab:plottermsdataset}.

\begin{table}[h]
\begin{center}
\small
\begin{tabular}{|c c c|} \hline
   Metadata Slot  & Ontology Classes & Object Property\\ \hline
    \textbf{processes/services:} & & \\
    carbon cycling & envthes:cabonCycle\tablefootnote{\url{http://vocabs.lter-europe.net/EnvThes/21229}}, obi:process & obi:hasParticipant\\
    nitrogen cycling & envthes:nitrogenCycle\tablefootnote{\url{http://vocabs.lter-europe.net/EnvThes/21231}}, obi:process & obi:hasParticipant\\
    water cycling & envthes:waterCycle\tablefootnote{http://vocabs.lter-europe.net/EnvThes/21243}, obi:process & obi:hasParticipant\\
    other nutrient cycling & envthes:nutrientCycling\tablefootnote{\url{http://vocabs.lter-europe.net/EnvThes/20960}}, obi:process & obi:hasParticipant\\
    mycorrhization & metpo:mycorrhization\tablefootnote{\url{https://w3id.org/metpo/1000198}}, obi:process &  obi:hasParticipant\\
    soil aggregation & inrae:soilAggregation\tablefootnote{\url{http://opendata.inrae.fr/thesaurusINRAE/c_1354}}, obi:process & obi:hasParticipant\\
    function/service - aesthetic & ex:serviceAesthetic, obi:process & obi:hasParticipant\\
    function/service - pest control & mesh:pestControl\tablefootnote{\url{http://id.nlm.nih.gov/mesh/D010571}}, obi:process & obi:hasParticipant\\
    other & ex:serviceOther, obi:process & obi:hasParticipant\\
    \hline

    \textbf{environmental:}& & \\
    habitat structure & envthes:habitatStructure\tablefootnote{\url{http://vocabs.lter-europe.net/EnvThes/21478}} & sio:isPropertyOf\tablefootnote{\url{http://semanticscience.org/resource/SIO_000224}}\\
    spatial & envthes:spatial\tablefootnote{\url{http://vocabs.lter-europe.net/EnvThes/22018}} & envthes:constrains\tablefootnote{\url{https://w3id.org/iadopt/ont/constrains}}\\
    other & ex:descriptorsOther & envthes:constrains\\
    \hline
\end{tabular} 
\end{center} 
\caption{Table of additional dataset metadata slots, ontology classes assigned to them, and the object property they are connected to the plot with.}
\label{tab:plottermsdataset}
\end{table}

For people associated with the creation of the dataset, different roles are distinguished: Metadata creator, project leader, data creator, data collector, and contact person. 
For these we create placeholder roles and use them like the author role for publications (Fig. \ref{fig:schemaPub}), where a person is instantiated as the \textit{dcterms:creator\footnote{\url{http://purl.org/dc/terms/creator}}} of a publication, and embodies a specific role like the \textit{iao:authorRole}\footnote{\url{http://purl.obolibrary.org/obo/IAO_0000442}} via \textit{ro:hasRole}\footnote{\url{http://purl.obolibrary.org/obo/RO_0000087}}.
Analogously, funding parties are instantiated as an \textit{foaf:organization}\footnote{\url{http://xmlns.com/foaf/spec/\#term_Organization}} with the \textit{efo:funder}\footnote{\url{http://www.ebi.ac.uk/efo/EFO_0001736}} role.

\begin{landscape}
\begin{figure}[p]
  \centering
  \includegraphics[width=\linewidth,height=\linewidth,keepaspectratio]{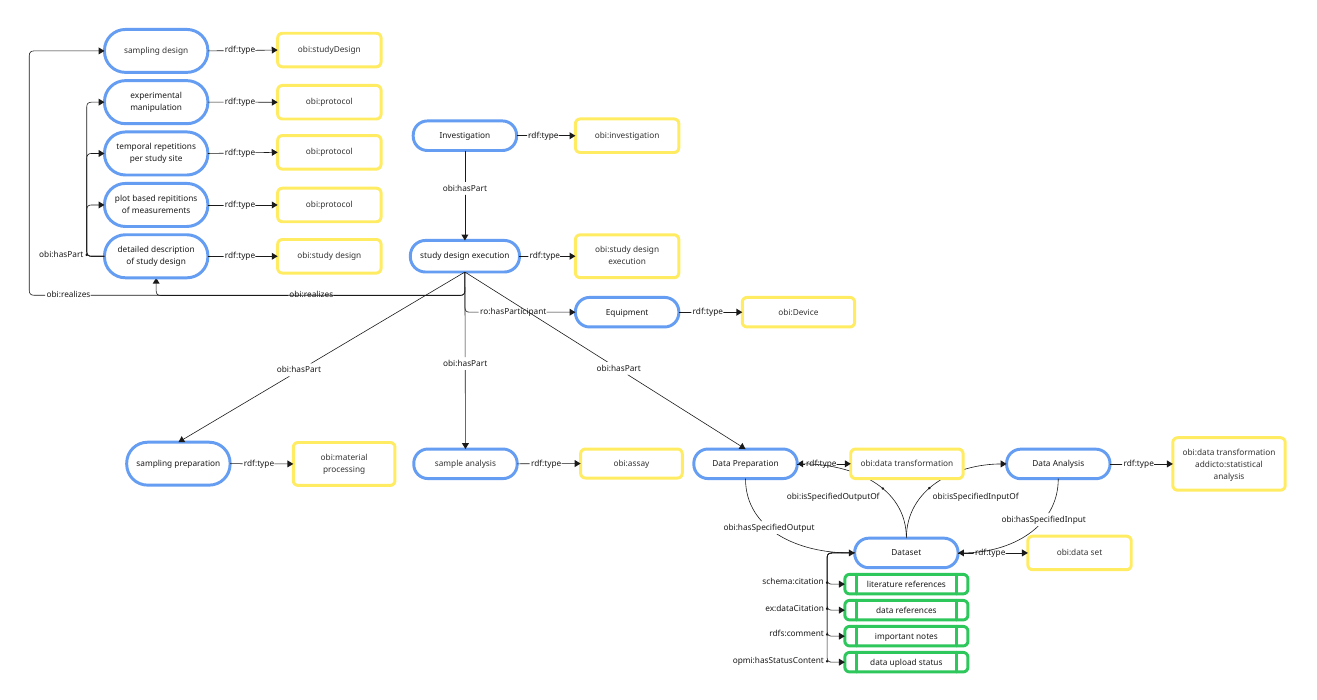}
  \caption{Schema for investigations that have as their output datasets of the BE. The metadata categories for datasets align well to the model of the ontology for biomedical investigations (obi), containing study design details, sampling preparation and analysis, as well as data preparation and analysis.}
  \label{fig:investigation}
\end{figure}
\end{landscape}

Finally, Fig. \ref{fig:investigation} shows the process of how a dataset is created and the study design followed in its creation.
In our modelling process, we match dataset metadata to the scheme \textit{OBI} provides for scientific investigations.
An \textit{obi:investigation}\footnote{\url{http://purl.obolibrary.org/obo/OBI_0000066}} is instantiated that has as its part an instance of \textit{obi:studyDesignExecution}\footnote{\url{http://purl.obolibrary.org/obo/OBI_0000471}}. 
This study design execution has two functions: It connects to study design information via \textit{obi:realizes}\footnote{\url{http://purl.obolibrary.org/obo/BFO_0000055}} and has as its \textit{obi:part} instances of the classes \textit{obi:materialProcessing}\footnote{\url{http://purl.obolibrary.org/obo/OBI_0000094}}, \textit{obi:assay}\footnote{\url{http://purl.obolibrary.org/obo/OBI_0000070}}, \textit{obi:dataTransformation}\footnote{\url{http://purl.obolibrary.org/obo/OBI_0200000}}, and \textit{addicto:statisticalAnalysis}\footnote{\url{http://addictovocab.org/ADDICTO_0001296}}.
We also attach the equipment used in the study as \textit{ro:hasParticipant} of the study design execution.
The dataset itself is \textit{obi:isSpecifiedOutputOf} of this data preparation and we attach further information as literals such as literature references via \textit{schema:citation}\footnote{\url{http://schema.org/citation}} and the data upload status using \textit{opmi:hasStatusContent}\footnote{\url{http://purl.obolibrary.org/obo/OPMI_0000622}}.

This concludes the presentation of the knowledge graph's schema and the vocabulary used. 
In the next sections we present modelling challenges we encountered and present our implementation approach.

\section{Modelling Challenges}\label{sec:modellingChallenges}
We described the process of modelling the knowledge graph in the previous section.
Now, we aim to expand upon the five greatest challenges we encountered in developing the graph's schema:
\subsection{BExIS Metadata Schemata}
The first and possibly most widely encountered challenge is concerned with the data used as source for the knowledge graph.
Several significant hindrances arise from it:
\begin{itemize}
    \item \textbf{Semi-structured input data:} The BE publication and dataset metadata is not fully structured. 
    Although there are type definitions, minimum and maximum lengths, descriptions, and definitions for several metadata fields, the existence of free-text fields for information on study design, data processing, equipment descriptions, etc., leads to significant modelling and/or processing overhead.
    \item \textbf{Metadata schemata for user convenience.} In BExIS, researchers and people associated with publications, datasets, and projects, are required to fill the metadata forms themselves.
    On one hand, this leads to high information content and completeness of the metadata, however, on the other hand, the online forms are designed to be as accessible and understandable as possible to gain these advantages, which sacrifices correctness and structure of the information within.
    In disseminating the metadata for this thesis, we encountered several issues because of this. 
    For example, some keywords are grouped together in the same category even though they have no relation to each other (for \textit{'Biotic Data Type'}, the answers \textit{'none', 'single species', 'multiple species', 'genetic'} and \textit{'total abundance'} are allowed).
    \item \textbf{Missing concept definitions.} In the metadata schemata, all categories have descriptions explaining what information the categories are supposed to contain (sometimes only sparse information is available), but the selection options users have when filling the forms have no definitions.
    Coming back to the example from the previous point, we may have a general understanding of what is meant by the term \textit{'genetic'}, but we can only guess what the intended meaning behind this term is.
    This ambiguity is not only a challenge for modelling a knowledge graph, as finding proper ontology terms for these concepts becomes more difficult, but it also decreases data quality, as users filling out the metadata forms can also only guess what is meant by certain terms, leading the source categories to contain information blurred from original user intent.
    \item \textbf{Schematic inconsistencies.} Even though the process of filling out both metadata schemata is the same, due to historical implementation details there are several discrepancies between the schemata for publication and dataset metadata.
    For example, both questionnaires collect information about plots, but only for datasets the exact number of plots per plot type is collected, while for publications, there are only the options \textit{few} or \textit{all}.
    \item \textbf{Anonymous plots.} Due to the points listed above, one considerable challenge we encountered concerned the inability to disseminate distinct plots from the data, therefore, workarounds were necessary to properly represent the information there is, and, unfortunately, some information that could exist in the knowledge graph does not because of it.
\end{itemize}
\subsection{Negations and the Open World Assumption - Modelling complete information}
As stated in Subsec. \ref{subs:challenge4}, the \textit{Open World Assumption} leads to modelling challenges for knowledge graphs.
In this case however, the data we instantiate contains complete information.
The metadata schemata used for BE publications and datasets contain the selection option \textit{'none'} for relevant categories. 
For example, if metadata for a publication states that no taxa were investigated, leading to no edge in the graph between any taxon and a plot from a plot collection, from the \textit{OWA} it follows that the publication might have investigated some taxa regardless.

Unfortunately, as we will discuss in the limitations section (Subs. \ref{sec:limitchall}), we were not able to find appropriate solutions to these two challenges due to time constraints.

\clearpage

\section{Implementation}
In the previous sections, we described the modelling process and its results in detail.
For the next step, we aim to instantiate the developed schemata to populate the knowledge graph.
To achieve this, a mapping is required that declares class definitions and properties and assigns them to the data that is supposed to be instantiated.
There are many applications that can support developers in writing mappings, even user interfaces like \textit{Ontotext Refine}\footnote{\url{https://www.ontotext.com/products/ontotext-refine/}} exist that allow developers to assign mappings using a visual mapper for tabular data.
In this work, we choose the \textit{R2RML}\footnote{\url{https://www.w3.org/TR/r2rml/}} mapping language. 
We detail its application in detail below.

\subsection{R2RML Mapping Language}
The main purpose of the R2RML mapping language is to provide mappings that turn relational databases into RDF datasets.
Our input data consists of tabular data that, after extensive processing, can be uploaded to a relational database.
With the use of an R2RML processing engine\footnote{\url{https://github.com/chrdebru/r2rml}}, we are able to turn this data into RDF triples by translating the mappings we create from the developed schemata.
There are three substantial advantages using this approach:
First, the R2RML mapping language is highly expressive and allows us to declare multiple conditionals in the mappings, that are difficult or impossible using other approaches (as we are using a relational database as the data source, R2RML even allows for results of SQL queries to be mapped).
Second, R2RML makes virtualized knowledge graphs possible that allow SPARQL queries to be executed on relational database by translating SPARQL to SQL using written mappings (we refer to ontop\footnote{\url{https://ontop-vkg.org/}}, a system that implements this), this allows integrating knowledge graphs together with relational databases.
Third, R2RML allows us to declare mappings for named subgraphs. 
We will present this functionality in detail in the next section as it is integral for implementing semantic units for the knowledge graph.

\subsection{Implementation of Semantic Units}
In order to implement semantic units, a solution for declaring named subgraphs in the knowledge graph is required.
Thankfully, the R2RML mapping language and its processing engines allow multiple output formats for mapped files.
The standard output option returns \textit{turtle}\footnote{\url{https://www.w3.org/TR/turtle/}} (ttl) files, an RDF serialization that outputs triples in the form of \textit{<subject> <predicate> <object>.}
A further output option that we use for this knowledge graph is \textit{TriG}\footnote{\url{https://www.w3.org/TR/trig/}}, a ttl extension that allows for \textit{graph maps}, mappings to declare named subgraphs inside of the main graph, and to declare what graphs specific triples belong to.

We refer to Listing \ref{lst:su} below that illustrates an example of how a semantic unit can be built using these graph maps.
The listing depicts two mapping entities, denoted by the notation <\#...> in lines 1 and 26.
The first entity is a named individual identification unit for the second entity, the publication.
Each mapping declares a logical table (lines 3 and 29).
These depict the table in the relational database that the data originates from.
Afterwards, a subject map is declared (lines 4 and 31). 
This is used to assign ontology classes to mapping entities.
In line 6, a graph map statement is declared in the subject mapping of the semantic unit.
This results in the creation of a new subgraph in the final knowledge graph with the IRI: 
\begin{sloppypar}
\textit{http://example.com/base/semunit/namedindividualidentificationunit/-Publication\_\{id\}}
\end{sloppypar}
This creates a new supgraph for each entry in the \{id\} column of the logical table.
Furthermore, when a graph map statement is used inside of a subject map, all triples belonging to the mapping entity will automatically be assigned to the graph map.
Below the graph map statement, object and data properties are declared using predicate object maps.
For example, the semantic unit has another type assignment (line 8), a label (line 13), and a \textit{semanticUnitSubject} property that connects it to the publication (line 21).

As for the publication itself, important declarations related to the semantic unit are its label (line 37) and its type assignment (line 55).
These statements are assigned to the semantic units named subgraph via graph map statements (line 42 and 61).

\begin{lstlisting}[style=prompt,
  caption={Excerpt of an R2RML mapping depicting mappings for publications and named individual identification units for publications.},
  label={lst:su},
  captionpos=b]
<#NamedIndividualIdentificationUnit_publication>
  a rr:TriplesMap ;
  rr:logicalTable   [ rr:tableName "be_publication_metadata" ] ;
  rr:subjectMap     [
    rr:template    "http://example.com/base/semunit/namedindividualidentificationunit/Publication_{id}" ;
    rr:graphMap    [ rr:template "http://example.com/base/semunit/namedindividualidentificationunit/Publication_{id}" ] ;
  ] ;
  rr:predicateObjectMap [
    rr:predicate rdf:type ;
    rr:objectMap  [ rr:constant semunit:namedindividualidentificationunit] ;
  ] ;
  
  rr:predicateObjectMap [
        rr:predicate rdfs:label ;
        rr:objectMap [
            rr:template "Publication {id} a fabio:{typeofpublication}" ;
            rr:termType rr:Literal ;
        ] ;
    ] ;  

  rr:predicateObjectMap [
    rr:predicate ex:semanticUnitSubject ;
    rr:objectMap  [ rr:parentTriplesMap <#Publication_id> ] ;
  ] .

<#Publication_id>
a rr:TriplesMap ;

    rr:logicalTable [ rr:tableName "be_publication_metadata"] ;

    rr:subjectMap [
        rr:template "http://example.com/base/Publication_{id}" ;
        rr:class iao:publication ;
        rr:termType rr:IRI;
    ] ;

    rr:predicateObjectMap [
        rr:predicate rdfs:label ;
        rr:objectMap [
            rr:template "Publication {id}, {title}" ;
            rr:language "en" ;
        ] ;
        rr:graphMap  [
            rr:template "http://example.com/base/semunit/namedindividualidentificationunit/Publication_{id}"
        ] ;
    ] ;

    rr:predicateObjectMap [
        rr:predicate obi:isSpecifiedOutputOf ;
        rr:objectMap [
            rr:template "http://example.com/base/Documenting_{id}" ;
        ] ;
    ] ;

    rr:predicateObjectMap [
        rr:predicate rdf:type ;
        rr:objectMap [
            rr:template "fabio:{typeofpublication}" ;
            rr:termType rr:IRI ;
        ] ;
        rr:graphMap  [
            rr:template "http://example.com/base/semunit/namedindividualidentificationunit/Publication_{id}"
        ] ;
    ] ;

    rr:predicateObjectMap [
        rr:predicate obi:partOf ;
        rr:objectMap [
            rr:template "http://example.com/base/BE_Infrastructure_{infrastructureclass}" ;
            rr:termType rr:IRI ;
        ] ;
    ] .
\end{lstlisting}


\subsection{Schema Validation - SHACL}
Finally, we present an example of a SHACL shape that can be used for schema validation on the instantiated knowledge graph.
We use the same example as in Listing \ref{lst:su}, publications and their named individual identification units.
Listing \ref{lst:shacl} shows the SHACL shapes that could be used for validating these triples.
Lines one, two, and three, declare an entity that is a node shape, specifying that it is a SHACL shape, for the class \textit{semunit:namedIndividualIdentificationUnit}
In lines five to thirteen, a SPARQL query is used to check for the validity of the IRI of the resource.
Further, two property validations follow in lines 15-20 and 22-28.
The first states that there must be exactly one outgoing property that declares a publication resource as the \textit{ex:semanticUnitSubject} and the second checks the validity of the label using a customized pattern.

The second SHACL shape begins in line 30, specifying that it targets entities of type \textit{iao:publication}.
The remaining node shape contains multiple property checks, one for additional type assignments (lines 42-57), for specifying that it is the specified output of a documenting process (lines 59-72), and that it is part of an infrastructure type (lines 74-87.
\begin{lstlisting}[style=prompt,
  caption={Two SHACL shapes that could be used to validate triples for publications and for semantic units identifying them.},
  label={lst:shacl},
  captionpos=b]
  ex:NamedIndividualIdentificationUnitShape
    a sh:NodeShape ;
    sh:targetClass semunit:namedindividualidentificationunit ;

    sh:sparql [
      sh:message "Semunit node IRI must start with /semunit/namedindividualidentificationunit/Publication_" ;
      sh:select """
        SELECT $this WHERE {
          FILTER ( !STRSTARTS(STR($this),
            "http://example.com/base/semunit/namedindividualidentificationunit/Publication_") )
        }
      """ ;
    ] ;

    sh:property [
      sh:path ex:semanticUnitSubject ;
      sh:minCount 1 ;
      sh:maxCount 1 ;
      sh:node ex:PublicationShape ;
    ] ;

    sh:property [
      sh:path rdfs:label ;
      sh:minCount 1 ;
      sh:datatype xsd:string ;
      sh:pattern "^Publication\\s+\\d+\\s+a\\s+fabio:.+$" ;
      sh:message "Label should look like: 'Publication {id} a fabio:{typeofpublication}'." ;
    ] .

ex:PublicationShape
    a sh:NodeShape ;
    sh:targetClass iao:publication ;

    sh:property [
      sh:path rdfs:label ;
      sh:minCount 1 ;
      sh:languageIn ("en") ;
      sh:uniqueLang true ;
      sh:message "Publication must have an English rdfs:label." ;
    ] ;

    sh:property [
      sh:path rdf:type ;
      sh:minCount 1 ;
      sh:nodeKind sh:IRI ;
      sh:sparql [
        sh:message "At least one rdf:type must be a FABIO IRI." ;
        sh:select """
          SELECT $this WHERE {
            FILTER NOT EXISTS {
              $this rdf:type ?t .
              FILTER ( isIRI(?t) && STRSTARTS(STR(?t), "http://purl.org/spar/fabio/") )
            }
          }
        """ ;
      ] ;
    ] ;

    sh:property [
      sh:path obi:isSpecifiedOutputOf ;
      sh:minCount 1 ;
      sh:nodeKind sh:IRI ;
      sh:sparql [
        sh:message "obi:isSpecifiedOutputOf must point to an IRI starting with /Documenting_." ;
        sh:select """
          SELECT $this WHERE {
            $this obi:isSpecifiedOutputOf ?v .
            FILTER ( !STRSTARTS(STR(?v), "http://example.com/base/Documenting_") )
          }
        """ ;
      ] ;
    ] ;

    sh:property [
      sh:path obi:partOf ;
      sh:minCount 1 ;
      sh:nodeKind sh:IRI ;
      sh:sparql [
        sh:message "obi:partOf must point to an IRI starting with /BE_Infrastructure_." ;
        sh:select """
          SELECT $this WHERE {
            $this obi:partOf ?v .
            FILTER ( !STRSTARTS(STR(?v), "http://example.com/base/BE_Infrastructure_") )
          }
        """ ;
      ] ;
    ] ;


\end{lstlisting}

\chapter{Method - Extracting Information from Metadata  using Large Language Models} \label{chapter7}

In the first part of our method, we presented how structured metadata may be used as a starting point for knowledge graph construction.
One big advantage of the use case discussed in this thesis  is that the publication and dataset metadata provided by the Biodiversity Exploratories already exists in a structured format.
Therefore, knowledge graph construction is simplified considerably, as a measure of quality control and structure has already been introduced as a result of human effort.
Another upside of this level of data quality is that it is not only useful for knowledge graph construction, but also necessary to publish FAIR-aligned datasets, simplify overall reuse, and enable further functionalities in BExIS like structured search.

Unfortunately, these considerable advantages come with a cost:
The human effort required to provide structured metadata of high quality is immense, and includes many people at different points of the data processing pipeline.
Therefore, in the second part of our method, we aim to investigate two avenues at which large language models may contribute to either reduce the effort in creating structured metadata, or to extract latent information.


\section{Extracting Metadata Categories}
One point of interest in providing FAIR metadata for the Biodiversity Exploratories could be to lessen the human effort required in metadata curation or entry to BExIS by extracting metadata categories from the titles and abstracts/summaries of publications and datasets automatically with the assistance of LLMs. 

Therefore, in this section we conduct small experiments to test the feasibility of this approach.
We first describe the input data and its format, select a list of relevant metadata categories that we believe could be extracted automatically, and then describe our approach in detail.
\subsection{Input Data}
The input data consists of publication and dataset titles and abstracts, concatenated in the form:

\verb|Title: {titlecolumn}\n Abstract: {abstractcolumn}|

We test publications and datasets separately as their metadata schemata used in BExIS differ from each other.

\subsection{Metadata Categories to extract}
For this experiment, we identify a set of promising metadata categories for publications and datasets.
To identify these categories, we manually investigated a sample of titles and abstracts, and analyzed which metadata categories we were able to fill from information contained within them.
For example, as a result we leave out categories that include personal information as this kind of information does not generally exist in titles or abstracts.
Therefore, we identify the following categories of interest:

\textbf{Publications:}
\begin{itemize}
    \item Keywords
    \item Exploratory: ALB, SCH, HAI
    \item Plot Level
    \item Grassland, Forest
    \item Aboveground, Belowground
    \item Field, Laboratory, Review
    \item Infrastructure
    \item Biotic Data Taxon
    \item Biotic Data Type
    \item Processes and Services
    \item Environmental Descriptors
    \item Experiment Type: REX1, REX2, LUX, FOX
\end{itemize}

\textbf{Datasets:}
\begin{itemize}
    \item Project
    \item Number of GP/EP/MIP/VIP
    \item Experimental manipulation
    \item Temporal repetitions per study site
    \item Plot based repetitions of measurements
    \item Study design
    \item Sampling design
    \item Sample preparation
    \item Sample analysis
    \item Equipment
    \item Data preparation
    \item Data analysis
    \item Content type
\end{itemize}
Note: Datasets contain all metadata (except 'infrastructure') categories that are listed for publications as well and are extracted as well.

For all of these categories, we either allow short, free text entries from the LLM, or allow a selection of answers depending on the category definitions in the BExIS metadata schemata.
For example, keywords may contain arbitrary text entries, while the forest and grassland categories only allow \textit{yes/no}.

\subsection{Metadata Reconstruction Schema}
For metadata reconstruction, we aim to supply a large language model with not only the titles and abstracts, but also with a schema to guide the extraction.
We write two csv files for the respective inputs, each containing three columns: \textit{category name}, \textit{category definition}, and \textit{allowed answers}.
In Table \ref{tab:extractionschema} below, we show an excerpt of the publication extraction schema:

\begin{table}[h]
\centering
\small
\begin{tabularx}{\linewidth}{|Y Y Y|}
\hline
\textbf{category name} & \textbf{category definition} & \textbf{allowed answers} \\ \hline
keywords & Keywords that describe the publication and its content & any \\&&\\
grassland & Did the study take place in grasslands? & yes, no \\&&\\
environmental descriptors & What environmental descriptors can be used to describe the study? 
Possible answers are: ‘managed land use’, ‘unmanaged land use’, ‘soil’, ‘climate’, ‘none’ &
‘managed land use’, ‘unmanaged land use’, ‘soil’, ‘climate’, ‘none’ \\
\hline
\end{tabularx}
\caption{Table that shows an excerpt of the schema used for metadata reconstruction for publication.}
\label{tab:extractionschema}
\end{table}
We allow four types of answers: \textit{any}, \textit{yes or no}, a list of selectable keywords, and \textit{any long} for free text answers.
If applicable, we also append the selectable keywords to the category definition to supply more context to the model.

\subsection{LLM Calls}
For this experiment we host an instance of \textit{Mistral-Small-3.2-24B-Instruct-2506}\footnote{\url{https://huggingface.co/mistralai/Mistral-Small-3.2-24B-Instruct-2506}}, an instruction fine-tuned model that is capable at processing both english and german texts.
We leverage this model's instruction-following capabilities by providing a system prompt that includes rules for data processing and output format in Listing \ref{lst:sys-prompt}:\\

\begin{lstlisting}[style=prompt,
  caption={System prompt for metadata reconstruction.},
  label={lst:sys-prompt},
  captionpos=b]
You are an expert metadata curator for the Biodiversity Exploratories (BE) project.
Your task: given a TITLE and ABSTRACT, propose metadata as JSON.
RULES
- Output STRICTLY valid JSON (no comments). Do not include any text before or after the JSON.
- Choose ALL applicable values for multi-select-like fields.
- Free-text style fields are marked with "any".
- For free-text style fields: return an array of short phrases (e.g., ["soil biodiversity","grassland"]). Avoid hallucinations.
- For free-text style fields marked with "any long", longer phrases are allowed like "All species data have been standardized. This was done by dividing the recorded cover value of each species through the sum of all species cover values of a subplot. ..."
- If information is missing, you MAY return empty arrays [] or nulls. Do not invent specifics.
SCHEMA KEYS:
{schema_keys}
MULTI-SELECT FIELDS & ENUMS (if a list is provided below, prefer those exact labels; otherwise use short phrases):
{enum_block}
Return only a single JSON object with keys drawn from the schema keys above. Do NOT include any explanation.
\end{lstlisting}

\subsection{Extraction Results}
We first convert the model's output from JSON back to csv, and then compare with the ground truth metadata.
In this experiment, we took 100 publications and 100 datasets, and present the evaluation results in four tables, two each for publications (Table \ref{tab:extractionresults} and Table \ref{tab:extractionresultspub2}) and datasets (Table \ref{tab:extractionresultsdatasets1} and Table \ref{tab:extractionresultsdatasets2}), one for \textit{yes/no} fields and one for \textit{multi-label} fields.

\begin{table}[h!]
\centering
\small
\begin{tabularx}{\linewidth}{|Y Y Y Y Y|}
\hline
\multicolumn{5}{|c|}{\textbf{Publications (yes/no)}}\\ \hline
\textbf{category} & \textbf{precision} & \textbf{recall} & \textbf{f1} & \textbf{accuracy}\\ \hline
alb&\textbf{.879}&.630&.734&.79\\
sch&.643&.486&.554&.71\\
hai&.818&.509&.628&.68\\
grassland&.84&.84&.84&.84\\
forest&.628&.871&.73&.8\\
aboveground&.705&.878&.782&.76\\
belowground&.761&\textbf{.972}&.854&.88\\
field&.653&.904&.758&.7\\
laboratory&.824&.955&\textbf{.884}&.89\\
review&.556&.455&.5&\textbf{.9}\\
\hline
\end{tabularx}
\caption{Table of precision, recall, F1-score, and accuracy scores for \textit{yes/no} fields for a test run of 100 publications.}
\label{tab:extractionresults}
\end{table}

For \textit{yes/no} fields, we collect the metrics precision (when the model predicts yes, how often was it correct?), recall (of the times the model should have returned yes, did it actually return yes?), F1-score (harmonic mean of precision and recall), and accuracy (overall proportion of model predictions that were correct).
Regarding publications, the highest precision was achieved when predicting whether the publication was related to the \textit{Schwäbische Alb} repository.
Highest recall was achieved when predicting belowground, highest F1-score for predicting a laboratory study, and best accuracy when predicting a review study.
The most expressive metric is the F1-score as it balances precision and recall.
The mean F1-score across all ten \textit{yes/no} categories is \verb|~|0.7264 which is a surprisingly good result and suggests some categories may be extracted automatically, while some categories should not.

\begin{table}[h!]
\centering
\small
\begin{tabularx}{\linewidth}{|Y Y Y Y Y Y|}
\hline
\multicolumn{6}{|c|}{\textbf{Publications (multi-label fields)}}\\ \hline
\textbf{category} & \textbf{precision} & \textbf{recall} & \textbf{f1} & \textbf{accuracy} & \textbf{avg jaccard}\\ \hline
biotic data taxon &.452&.555&.498&\textbf{.43}&.521\\
biotic data type&\textbf{.551}&\textbf{.609}&\textbf{.579}&.42&\textbf{.552}\\
environmental descriptors&.210&.285&.242&.08&.206\\
infrastructure&.349&.409&.377&.29&.361\\
keywords&.103&.207&.137&0&.076\\
plot level&.348&.4&.372&.4&.4\\
processes&.269&.397&.321&.24&.332\\
\hline
\end{tabularx}
\caption{Table of precision, recall, F1-score, accuracy, and jaccard similarity scores for \textit{multi-label} fields for a test run of 100 publications.}
\label{tab:extractionresultspub2}
\end{table}

Regarding \textit{multi-label} fields, we evaluate across precision, recall, F1-score, accuracy, and the average jaccard similarity that is used to determine the similarity between sets.
For Publications, the best performing category seems to be biotic data types, with an F1-score of 0.579 and average jaccard similarity of 0.552.
As we can see in Table \ref{tab:extractionresultspub2}, only biotic data taxon and type come close to \verb|>=|0.5 and the keyword category presents a clear outlier, which we will discuss further in Ch. \ref{chapter9}.

\begin{table}[h!]
\centering
\small
\begin{tabularx}{\linewidth}{|Y Y Y Y Y|}
\hline
\multicolumn{5}{|c|}{\textbf{Datasets (yes/no)}}\\ \hline
\textbf{category} & \textbf{precision} & \textbf{recall} & \textbf{f1} & \textbf{accuracy}\\ \hline
alb&\textbf{1}&.09&.148&.08\\
sch&\textbf{1}&.05&.095&.05\\
hai&\textbf{1}&.21&.347&.21\\
grassland&.966&.757&.848&.9\\
forest&.955&.851&.9&.86\\
aboveground&.961&.88&.918&.87\\
belowground&.938&\textbf{.882}&\textbf{.909}&\textbf{.97}\\
temporal repetitions&.776&.59&.67&.59\\
plot repetitions&.934&.71&.807&.71\\
\hline
\end{tabularx}
\caption{Table of precision, recall, F1-score, and accuracy scores for \textit{yes/no} fields for a test run of 100 datasets.}
\label{tab:extractionresultsdatasets1}
\end{table}

As for the \textit{yes/no} fields of dataset metadata, we note an interesting trend where the model seems to predict the \textit{alb}, \textit{sch}, and \textit{hai} categories very sparsly, attaining precision scores of 1 with low recall.
We assume that this stems from dataset titles and descriptions being less descriptive than those of publications.
However, effectiveness for the remaining categories is very high, regularly reaching F1-scores in the 0.80 or 0.90 range. 

\begin{table}[h!]
\centering
\small
\begin{tabularx}{\linewidth}{|Y Y Y Y Y Y|}
\hline
\multicolumn{6}{|c|}{\textbf{Datasets (multi-label fields)}}\\ \hline
\textbf{category} & \textbf{precision} & \textbf{recall} & \textbf{f1} & \textbf{accuracy} & \textbf{avg jaccard}\\ \hline
biotic data taxon&.571&.619&.594&.53&.565\\

biotic data type&.221&.438&.294&.22&.22\\

content type&\textbf{.978}&\textbf{.88}&\textbf{.926}&\textbf{.88}&\textbf{.88}\\

environmental descriptors&.041&.07&.052&.06&.065\\

keywords&.034&.289&.062&0&.045\\

plot level&.148&.236&.181&.24&.245\\

processes&.051&.068&.058&.04&.048\\

project&.012&.01&.011&.01&.01\\

\hline
\end{tabularx}
\caption{Table of precision, recall, F1-score, accuracy, and jaccard similarity scores for \textit{multi-label} fields for a test run of 100 datasets.}
\label{tab:extractionresultsdatasets2}
\end{table}

For \textit{multi-label} fields of dataset metadata, scores are generally low with two outliers:
The content type category reaches an F1-score of 0.926 and an average jaccard similarity of 0.88 and the biotic data taxon category has an F1-score of 0.594 and an average jaccard similarity of 0.565.
The worst performing category contains project names, achieving scores close to zero for all metrics.

Finally, for datasets we attempted to extract information for free-text metadata categories as well (for example study design and sample preparation).
However, for these categories our evaluation did not seem to return meaningful scores, as the value zero was returned across the board.
We believe that this is due to generally low coverage of these categories in the test set (and the dataset as a whole) and difficulty in evaluating such answers with this approach, as predictions containing free-text answers are difficult to capture with these metrics.

\section{Finding Overarching Topics}
Another possible application of large language models on the source data, that might also be useful for a knowledge graph in its usage and application, leverages text embeddings to find latent, overarching topics of datasets and publications via their titles and abstracts/summaries.

Oftentimes, titles and abstracts do not explicitly state every concept or keyword a publication or dataset is about.
Therefore, it might be possible to find latent concepts using text embeddings that group meaningful texts together or move them apart, depending how semantically related their content is.
Additionally, this allows us to investigate whether there are any clusters of publications and datasets in the embedding space, and what topics these clusters may be about.

In a further investigation, we provide anchor concepts in the embedding space, and investigate how publications and datasets group around them. 
We use concepts from the \textit{NASA Earthdata Global Change Master Directory (GCMD) Keywords}\footnote{\url{https://gcmd.earthdata.nasa.gov/KeywordViewer/}}, a controlled vocabulary about earth science, earth science services, platforms, and more. 
We choose this vocabulary as it provides a wide range of terms that have a higher probability to cover latent concepts.
Additionally, its hierarchical structure allows us to embed anchors on different levels of granularity.
\subsection{Embedding Approach}

Our input data consists of publication and dataset titles and abstracts, concatenated in the form:

\verb|Title: {titlecolumn}\n Abstract: {abstractcolumn}|

For the anchor concepts, we export Keywords in RDF format from the keyword viewer\footnote{\url{https://gcmd.earthdata.nasa.gov/KeywordViewer/}}, SPARQL query for concept labels and definitions, export as csv, and concatenate them as:

\verb|Label: {labelcolumn}\n Definition: {definitioncolumn}|
\\

We split this task into five different steps:
\begin{itemize}
    \item (1) Concatenate dataset and publication input data
    \item (2) Embed documents using an embedding model
    \item (3) Cluster points in the embedding space and label clusters
    \item (4) Embed anchor concepts
    \item (5) Assign documents to anchors    
\end{itemize}

We use the \textit{BGE-M3} embedding model \cite{chen2024bge} available on Hugging Face\footnote{\url{https://huggingface.co/BAAI/bge-m3}}, which is particularly useful in this case because of its multilingual capabilities, as titles and abstracts are written in either german or english and
self host the model on \textit{Draco}\footnote{\url{https://www.uni-jena.de/en/22210/scientific-computing-and-data-science}}, one of the computing clusters of the \textit{Friedrich Schiller Universiy Jena}.
We cluster documents in the embedding space using \textit{HDBSCAN}\footnote{\url{https://hdbscan.readthedocs.io/en/latest/index.html}} \cite{campello2013density} using a minimum cluster size of 20 and compute \textit{c-tf-idf}, the tf-idf per cluster, to label each cluster.

\begin{table}[h]
\centering
\small
\begin{tabularx}{\linewidth}{|c c >{\raggedright\arraybackslash}X|} \hline
   n docs  & id & top terms\\ \hline
    194& 6& soil; use; land; land use; plant; biodiversity exploratories; intensity; an; grassland; biodiversity\\
    70&4&effects; use; land use; land; vegetation; grassland; eps; all; diversity; dataset\\
    37&2&soil; grassland; community; the data; community composition; composition; dataset; we; specific; can be\\
    28&1&effects; biodiversity exploratories; diversity and; diversity; was; study; biodiversity; plant; the data; tree\\
    22&5&biodiversity exploratories; tree; diversity and; duen; forests; hainich duen; the data; eps; collected; study\\
    22&3&plant; soil; eps; all; on all; community; community composition; composition; land; within\\
    21&7&soil; diversity and; community; eps; composition; also; community composition; diversity; on all; within\\
    20&0&vegetation; alb; data of; 2011; 2017; 2021; 2021 and; 150; all forest; abundance\\
    \hline
\end{tabularx}
\caption{Table of topic clusters from embedded publication and dataset titles and abstracts. Column one contains the number of documents per cluster, the second column contains the cluster ID, and the third contains the top terms that occur within the clusters via c-tf-idf.}
\label{tab:cluster1}
\end{table}

The resulting clusters are shown in Table \ref{tab:cluster1}.
There are eight clusters with 20 or more documents belonging to them. 
The biggest cluster contains 194 documents, which is about five percent of the \verb|~|4000 input documents.
Unfortunately, there are no cluster labels, which makes it difficult to distinguish the clusters from each other.
For this reason, we choose to include a range of possible concepts as anchors in the embedding space (GCMD Keywords).
When clustering around these anchors, we get the following results, shown in Table \ref{tab:cluster2}.
Using adaptive clustering, an approach that allows a clustering algorithm to discover clusters of varying shape and densities, yields the clusters shown in Table \ref{tab:cluster3}.

\begin{table}[h!]
\centering
\small
\begin{tabular}{|c c|} \hline
   n docs  & anchor label\\ \hline
    563&nutritional constraints\\
    442&species distribution models\\
    142&morphological adaptation\\
    128&metagenomics\\
    95&plant characteristics\\
    82&soil salinity/soil sodicity\\
    78&land use/land cover change\\
    78&microfossil\\
    71&silviculture\\
    51&biodiversity functions\\
    50&forest mensuration\\
    39&fungi\\
    35&microclimate\\
    \hline
\end{tabular}
\caption{Table of clusters around anchor concepts.}
\label{tab:cluster2}
\end{table}

\begin{table}[h!]
\centering
\small
\begin{tabular}{|c c|} \hline
   n docs  & anchor label\\ \hline
    113&nutritional constraints\\
    89&species distribution models\\
    33&tree rings\\
    29&morphological adaptation\\
    26&metagenomics\\
    24&soil rooting depth\\
    19&carbon dioxide\\
    19&fatty acid desaturase\\
    19&plant characteristics\\
    18&pollinator species\\
    18&mushrooms\\
    \hline
\end{tabular}
\caption{Table of clusters around anchor concepts with adaptive clustering that allows clusters of varying shapes and densities, leading to an overall smaller number of points per cluster.}
\label{tab:cluster3}
\end{table}

\subsection{Clustering application for the Knowledge Graph}
In this subsection we expand upon the possible upside of this embedding based clustering of publications and datasets for the knowledge graph.

First, due to these results it becomes possible to filter resources in the KG depending on the cluster they belong to.
This introduces a further facet of how data in the KG can be viewed through, and may assist users in getting an overview of the contents of the KG.
Thus, it also may assist users in their ability to search for topic-specific literature for a synthesis effort or meta study.

Second, these clusters form around concepts, keywords, or topics, that are latent in the input data, or are provided from outside of the original input data. 
On the one hand, this may discover research avenues of the Biodiversity Exploratories that are not mentioned explicitly in the title or abstract of publications and datasets.
On the other hand this introduces the possibility to compare topic clusters to those of other taxonomies, ontologies, or other concept hierarchies, contributing to matching efforts between hierarchies, or the integration of BE data to existing sources.

To this end, semantic units can be included in the knowledge graph that represent clustered concepts.
By implementing a compound unit for every concept, publications and datasets, along with their relatedness score to that concept, can be associated and queried for in the knowledge graph.

\subsection{Linking resources to the main objectives of the BE using embeddings}
Taking the idea of using anchors in the embedding space further, we can chose concepts more related to the Biodiversity Exploratories as well.
As stated in Sec. \ref{sec:BE} (see Fig. \ref{fig:BE}) the research goals of the BE investigate the interactions between land use types and intensities, biodiversity and community composition, and ecosystem processes and functioning.
In this section we investigate the idea of using these research goals as anchors in the embedding space:
\begin{itemize}
    \item \textbf{G1:} How does land use type and intensity affect biodiversity?
    \item \textbf{G2:} How does land use type and intensity affect ecosystem process?
    \item \textbf{G3:} How do different components of biodiversity interact?
    \item \textbf{G4:} How does biodiversity influence ecosystem processes and services?
\end{itemize}

For better coverage across these goals we reformulate these goals with \textit{GPT-5}\footnote{\url{https://openai.com/index/introducing-gpt-5/}} as follows:
\begin{itemize}
    \item \textbf{G1':} Investigate how land-use configuration and management intensity shape species richness and community composition.
Quantify biodiversity responses along gradients of land-use type and use intensity.
Determine how different land-use practices and their intensity alter the diversity of taxa and functional groups.
Assess the effects of landscape structure and land-use pressure on biodiversity patterns.
Examine how the form and degree of land use drive changes in organisms and their diversity.
    \item \textbf{G2':} Evaluate how land-use types and their intensity influence key ecosystem processes such as productivity and nutrient cycling.
Test how management regime and land-use configuration modify rates of decomposition, carbon fluxes, and energy flow.
Analyse process-level responses (e.g., soil respiration, primary production) along land-use intensity gradients.
Determine how spatial arrangement and intensity of land use affect ecosystem functioning.
Examine the sensitivity of ecosystem processes to differences in land-use form and management intensity.
    \item \textbf{G3':} Explore interactions among biodiversity components (genes, species, functional traits, trophic levels) within ecosystems.
Investigate how species, functional groups, and trophic guilds influence each other through competition, facilitation, and predation.
Determine the network structure of biodiversity interactions and how they vary across contexts.
Analyse cross-component linkages (taxonomic, functional, phylogenetic) and their ecological consequences.
Assess how changes in one biodiversity facet propagate to others through biotic interactions.
    \item \textbf{G4':} Quantify how biodiversity drives ecosystem functioning and the delivery of services such as pollination, carbon storage, and water regulation.
Test the relationships between biodiversity (richness, evenness, trait diversity) and ecosystem process rates and service provision.
Determine how losses or gains in biodiversity alter ecosystem processes and the benefits people receive.
Evaluate the pathways by which biodiversity affects ecosystem functioning and service outcomes.
Examine biodiversity–function relationships and their implications for ecosystem services across landscapes.
\end{itemize}

We embed these anchors and compare their cosine similarity scores to those of the document embeddings. 
This method allows us to define a threshold score: If a document candidate has a higher similarity to one of the goals than the threshold, then we assign it to that goal.
Therefore, the threshold represents a minimum acceptance barrier that allows us to fine-tune the tradeoff between overall coverage (of all documents, how many get assigned to atleast one goal) and absolute cosine similarity values (how confident we are in the assignments).

Another important metric is the margin between the top two candidates regarding similarity score.
This metric quantifies for a
For an embedded document, this metric quantifies how close the top candidate research goals are to each other regarding cosine similarity, an important metric as for many documents in our data, the cosine similarities of the highest scoring research goals have a margin between 0 and 0.05.

The clustering, goal assignment, and analysis can be run with a variety of settings.
Below, we compile the results with the following parameters: cosine similarity $>$ 0.3, no adaptive clustering, multiple goal assignments when top candidates are $<=$ 0.03 apart, and a maximum of two labels per document.

\begin{figure}[t]
\includegraphics[width=\linewidth,height=\linewidth,keepaspectratio]{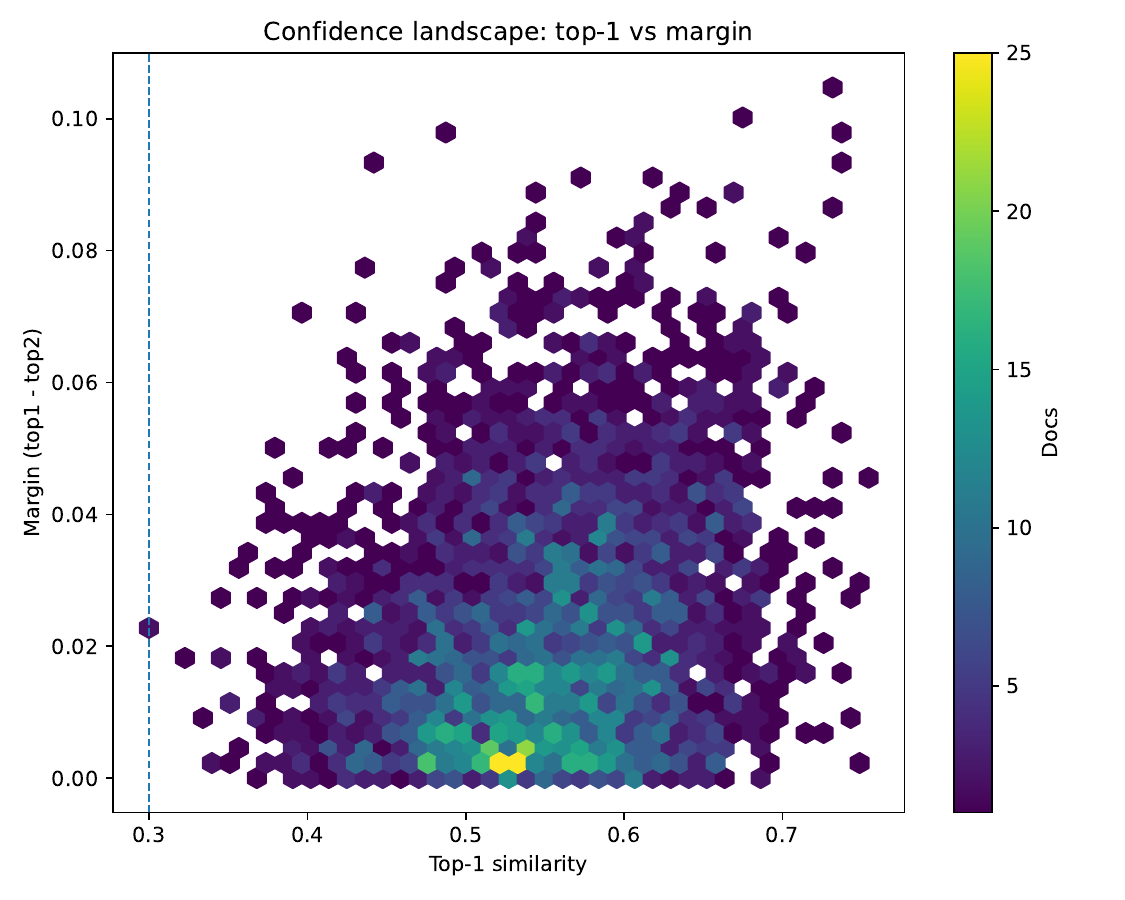}
  \caption{Plot of cosine similarity scores (x-axis) of the top candidates and the margin between top1  and top2 candidates (y-axis).
  Instead of dots, colored hexes are used to represent the number of documents per coordinate, with darker colors (purple) representing lower number of documents, and lighter colors (yellow) representing higher number of documents. The light blue dotted line on the left side of the plot represents the cosine similarity threshold set for our runs of the scripts, accepting candidates at a score of $>=$0.3.}
  \label{fig:top1vsmargin}
\end{figure}

\begin{figure}[h]
  \centering
\includegraphics[width=1\linewidth,height=\linewidth,keepaspectratio]{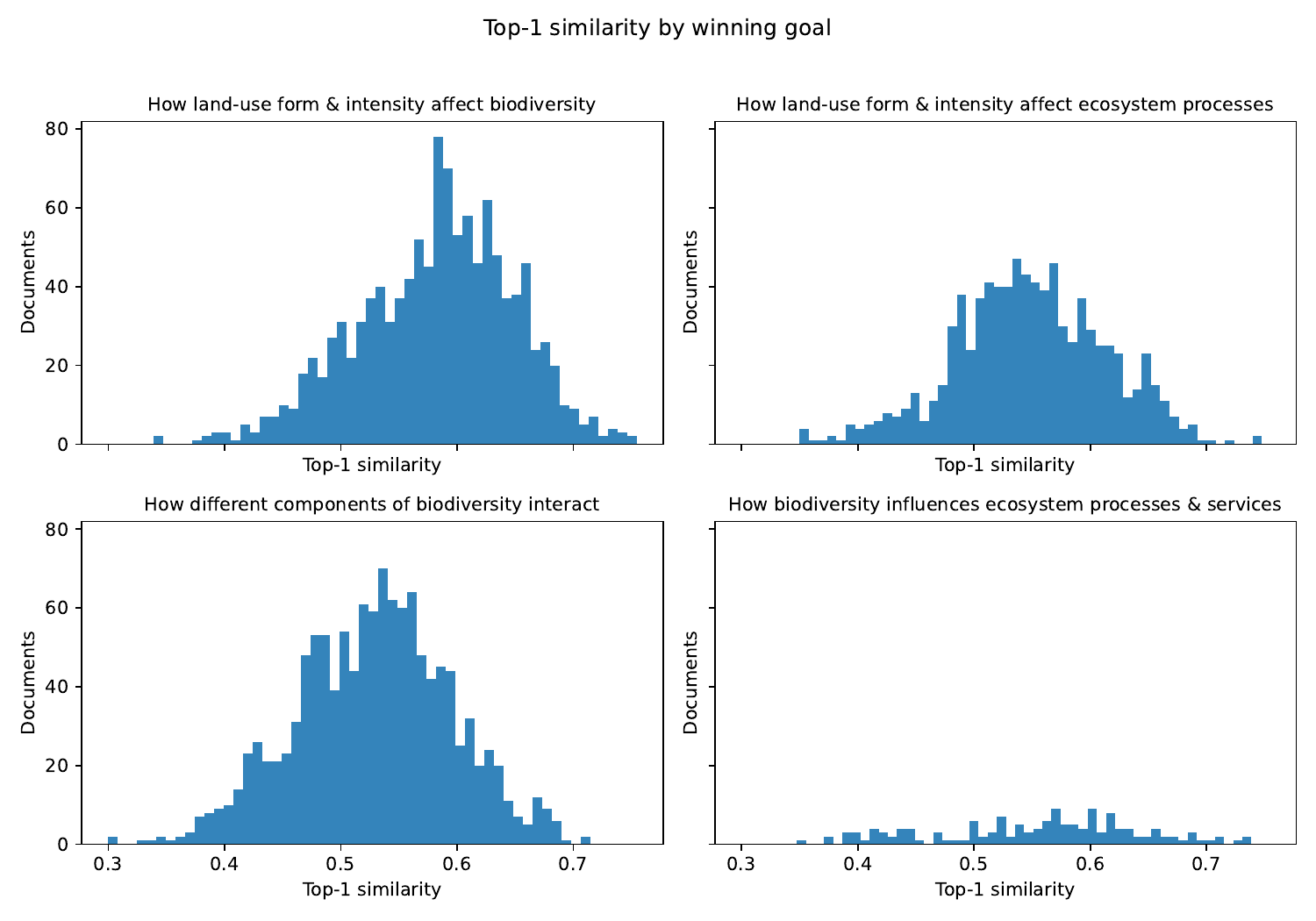}
  \caption{Collection of plots for the four research goals of the Biodiversity Exploratories. Plots show the distribution of cosine similarity scores (x-axis) and total number of documents per score (y-axis).}
  \label{fig:allgoals}
\end{figure}

First, in Fig. \ref{fig:top1vsmargin} we compile an overview of top-1 similarity scores (cosine similarity) on the x axis, the margin between top-1 and top-2 results on the y axis, and colored hexes on the plot for the number of documents, lighter colors referring to more occurrences.
As the figure shows, most documents score between $~$0.4 and $~$0.7, and the margin between the top candidates is more abundant at lower margin values, leading to our decision to allow multiple labels per document.
Also, a hot spot seems to emerge between similarity 0.5 - 0.55 and margin 0.00 - 0.005.
Further, the distribution of similarity scores and document numbers per research goal is shown in Fig. \ref{fig:allgoals}, in which we observe that research goals \textbf{G2} and \textbf{G4} have less total documents assigned to them as the other goals, with \textbf{G4} being a clear outlier.
This means that considerably less publications and datasets investigate the influences of biodiversity on ecosystem processes/services.

\begin{figure}[h]
  \centering
\includegraphics[width=1\linewidth,height=\linewidth,keepaspectratio]{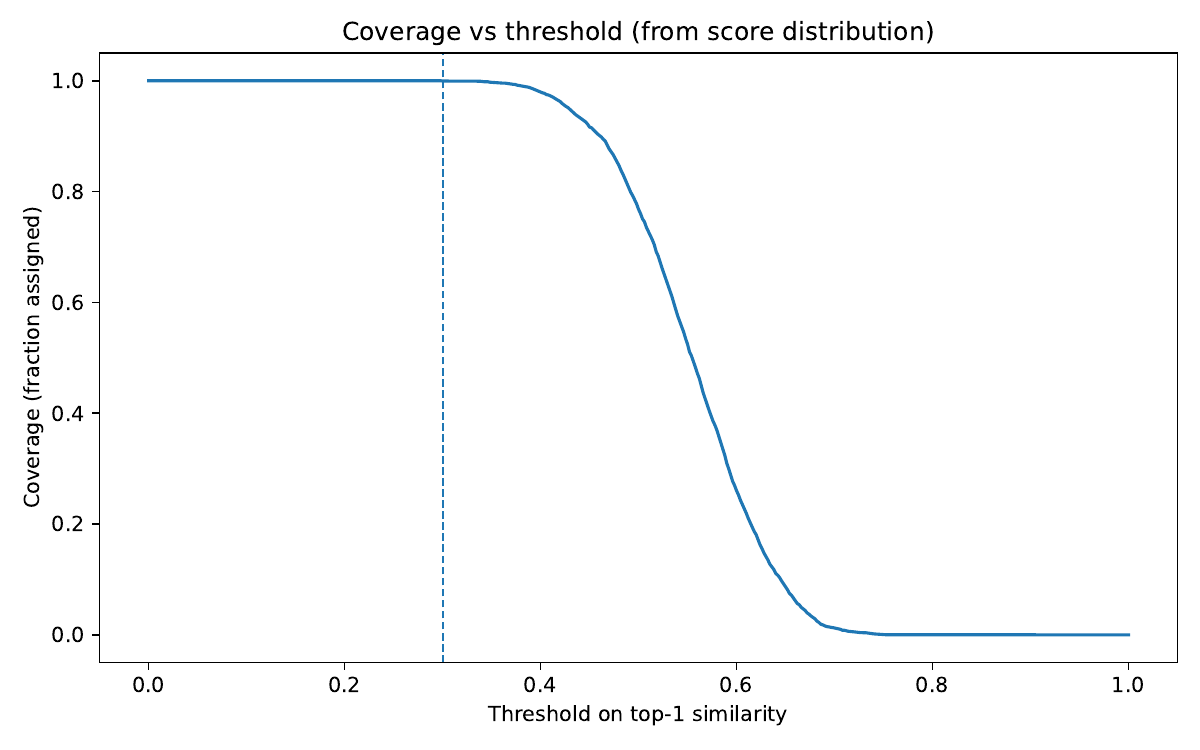}
  \caption{Plot showing the tradeoff between threshold of cosine similarity scores to assign candidates to research goals (x-axis) and the overall document coverage this results in (y-axis). The blue vertical line represents the threshold of $>=$ 0.3 used in our runs.}
  \label{fig:cov}
\end{figure}

Finally, in Fig. \ref{fig:cov} we can observe the tradeoff between coverage and threshold values for similarity scores on the collection of documents. 
We note a sharp decrease in coverage when the threshold is raised to $>$0.4.
These results may be used to infer what threshold may be best if labels resulting from this analysis were to included in the knowledge graph.

With this analysis we conclude our investigation of embedding based methods for topic discovery, keyword clustering, and concept assignment.

\clearpage
\section{Concluding Remarks}
In the this chapter, we presented our investigations of information extraction methods utilizing large language models that may be used to lessen the workload of creating FAIR and structured metadata, aid researchers of the Biodiversity Exploratories in entering metadata to BExIS, and the use of embeddings to find latent information in publication and dataset titles and abstracts to enrich the metadata further.

While these experiments may have only scratched the surface of what could be achieved with these methods, we believe that these preliminary results can still lead to actionable insights that we will discuss in the following chapters, containing the presentations of our results, evaluation of the research questions, and their discussion.
\chapter{Results} \label{chapter8}
In this chapter we present the knowledge graph as the result of the modelling process detailed in Ch. \ref{chapter6}.
We will first give an overview of the graph, explain how it can be queried using SPARQL, present queries to answer the competency questions, and present visualizations of the semantic units constructed on the KG.

\section{Overview}
To fill the developed knowledge graph schema with data, we wrote three main R2RML mapping files, for publications, datasets, and links between the two. 
These mappings span \verb|~|11000 lines total, and are used to turn the contents of three \textit{PostgreSQL}\footnote{\url{https://www.postgresql.org/}} database tables into RDF files.
The database tables consist of SQL exports from Ontotext Refine projects that contain raw data from the BExIS API, and hundreds of transformation steps, for which provenance data is saved in the project files.
We upload the RDF files to the graph store \textit{Ontotext GraphDB}\footnote{\url{https://www.ontotext.com/products/graphdb/}}.
The final knowledge graph submitted as the result of this work contains 763987 triples total.

\section{Querying Semantic Units}
Before presenting the results for the competency questions, we believe there is an advantage in explaining the patterns that can be used to query semantic units such that the following sections can be understood easier.
\subsection{First Query Pattern}
In the first pattern, shown in Listing \ref{lst:pattern1} below, shows a simple pattern that lists all triples that are contained in a specific semantic unit.
Via the \textbf{GRAPH} keyword (line 3), we are able to specify the IRI of a semantic unit, and query over all triples via ?s ?p ?o as with other known graph queries.

\begin{lstlisting}[style=prompt,
  caption={SPARQL query to retrieve all triples contained in a specific semantic unit.},
  label={lst:pattern1},
  captionpos=b]
SELECT ?s ?p ?o
WHERE {
  GRAPH <SU_IRI> { ?s ?p ?o }
}
\end{lstlisting}
\begin{figure}[h]
  \centering
\includegraphics[width=1\linewidth,height=\linewidth,keepaspectratio]{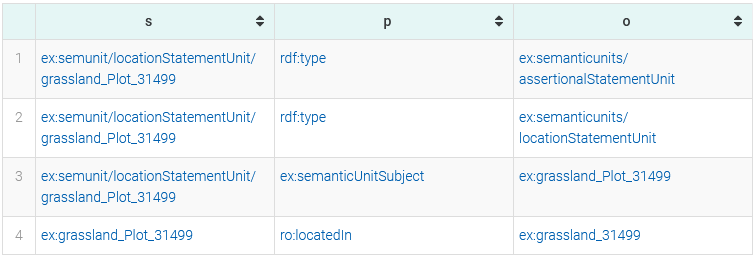}
  \caption{Results for the first semantic unit query pattern.}
  \label{fig:respattern1}
\end{figure}

\begin{sloppypar}
Inputting the IRI of a specific semantic unit in this pattern, for example a location statement unit that states that a resource is located in a different resource, returns results as shown in Fig. \ref{fig:respattern1}.
The location statement unit for grassland plots of Publication 31499 contains two type triples (the semantic unit assignments), has a semantic unit subject triple, and the semantically meaningful information from the data graph: grassland\_Plot\_31499 is located in grassland\_31499.
\end{sloppypar}
\subsection{Second Query Pattern}
We introduce a second query pattern in Listing \ref{lst:pattern2} that is useful for graph exploration when inspecting compound units.
The only difference to the first query pattern is that instead of returning all triples contained within a semantic unit, we only retrieve triples over the hasAssociatedSemanticUnit property.
\begin{lstlisting}[style=prompt,
  caption={SPARQL query to retrieve all semantic units that are associated with a compound unit.},
  label={lst:pattern2},
  captionpos=b]
PREFIX rdfs: <http://www.w3.org/2000/01/rdf-schema#>
SELECT ?Links
WHERE {
  GRAPH <Compound_IRI> {
    ?s <http://example.com/base/semanticunits/hasAssociatedSemanticUnit> ?o .
  }
}
\end{lstlisting}

\begin{figure}[h]
  \centering
\includegraphics[width=1\linewidth,height=\linewidth,keepaspectratio]{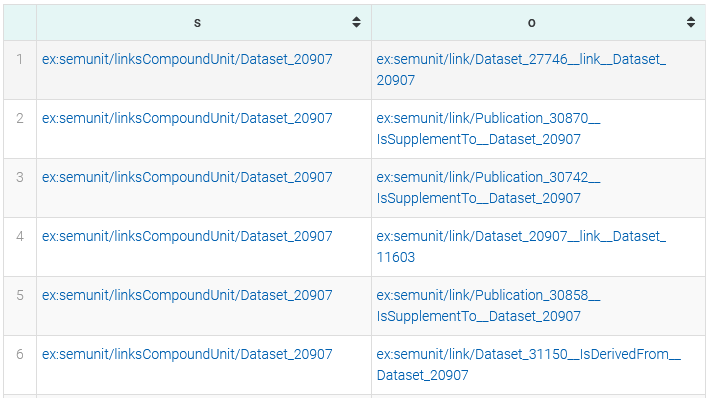}
  \caption{Excerpt of the results for \textbf{CQ D2}, queried with semantic units.}
  \label{fig:respattern2}
\end{figure}

Fig. \ref{fig:respattern2} shows the results for the second query pattern.
The subject in the result list is a compound unit for the links to or from Dataset 20907, and the objects are semantic units, that contain single links.
For example, in result two we can observe that Publication 30879 is supplement to Dataset 20907, and in result six we observe that Dataset 31150 is derived from Dataset 20907.







\section{Evaluation using competency questions}\label{sec:evalCQ}
In this section we evaluate whether the competency questions for the knowledge graph can be answered, and the queries necessary to do so.
For better readability, we restate the competency questions:

\textbf{Competency Questions}
\begin{itemize}
    \item \textbf{CQ P1:} Who authored the publication \textbf{\emph{P}}?
    \item \textbf{CQ D1:} Who authored dataset \textbf{\emph{D}}?
    
    \item \textbf{CQ P2:} What datasets is publication \textbf{\emph{P}} linked to and what are the types of connections? 
    \item \textbf{CQ D2:} What publications is dataset \textbf{\emph{D}} linked to and what are the types of connections?
    
    \item \textbf{CQ P3:} For publication \textbf{\emph{P}}, what are the plot levels of the plots it investigated, were they forest or grassland plots, what experiment do they belong to, and what exploratory or multiple exploratories do those plots cover?
    \item \textbf{CQ D3:} For dataset \textbf{\emph{D}}, what are the plot levels of the plots it investigated, were they forest or grassland plots, what experiment do they belong to, what exploratory or multiple exploratories do those plots cover, and how many plots per plot level are investigated?
    
    \item \textbf{CQ P4:} What are all semantic units that have publication \textbf{\emph{P}} as subject, or that are associated with it?  
    \item \textbf{CQ D4:} What are all semantic units that have dataset \textbf{\emph{D}} as subject, or that are associated with it? 
    \item \textbf{CQ 5:} List all publications that belong to the infrastructure \textit{Instrumentation/Remote sensing} that investigate plant productivity on grassland plots. What are the datasets they reference and what are their reference types?
\end{itemize}

\subsection{CQ P1 and CQ D1}
To answer the first set of questions, the knowledge graph has to be able to connect publications and datasets to the people involved in their creation and assign appropriate roles to them.
We represent this information using several semantic units:
\begin{itemize}
    \item (1) Named individual identification units for all entities involved,
    \item (2) several statement units that declare individual people as either first authors or co-authors of a publication, \textit{creator statement units} that link authors to publications, and
    \item (3) a compound unit that bundles this information together
\end{itemize}

\begin{figure}[h]
  \centering
\includegraphics[width=1\linewidth,height=\linewidth,keepaspectratio]{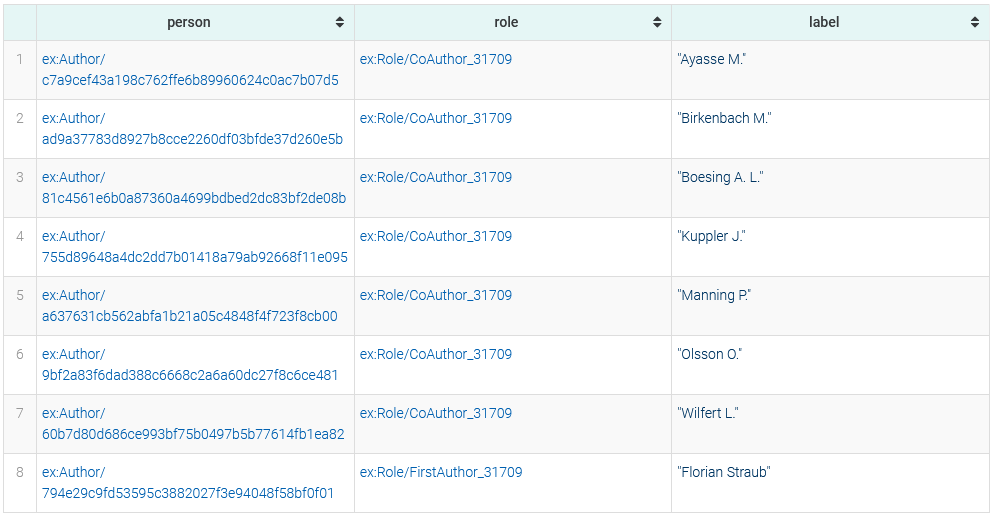}
  \caption{Results for \textbf{CQ P1}, queried without semantic units.}
  \label{fig:res1}
\end{figure}

\begin{figure}[h]
  \centering
\includegraphics[width=0.8\linewidth,height=0.8\linewidth,keepaspectratio]{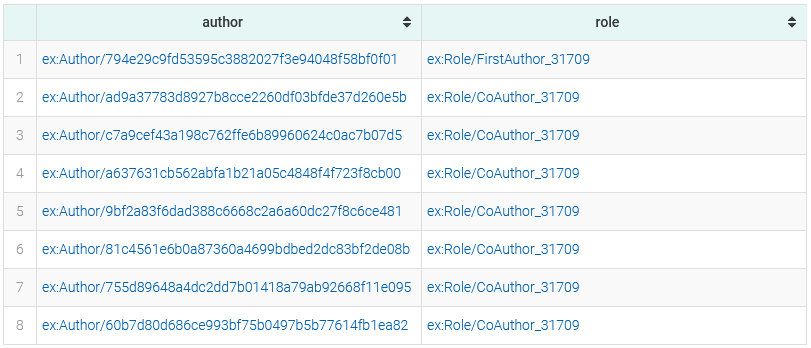}
  \caption{Results for \textbf{CQ P1}, queried with semantic units.}
  \label{fig:res1su}
\end{figure}

We then answer the questions using SPARQL queries, and list queries without and with semantic units below in Listings \ref{lst:sparql1} and \ref{lst:sparqlsu1}, and the results of those queries in Fig. \ref{fig:res1} and \ref{fig:res1su}.
Regarding the first query, we select persons, roles, and labels, and filter for roles that belong to the publication with the ID 31709. For this we must know the IRI of the first author and co-author roles.

The second query searches for all creator and role statements made inside of the authors and role compound unit of Publication 31709.
This returns authors and roles, but not labels (we can however include labels in compound units).

\begin{lstlisting}[style=prompt,
  caption={SPARQL query to answer \textit{Who authored the publication \textbf{P}?} without using semantic units.},
  label={lst:sparql1},
  captionpos=b]
PREFIX rdfs: <http://www.w3.org/2000/01/rdf-schema#>
PREFIX ro:   <https://www.ebi.ac.uk/ols4/ontologies/ro>

SELECT ?person ?label ?role
WHERE {
  ?person ro:hasRole ?role ;
          rdfs:label ?label .
  VALUES ?role {
    <http://example.com/base/Role/FirstAuthor_31709>
    <http://example.com/base/Role/CoAuthor_31709>
  }
}
ORDER BY ?role ?label
\end{lstlisting}

\begin{lstlisting}[style=prompt,
  caption={SPARQL query to answer \textit{Who authored the publication \textbf{P}?} utilizing semantic units.},
  label={lst:sparqlsu1},
  captionpos=b]
PREFIX rdfs: <http://www.w3.org/2000/01/rdf-schema#>
SELECT ?author ?role
WHERE {
  GRAPH <http://example.com/base/semunit/authorsAndRolesCompoundUnit/Publication_31709> {
    ?publication <http://purl.org/dc/terms/creator> ?author .
    ?role <https://www.ebi.ac.uk/ols4/ontologies/roroleOf> ?author .
  }
}
\end{lstlisting}

\clearpage
\subsection{CQ P2 and CQ D2}
For the second set of competency questions, we investigate links and link types between datasets and publication.
We take as example the Dataset 20907, that is connected to various other datasets and publications through various link types.
Listing \ref{lst:sparql2} shows how we query for this information without semantic units.
We query over the predicates \textit{ex:linksTo} and \textit{ex:linksFrom} to return all targets and sources, and then query for predicates that further specify the connection type. 

\begin{figure}[h]
  \centering
\includegraphics[width=1\linewidth,height=\linewidth,keepaspectratio]{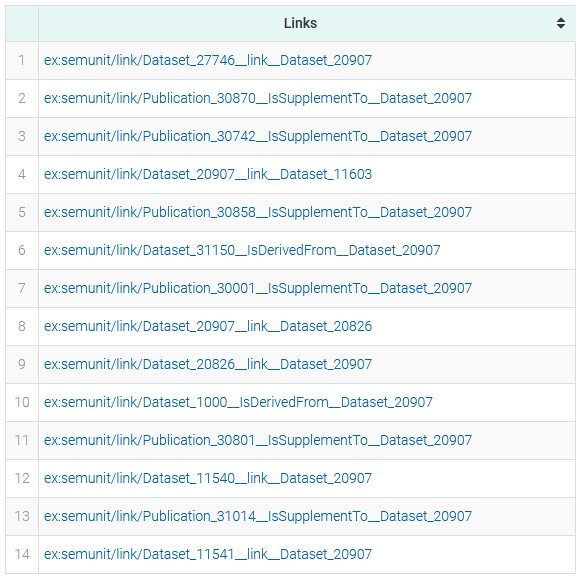}
  \caption{Results for \textbf{CQ D2}, queried with semantic units.}
  \label{fig:res2su}
\end{figure}

To make the semantic unit approach possible for this competency question, we create identification units for all publications and datasets, create statement units for all types of links that are possible (\textit{linksTo, linksFrom, isSupplementTo, link, etc.}), and create compound units that collect all link statement units for a given subject.
Once the semantic units are created, we can query the knowledge graph as shown in Listing \ref{lst:sparql2su} to return an overview of all connections, and are able to query those units for further information.
We omit the results for the query without semantic units because the result table is very large, but show the results of the associated semantic unit approach in Fig. \ref{fig:res2su}.
Note that we are also able to query for the exact same results with semantic units as without them, if we query for link properties as shown in Listing \ref{lst:sparql2}, but gain the upside of querying for associated semantic units first to gain an overview.
\begin{lstlisting}[style=prompt,
  caption={SPARQL query without semantic units that returns all links and link types coming to and from Dataset 20907.},
  label={lst:sparql2},
  captionpos=b]
PREFIX ex: <http://example.com/base/>
select * where { 
    ex:Dataset_20907 ex:linksTo ?target .
    ex:Dataset_20907 ?propertyTarget ?target .
    FILTER(?propertyTarget != ex:linksTo)
    FILTER(?propertyTarget != ex:linksFrom)
    
    ex:Dataset_20907 ex:linksFrom ?source .
    ex:Dataset_20907 ?propertySource ?source .
    FILTER(?propertySource != ex:linksFrom)
    FILTER(?propertySource != ex:linksTo)
} limit 100 

\end{lstlisting}

\begin{lstlisting}[style=prompt,
  caption={SPARQL query with semantic units that returns all links and link types coming to and from Dataset 20907.},
  label={lst:sparql2su},
  captionpos=b]
PREFIX rdfs: <http://www.w3.org/2000/01/rdf-schema#>
SELECT ?Links
WHERE {
  GRAPH <http://example.com/base/semunit/linksCompoundUnit/Dataset_20907> {
    ?dataset <http://example.com/base/semanticunits/hasAssociatedSemanticUnit> ?Links.
  }
}
\end{lstlisting}

\clearpage
\subsection{CQ P3 and CQ D3}
To answer the third set of competency questions, information covering multiple edges in the knowledge graph must be retrieved.
We need to query for the publication or dataset that is the source of a plot collection, retrieve what plots are parts of the plot collection and whether they are located in forest or grassland plots, and find the plot levels of the plot collection and the exploratory it belongs to.

For these purposes, we once again construct named individual identification units for all entities in the knowledge graph, and define three types of statement units: source statement units that connect plots collections to publications and datasets, location statement units that contain the information that certain plots are located in forest or grassland environments, and parthood statement units, that convey that plots are parts of plot collections and that plot collections are parts of plot levels and exploratories.
Finally, we create a compound unit that is associated with all listed statement units and has as its subject a plot collection.

We list the SPARQL query that retrieves this information without querying semantic units in Listing \ref{lst:sparql3} and the query using semantic units is shown in Listing \ref{lst:sparql3su}.

\begin{figure}[h]
  \centering
\includegraphics[width=1\linewidth,height=\linewidth,keepaspectratio]{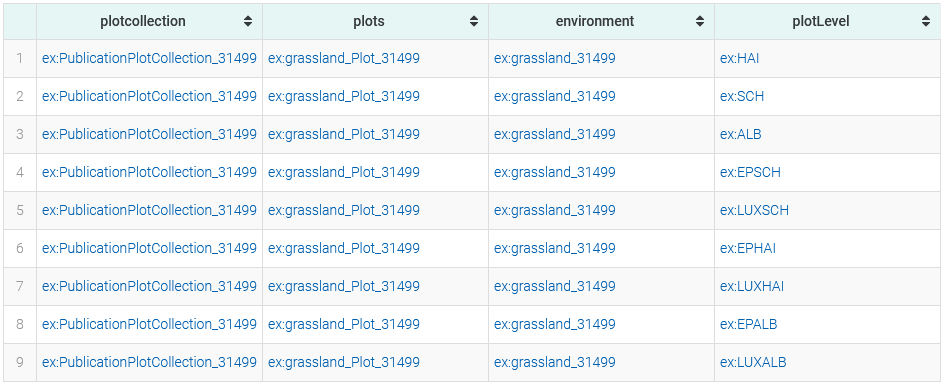}
  \caption{Results for \textbf{CQ D3}, queried without semantic units.}
  \label{fig:res3}
\end{figure}

The retrieved results are shown Fig. \ref{fig:res3}, where we observe that Publication 31499 investigates grassland plots across experimental plots of all exploratories, and is part of the LUX experiment.

\begin{lstlisting}[style=prompt,
  caption={SPARQL query without semantic units that returns the plot collection, plots and the environments they are located in, and the plot level and exploratory the plot collection is part of for Publication 31499.},
  label={lst:sparql3},
  captionpos=b]
PREFIX ex: <http://example.com/base/>
PREFIX iao: <http://purl.obolibrary.org/obo/iao.owl>
PREFIX dcterms: <http://purl.org/dc/terms/>
PREFIX obi: <http://purl.obolibrary.org/obo/OBI/>
PREFIX ro: <https://www.ebi.ac.uk/ols4/ontologies/ro>
select * where { 
    <http://example.com/base/Publication_31499> a iao:publication .
    ?plotcollection dcterms:source <http://example.com/base/Publication_31499> .
    ?plots obi:partOf ?plotcollection .
    ?plots ro:locatedIn ?environment .
    ?plotcollection obi:partOf ?plotLevel .
} limit 100 
\end{lstlisting}

\begin{lstlisting}[style=prompt,
  caption={SPARQL query using a compound unit that returns the plot collection, plots and the environments they are located in, and the plot level and exploratory the plot collection is part of for Publication 31499.},
  label={lst:sparql3su},
  captionpos=b]
PREFIX rdfs: <http://www.w3.org/2000/01/rdf-schema#>
PREFIX obi: <http://purl.obolibrary.org/obo/OBI/>
PREFIX dcterms: <http://purl.org/dc/terms/>
PREFIX ro: <https://www.ebi.ac.uk/ols4/ontologies/ro>

SELECT ?plotCollection ?plots ?plotLevel ?environment WHERE {
  GRAPH <http://example.com/base/semunit/Exploratory_PlotLevel_Collection_Plots_Environment_ Publication_CompoundUnit/Publication_31499> {
    ?plotCollection obi:partOf ?plotLevel .
	?plots obi:partOf ?plotCollection .
	?plots ro:locatedIn ?environment .
	?plotCollection dcterms:source ?publication .
  }
} 
\end{lstlisting}

\clearpage
\subsection{CQ P4 and CQ D4}
\begin{figure}[h]
  \centering
\includegraphics[width=1\linewidth,height=\linewidth,keepaspectratio]{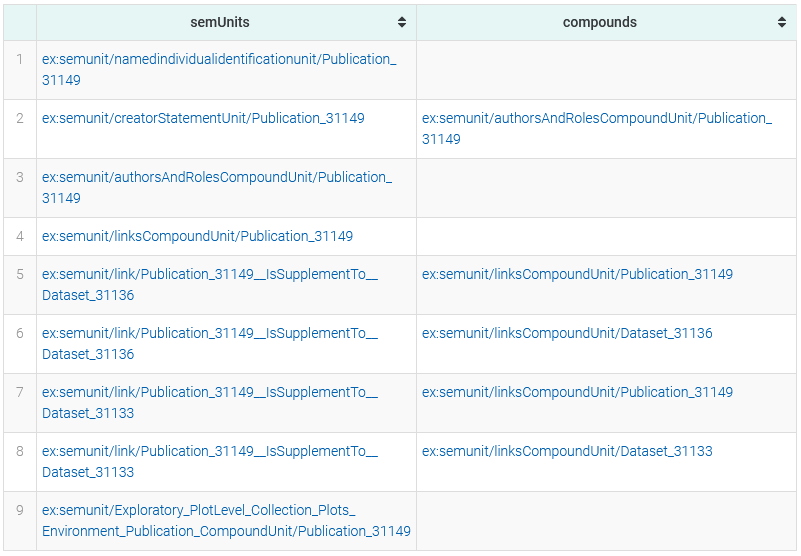}
  \caption{Results for \textbf{CQ P4}.}
  \label{fig:CQ4SU}
\end{figure}

To answer the fourth set of competency questions, we return all semantic units that have a specific publication or dataset as its subject, and also return all semantic units that the results of this query are associated with.
Listing \ref{lst:sparql4} contains this SPARQL query.
We first query via the semantic unit subject property in line four, and then append further optional associations in line five.
This query returns a list of nine results that we present in Fig. \ref{fig:CQ4SU}, containing an identification unit, statement units related to creators, authors, links between datasets and publications, and the compound unit created to answer the previous competency question.

\begin{lstlisting}[style=prompt,
  caption={SPARQL query that returns all semantic units that have the Publication 31149 as its subject, and all further associated semantic units.},
  label={lst:sparql4},
  captionpos=b]
PREFIX ex: <http://example.com/base/>
PREFIX iao: <http://purl.obolibrary.org/obo/iao.owl>
select * where { 
    ?semUnits <http://example.com/base/semanticUnitSubject> <http://example.com/base/Publication_31149> .
    OPTIONAL{?compounds <http://example.com/base/semanticunits/hasAssociatedSemanticUnit>+ ?semUnits .}
} limit 100  
\end{lstlisting}

\clearpage
\subsection{CQ 5}
To answer the final competency question we must arrange a variety of information across the knowledge graph, querying for both publications, datasets, links, infrastructures, plots, environments, species, and processes and services.
The SPARQL query necessary to return this information therefore is more complicated than previous queries.
We list the query that does not utilize semantic units in Listing \ref{lst:sparql5}, and omit prefixes in this listing to save space.

\begin{lstlisting}[style=prompt,
  caption={SPARQL query to answer CQ5 that does not query over semantic units.},
  label={lst:sparql5},
  captionpos=b]
select DISTINCT ?publication ?title ?titleTo ?linkPropTo ?projectToLabel ?titleFrom ?linkPropFrom ?projectFromLabel where { 
    ?publication a iao:publication .
    ?publication obi:hasPart ?titleEntity .
    ?titleEntity a iao:documentTitle .
    ?titleEntity dcterms:description ?title .
    ?publication obi:partOf <http://example.com/base/BE_Infrastructure_Instrumentation_RemoteSensing> .
    OPTIONAL{
    ?linkTo a iao:dataset .
    ?publication ex:linksTo ?linkTo .
    ?linkTo dcterms:title ?titleTo .
    ?publication ?linkPropTo ?linkTo .
    FILTER(?linkPropTo != ex:linksTo)
    }
    OPTIONAL{
    ?linkFrom a iao:dataset .
    ?publication ex:linksFrom ?linkFrom .
    ?linkFrom dcterms:title ?titleFrom .
    ?publication ?linkPropFrom ?linkFrom .
    FILTER(?linkPropFrom != ex:linksFrom) .
    }
    ?plotcollection dcterms:source ?publication .
    ?plots obi:partOf ?plotcollection .
    ?plots ro:locatedIn ?env .
    ?env a envo:grasslandBiome .
    ?plots obi:hasPart <http://www.gbif.org/species/6> .
    ?process ro:hasParticipant ?plots .
    ?process a ?processtype .
    ?process a <http://vocabs.lter-europe.net/EnvThes/21417> .
} limit 100
\end{lstlisting}
Further, the query over the compound unit implemented for this competency question is shown in Listing \ref{lst:sparql5su} below.
Unfortunately, this query is considerably longer than its counterpart and does not utilize the compound units well as we still have to fetch a lot of information from outside the named subgraph, as not all necessary triples are contained within it.
The query consists of four blocks, the first (lines 2 to 9) collect all compound units created for this competency question and collect outside information while the second and third (lines 10 to 27) collect the titles of linked datasets.
The final fourth block (lines 36 to 44) contains a GRAPH query over the compound unit, that collects triples that are contained within it.
This functions similarly to the query without semantic units and uses the same triple patterns.

To fetch all information from the compound unit alone, it would have to be added to the compound unit in its entirety, including publication and dataset titles of linked resources, or various type information.
We believe that the semantic unit query is subpar than its counterpart in this example because we are searching for a list of publications instead of returning information for a resource that is already known (we can query using its IRI).

\begin{lstlisting}[style=prompt,
  caption={SPARQL query to answer CQ5 that queries over semantic units.},
  label={lst:sparql5su},
  captionpos=b]
select DISTINCT ?publication ?title ?titleTo ?linkPropTo ?titleFrom ?linkPropFrom where { 
    ?compoundUnit a <http://example.com/base/semanticunits/InfrastructureProcessAndServiceEnvironmentPublication LinkProjectsCompoundUnit> .
    ?compoundUnit <http://example.com/base/semanticUnitSubject> ?publication .    
    ?publication obi:hasPart ?titleEntity .
    ?titleEntity a iao:documentTitle .
    ?titleEntity dcterms:description ?title .
    ?env a envo:grasslandBiome .
    ?process a ?processtype .
    ?process a <http://vocabs.lter-europe.net/EnvThes/21417> .	
    OPTIONAL {
        GRAPH ?compoundUnit { ?publication ex:linksTo ?linkTo . }
        ?linkTo a iao:dataset ;
                dcterms:title ?titleTo .
        OPTIONAL {
            GRAPH ?compoundUnit { ?publication ?linkPropTo ?linkTo . }
            FILTER(?linkPropTo NOT IN (ex:linksTo, ex:linksFrom))
        }
    }
    OPTIONAL {
        GRAPH ?compoundUnit { ?publication ex:linksFrom ?linkFrom . }
        ?linkFrom a iao:dataset ;
                  dcterms:title ?titleFrom .
        OPTIONAL {
            GRAPH ?compoundUnit { ?publication ?linkPropFrom ?linkFrom . }
            FILTER(?linkPropFrom NOT IN (ex:linksTo, ex:linksFrom))
        }
    }
    GRAPH ?compoundUnit {
    	?publication obi:partOf <http://example.com/base/BE_Infrastructure_Instrumentation_RemoteSensing> .
    	?plotcollection dcterms:source ?publication .
        ?process ro:hasParticipant ?plots .
    	?plots obi:partOf ?plotcollection .
    	?plots ro:locatedIn ?env .
        ?plots obi:hasPart <http://www.gbif.org/species/6> .
        
        OPTIONAL{
    	?publication ex:linksTo ?linkTo .
    	?publication ?linkPropTo ?linkTo .
    	FILTER(?linkPropTo != ex:linksTo)
    	}
    	OPTIONAL{
    	?publication ex:linksFrom ?linkFrom .
    	?publication ?linkPropFrom ?linkFrom .
    	FILTER(?linkPropFrom != ex:linksFrom) .
    	}     
	}
} limit 100 

\end{lstlisting}

\begin{figure}[h]
  \centering
\includegraphics[width=1\linewidth,height=\linewidth,keepaspectratio]{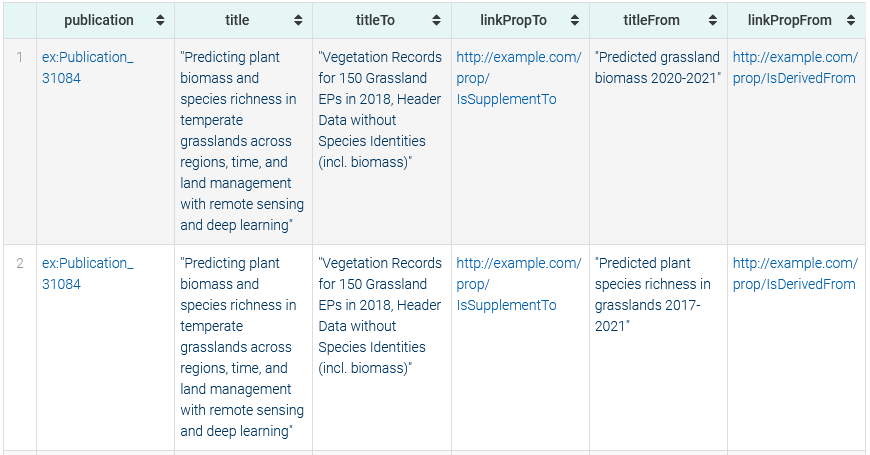}
  \caption{Results for \textbf{CQ 5}.}
  \label{fig:CQ5}
\end{figure}

We present an excerpt of the results from this query in Fig. \ref{fig:CQ5} that shows the first two results in which Publication 31084 fits the query requirements and is supplement to a dataset about vegetation records and two further datasets that are derived from the publication about predicted grassland biomass and predicted plant species richness in grasslands.
In total this query returns 24 results.

\section{Semantic Unit Visualization}
An advantage of semantic units we discovered when testing different SPARQL queries to answer the competency questions, is the visualization of their data graphs, which differs from the visualization of the whole data graph as it contains only relevant triples and is less cluttered as a result.
We believe this might be of great use for users, especially for graph exploration and result representation.
In this section we present visualizations of the compound units used to answer the competency questions.

We are able to create these visualizations through simple CONSTRUCT queries of compound units, as presented in Listing \ref{lst:construct}:

\begin{lstlisting}[style=prompt,
  caption={Listing of a SPARQL CONSTRUCT query that returns a visualization of a links compound unit.},
  label={lst:construct},
  captionpos=b]
CONSTRUCT { ?s ?p ?o }
WHERE {
  GRAPH <http://example.com/base/semunit/linksCompoundUnit/Dataset_20907> { ?s ?p ?o }
}
LIMIT 100  
\end{lstlisting}

\begin{figure}[h]
  \centering
\includegraphics[width=1\linewidth,height=\linewidth,keepaspectratio]{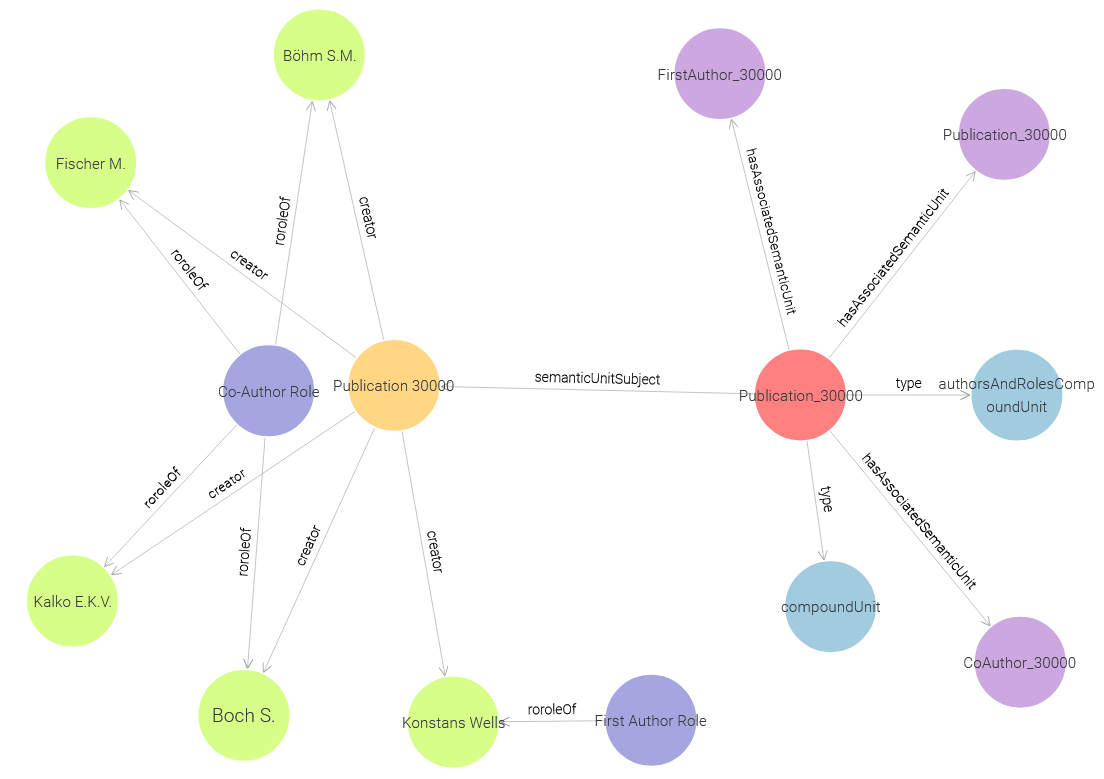}
  \caption{Visual representation of the authors and roles compound unit for Publication 30000.}
  \label{fig:vis1}
\end{figure}

In the first example, shown in Fig. \ref{fig:vis1}, we visualize the compound unit that contains authors of a publication and their roles.
The right hand side of the figure contains the compound unit as a red node.
Attached to it are its types and associated semantic units, and its subject, the Publication 30000 on the left hand side.
Linked to the publication are all creators of it, and the first author or co-author role, depending on their involvement. 
The visual representation even attaches the resource labels dynamically instead of displaying raw IRIs, so users can gain a better understanding of a nodes contents.

\begin{figure}[t]
  \centering
\includegraphics[width=1\linewidth,height=\linewidth,keepaspectratio]{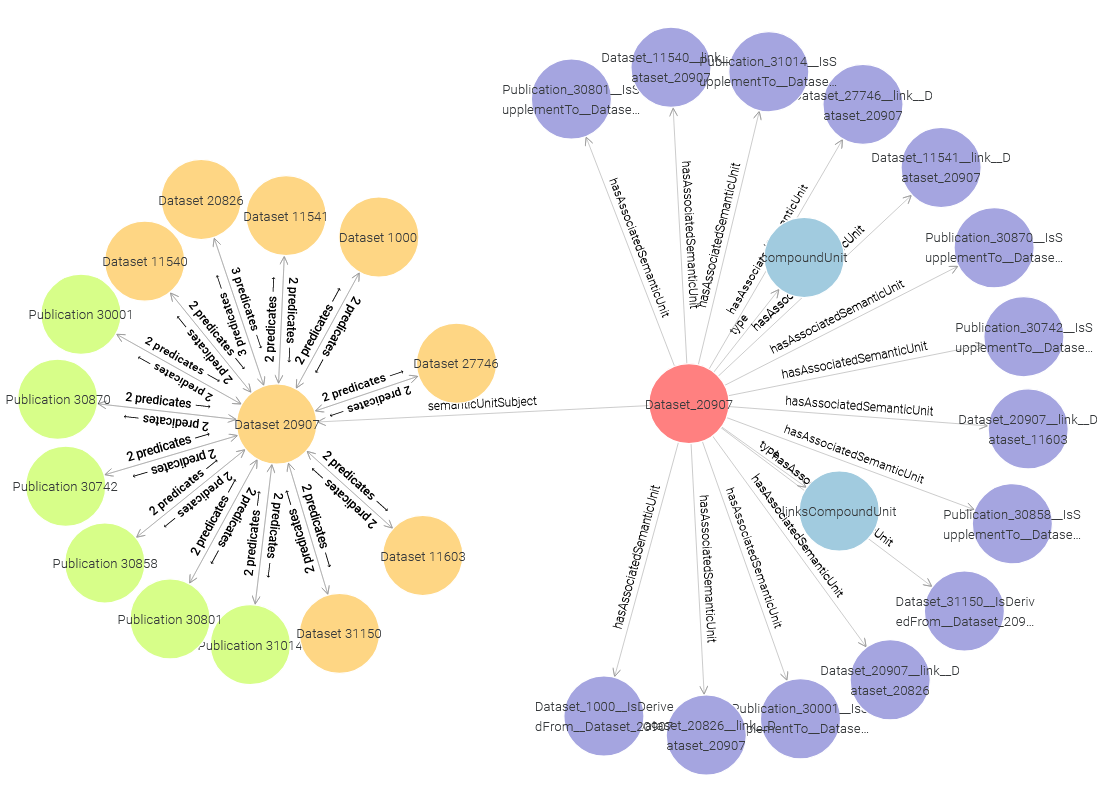}
  \caption{Visual representation of the links compound unit for Dataset 20907.}
  \label{fig:vis2}
\end{figure}

The second example (Fig. \ref{fig:vis2}), shows Dataset 20907 and all of its links and link types (masked in the figure by the \textit{'2 predicates'} statements to other publications and datasets, as well as all semantic units that are associated to the dataset links compound unit.

\begin{figure}[h!]
  \centering
\includegraphics[width=1\linewidth,height=\linewidth,keepaspectratio]{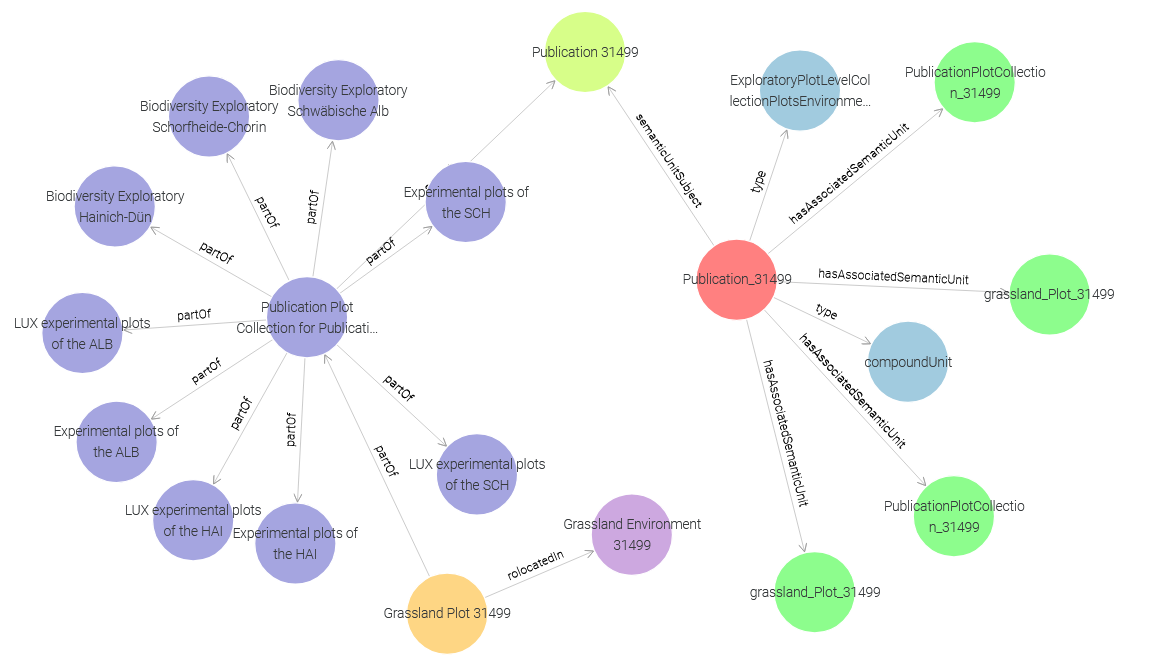}
  \caption{Visual representation of the compound unit developed for the third competency question.}
  \label{fig:vis3}
\end{figure}

\begin{landscape}
\begin{figure}[p]
  \centering
\includegraphics[width=\linewidth,height=\linewidth,keepaspectratio]{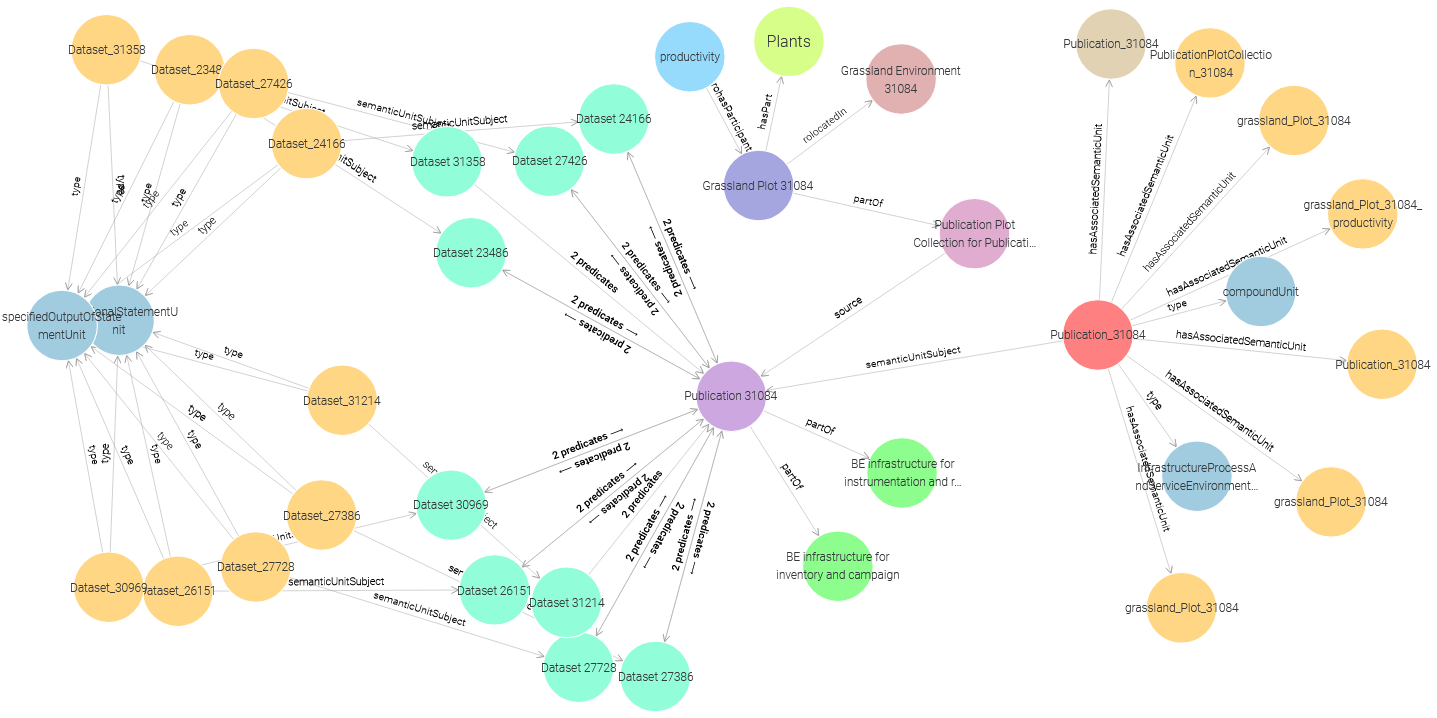}
  \caption{Visual representation of the compound unit developed for the fifth competency question.}
  \label{fig:vis4}
\end{figure}
\end{landscape}

The final two visualizations (Fig. \ref{fig:vis3} and Fig. \ref{fig:vis4}) present the contents of the compound units used to answer competency questions CQ P4, CQ D4, and CQ5.
While the former has a smaller amount of nodes and edges, containing information about plot levels and exploratories, plot collections, grassland plots and environments, and Publication 31499 as the semantic unit subject,
the latter shows the path lengths covered to answer the competency question.
As in previous figures, the right hand side contains information related to semantic units and the seven associated semantic units necessary to answer the competency question.
The center contains Publication 31084 as the semantic unit subject, its infrastructure types, information about the publications content (productivity of plants in grassland environments), and links to other datasets.

With this presentation of semantic unit visualizations, we conclude the results chapter of this thesis in which we covered the constructed knowledge graph, semantic unit querying, and competency question evaluation.
We will discuss these results and the insights we can gain from them in the following chapter, and include further discussions about the limitations of our approach, and present ideas for future work.

\chapter{Discussion} \label{chapter9}
In the previous chapter we presented the knowledge graph that is the main contribution of this work and evaluated whether the competency questions can be answered by querying the knowledge graph with and without the use of semantic units.
We also presented visualizations of compound units that we believe may support users in some application scenarios.

In this chapter, we evaluate whether we are able to answer the research questions with the results gathered in this work, discuss the impact of the competency questions we were able to answer using semantic units and present visualization scenarios for users.
Further, we discuss the limitations of our work and end the chapter with directions for future work.

\section{Evaluation of Research Questions}
Below, we list the research questions and their corresponding sub-questions once more and evaluate whether we are able to answer them using the methods employed in this thesis.

\begin{itemize}
    \item \textbf{RQ1:} Does semantic unit modelling reduce SPARQL query complexity for representative tasks compared to querying the data graph?
    \item \textbf{RQ2:} Can large language models be used to extract metadata categories from publication and dataset titles and abstracts?
    \item \textbf{RQ3:} Can embedding models be used to extract latent information from publication and dataset titles and abstracts to enrich BE metadata?
\end{itemize}
We further split the research questions into sub-questions:
\begin{sloppypar}
\begin{itemize}
    \item \textbf{SQ1:} Can SPARQL queries on the constructed knowledge graph answer sets of competency questions?
    \item \textbf{SQ2:} Can the same SPARQL queries be formulated over semantic units to answer the same sets of competency questions?
    \item \textbf{SQ3:} Is SPARQL query complexity lower for queries that answer competency question when querying over semantic units compared to queries on the data graph only?
    \item \textbf{SQ4:} Which metadata categories can the LLM extraction approach predict reliably (F1-score of 0.8 or higher)?
    \item \textbf{SQ5:} Can an embedding approach assign publication and dataset titles and abstracts to the research goals of the BE?
\end{itemize}
\end{sloppypar}

\subsection{RQ1: Discussion of Competency questions}
We were able to successfully answer all competency questions with the knowledge graph created in this work by querying the base graph and by querying semantic units.
What we aim to discuss further in this section however, is how both query types compare and whether the introduction of semantic units impacts queries.
For this purpose we distinguish two query types:
In the first, we query information about a know resource as in the first four sets of competency questions and in the second, we are searching for resources that fit a specific description like in the last competency question, where we query for all publications of a specific infrastructure, environment type, etc. 

Regarding the first query type, we notice that the queries arguably do not differ much or at all.
Using the first set of competency questions as an example, in which we query for authors and their roles in relation to a publication or dataset, we require the IRI of the first and co-author roles for a specific Publication to filter their roles and labels.
When querying for this information using the authors and roles compound unit, we need the IRI of the compound unit itself, and can then retrieve relevant triples without requiring more specific information.
Additionally, besides the use of a GRAPH keyword in the SPARQL query, the actual query inside that subgraph is rather simple, stating (1) ?publication dcterms:creator ?author and (2) ?role ro:roleOf ?author.
A further result of this is that the semantic unit query is shorter than the data graph query, which is noteworthy but does not allow for any conclusions about actual query complexity.

For the second query type however, we notice substantial difference between the two approaches.
From the perspective of a researcher that created the knowledge graph's schema and its data graph, answering the fifth competency question that connects information across the whole knowledge graph, was rather simple and its complexity derives mostly from two OPTIONAL statements for possible links to and from datasets, which an experienced SPARQL user may already be able to solve without much difficulty.

The experienced difficulty for the query using semantic units however, was much higher. 
In fact it was even more difficult than we expected beforehand, exposing limitations of the semantic unit approach implemented for this knowledge graph.
The approach to answering the competency question follows the same intuition: 
Filter for publications that fit certain criteria and then retrieve further information from them.
However, as the semantic units are implemented in the graph currently, it was impossible to retrieve all information via the compound unit without accessing the data graph directly, as not all information that we expected to require to answer the question was contained in the compound unit.
This resulted in a data graph query with the same complexity as the query without semantic units, that has additional complexity via the GRAPH statement, and added OPTIONAL queries to bind variables reliably.

We argue this complexity derives from two sources, the first being the actual difficulty of answering this query when semantic units are supposed to be used in its query due to the current implementation, and the second being a result of the learning curve that is to be expected when introducing new concepts.
We expect difficulties from the first source to be alleviated by adding additional triples to compound units, such as labels or type definitions to leverage them in the query, but we are currently unsure if this would contradict either the intentions behind or explicit definitions for semantic units.

We believe the learning curve introduced when querying a knowledge graph that contains semantic units should be investigated properly with a user evaluation, which we describe further in the limitations section below.
However, we assume that the experienced query difficulty for users for the first query type should be about even when comparing queries with and without semantic units.
For the second type of queries, we believe that in their current implementation, information should be queried without semantic units, or that query templates should be provided (a functionality of GraphDB).

To conclude, the answers to both \textbf{SQ1} and \textbf{SQ2} is yes on the basis that all competency questions could be answered on the constructed knowledge graph with and without semantic units.
Regarding the third sub-question, \textbf{SQ3}, our findings suggest that for questions to retrieve information about a known resource, experienced SPARQL query complexity is comparable to conventional queries.
For the second query type however, our findings suggest that query complexity is higher compared to conventional queries for the reasons listed above, which leads us to the following answer: 
No, SPARQL query complexity when querying over semantic units is likely not lower than the complexity of conventional queries. (\textbf{Important Note:} As we will discuss in the limitations section below, we were not able to conduct a proper user evaluation, leading these results to not be indicative, but not conclusive, as the sample size for experienced query complexity is only n=1.)

Therefore, our answer to \textbf{RQ1}, the first research question, is that while semantic unit modelling produces a knowledge graph that is able to answer all competency questions for representative tasks, SPARQL query complexity is not reduced compared to conventional queries on the data graph.

\subsection{RQ2: Discussion of Metadata Category Extraction}\label{subs:rq2}
In the method chapter that details our implementation of the extraction approach (Ch. \ref{chapter7}), we list the F1-scores for all extracted categories.
There, we observe that extraction quality differs between categories that only have two possible answers (yes or no), and those categories that allow multiple answers from a selection of keywords.
To answer \textbf{SQ4}, we list the following categories with F1-Scores above 0.8 below. We choose F1-scores as the threshold metric as it represents a balance between precision and recall, and choose the value of 0.8 as we deem this to be a sufficient indicator for the feasibility of this approach if it were to be improved further in the future.
\begin{itemize}
    \item Publications (yes/no): belowground, laboratory
    \item Datasets (yes/no): grassland, forest, aboveground, belowground, plot repetitions
    \item Publications (multi-label): N/A
    \item Datasets (multi-label): content type
\end{itemize}

The extraction method we implemented in this work leaves much room for improvement as we implemented a zero-shot approach.
Therefore, with fine-tuning on test, validation, and evaluation data, and further improvements to the prompting strategy, the results of this extraction could be improved further.
We also anticipate that the evaluation strategy could be enhanced further.
At this point, we would also like to note that for datasets, coverage for many metadata fields and especially dataset abstracts/summaries is rather low, which will make it more difficult to improve this approach further in the future.

This leads us to answer yes to the second research question, \textbf{RQ2}, because we were able to find at least nine metadata categories we were able to extract with an F1-score above 0.8 using the implemented approach.
This however only covers a small amount of total metadata categories, leading us to the conclusion that the approach should be developed further before its used in practice.

To conclude, our results show that this extraction approach has the potential to create structured metadata from publication and dataset titles and abstracts, and support workflows to creating FAIR-aligned metadata, and reduce human effort in creating structured metadata.

\subsection{RQ3: Discussion of Embedding Approach}\label{subs:rq3}
The final research question is concerned with the third part of the implemented method: Utilizing an embedding model to to extract latent information from publication and dataset titles and abstracts.

In the first embedding task we investigated, we were able to find two latent topic clusters that were highly populated:
\textit{land use intensity of grassland soil} (194 documents), and the \textit{effects of land use on vegetation and diversity on grassland experimental plots} (70 documents).

Regarding the second task, clustering documents around anchor concepts, we were able to find a variety of clusters, with non-adaptive clustering showing more promising results than adaptive clustering. The biggest clusters formed around the concepts \textit{nutritional constraints} (563), \textit{species distribution models} (442), \textit{morphological adaptation} (142), \textit{metagenomics} (128), \textit{plant characteristics} (95), \textit{soil salinity/soil sodicity} (82), \textit{land use/land cover change} (78), \textit{microfossils} (78), and more.

In the third task, we embed the main research goals of the Biodiversity Exploratories in the embedding space and compute the cosine similarity of documents to the embedded research goals.
The implementation allows us to tune the results of the assignments using a variety of metrics. We are able to specify a percentage of coverage for the documents, for example 80\%, to adaptively assign a cosine similarity threshold to enforce 80\% of documents to be assigned to a research goal. We can also specify an acceptance threshold for cosine similarity independent of percentage coverage, and we can allow multi-label assignments depending on a specified margin between top candidate research goals.

We believe that these approaches can be improved further, mainly by fine-tuning the embedding model. 
By creating datasets of positive and negative pairs, we would be able to to train the embedding model to place related concepts and documents closer together in the embedding space, and pull unrelated ones further apart.
Also, an evaluation is necessary to judge how accurate the cluster, concept, and research goal assignments are.

Considering these results, we conclude that the answer to both \textbf{SQ5} and \textbf{RQ3} is yes.
The embedding approach is able to assign publication and datasets titles and abstracts to the research goals of the Biodiversity Exploratories (though further evaluation is necessary to judge how good the assignments are), and embedding models can be used to extract latent information from publication and dataset titles and abstracts to enrich BE metadata.

\section{Visualization}
A noteworthy finding in this work regarding semantic units is their ability to provide a basis for visualization of semantically meaningful triples from the knowledge graph, that can be implemented for multiple use cases and integrated easily in an application or user interface.

\begin{figure}[h]
  \centering
\includegraphics[width=1\linewidth,height=\linewidth,keepaspectratio]{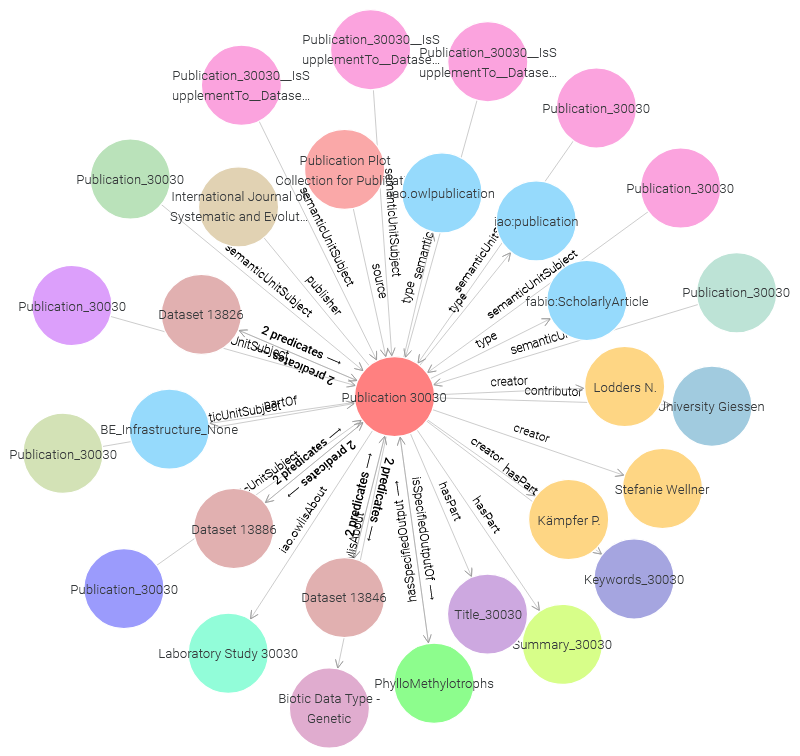}
  \caption{Visual of all nodes and edges connected to Publication 30030.}
  \label{fig:discvis}
\end{figure}

We demonstrate this point by comparing visualizations.
Fig. \ref{fig:discvis} shows the visual graph for all nodes and edges that are connected to Publication 30030 in the center of the figure.
Including data properties, there are 52 connections to or from the publication that when visualized at the same time, are difficult to understand for users.
Compare this to the visualizations from the previous chapter, for example Fig. \ref{fig:vis3} and Fig. \ref{fig:vis4}. 
Even though both visualizations, especially the latter, contain many nodes and edges, we believe that they are still easier to understand than the conventional visualization in the first figure.
This approach also has the added benefit of showing semantically meaningful connections across the knowledge graph, relating information across publications, datasets, plots, environments, etc., and additionally isolating only those connections that are relevant for a given scenario (for example queries relating to competency questions). 
As an added benefit, GraphDB allows users to save these visualizations for themselves or share them with other users as graph \textit{snapshots}.

These snapshots of compound units could be used in applications and user interfaces to give users greater understanding of the knowledge graph and its contents.
One can also imagine a scenario in which these visualizations are created dynamically next to a SPARQL query interface.
In such an application, users could write SPARQL queries using semantic units, and have visualizations of the semantic units on the side to assist in navigating queries, or to gain an overview of information that can be queried within a semantic unit.

For a further application scenario, we refer to \cite{vogt2025rethinkingowlexpressivitysemantic}. 
In the eighth modelling challenge, Vogt presents an approach to creating topic-specific subgraphs using compound units called \textit{standard information units} that can be used to retrieve semantically coherent subgraphs, create views for domain-specific information expectations, support modular graph construction, and more \cite{vogt2025rethinkingowlexpressivitysemantic}.

\section{Remark - Compound Unit Implementation}
At this point in the discussion we add a remark regarding the technical implementation of compound units for readers that might implement a semantic unit approach in a knowledge graph:

We encountered a technical issue at the stage of triple generation in which mappings are executed on the source data that is specific to compound units or those semantic units, that reuse content from others.
When assigning graph maps from one semantic unit to a compound unit, not all triples from the data graph are added to the data graph of the compound unit as expected, as the mappings are not executed transitively.
This leads to the mappings requiring the specific triples that should appear in the compound unit to be assigned separately, introducing additional implementation overhead, as specific triples must be mapped to each semantic unit they should appear in.

\section{Limitations}
In this section we discuss the shortcomings of this work and the reasons behind them.
We encountered many difficulties at various points during modelling and implementation, and believe that there are insights and directions for future work to be gained by examining them further.

\subsection{Time constraints}
The possibly largest limitation we faced in this work were time constraints.
We discuss many different ideas in this thesis, for example general KG modelling, semantic unit modelling, and two LLM tasks, and believe that especially the second part of our method regarding large language models only scratches the surface of what may be possible to achieve for the tested applications, but also for other tasks.

\subsection{Hand-written Mappings / Semantic Unit Mappings}
The R2RML mappings for the knowledge graph and semantic units developed in this thesis were entirely written by hand.
This results in \verb|~|11000 lines of mappings, an immense effort that is currently not scalable in any way and is prone to errors.
Further, the semantic units we implemented for the graph to answer the competency questions were implement for this specific purpose, and therefore, semantic unit coverage on the graph is low.
If semantic units are supposed to be implemented on larger knowledge graphs in the future, although the application in this work already surfaced significant application complexities and challenges, an automated way for mapping generation or significant aids for mapping writing are necessary.
\subsection{Modelling challenges}\label{sec:limitchall}
As we described in Sec. \ref{sec:modellingChallenges}, we encountered several modelling challenges regarding the source data, the open world assumption and how to model complete information, and negations.
We were able to solve most modelling challenges that derive from the source data, resulting in a knowledge graph that is able to answer all competency questions and contains as much knowledge from the publication and dataset metadata as possible, however, the other challenges remain open questions as we were not able to find further answers due to time limitations.
\subsection{User and Performance Evaluation}
One goal of semantic units are to enhance cognitive interoperability for users of the knowledge graph, which we were not able to evaluate as we did not have the time necessary to conduct a user evaluation.
However, we still want to provide a plan for how semantic units could be evaluated using the graph developed in this work:

The evaluation should focus on multiple points, the first being reported query complexity by users when comparing queries with and without semantic units.
The second should focus on graph exploration and is concerned with the question whether semantic units can support users in navigating the knowledge graph, or to get a grasp on the contents of the knowledge graph and the existing connection between data points contained within it.
Further points for a general performance evaluation should focus on actual query performance, meaning how many triples are evaluated per query with and without semantic unit queries, and how that impacts query execution times.
Another important point regards scalability:
Scaling knowledge graphs from thousands to millions to billions of triples is already a difficult technical task and semantic units introduce new triples to a knowledge graph, the impacts of which have not been evaluated previously.

\subsection{LLM Applications}
The two LLM applications covered in this work present valuable insights, but can be improved further from many different perspectives as time limitations did not allow us to improve upon the first results shown in Ch. \ref{chapter7} further.
We outlined possible improvements in the discussion of their respective research questions above (Subsections \ref{subs:rq2} and \ref{subs:rq3}).

\subsection{Closed Source Data}
Currently, publication and dataset metadata of the Biodiversity Exploratories is not available publicly.
Therefore, the instantiated knowledge graph presented in this work can not be made available online.
When the data becomes available, we provide access to the graph in this GitHub repository\footnote{\url{https://github.com/fusion-jena/Biodiversity-Exploratories-Knowledge-Graph}}.

\section{Future Work}
To close out the discussion of the contents and results of this thesis, we discuss directions for future work we believe have potential to contribute to the research field further.
\subsection{Integration into BE Infrastructure}
One point of interest is the integration of this knowledge graph into existing infrastructures and workflows of the Biodiversity Exploratories.
For this the knowledge graph would need complete semantic unit coverage, a SPARQL query interface, and an application to facilitate usage of the graph.

Further, we believe that the challenges and issues we have encountered with the source publication and dataset metadata can provide valuable insights for BExIS and the metadata schemata used within it.
The T-Box developed for this knowledge graph may be used to provide semantic meaning for metadata categories, and the modelling challenges may inform possible adaptations to BE metadata schemata.

\subsection{Automated Semantic Unit Mappings}
As we discussed in the limitations section, semantic unit mappings as implemented in this work require immense manual effort and are therefore prone to include errors.
Further research to support the creation of R2RML mappings and semantic unit mappings might therefore be of value.
\subsection{Reasoning Applications}
Unfortunately, we were not able to explore reasoning applications both on the data graph and the semantic unit graph developed in this work.
Therefore, applications of reasoning and SHACL validation could contribute to enhance the quality of the knowledge graph.
Further, reasoning applications on semantic units have not been investigated before and may contribute to arguments for including semantic units in a knowledge graph. 
\subsection{Further LLM Applications}
As discussed above, the two LLM applications we implemented in this work can be improved further.
Additionally, LLMs bear great potential for other knowledge graph applications, for example, assisting KG construction workflows, SPARQL query support, and knowledge graph embeddings. 
\chapter{Conclusion} \label{chapter11}
Knowledge graphs bear great potential as a tool to support biodiversity research, especially the creation of FAIR datasets, their integration with existing knowledge bases, and assistance in synthesis efforts and meta studies.
Their biggest disadvantage however, is how challenging their usage is. 
Querying knowledge graphs and understanding the way they represent knowledge is especially difficult for many user groups, and in most cases, also the researchers we aim to support by providing knowledge graphs for their domains.

In this thesis, we modelled and constructed a knowledge graph for publication and dataset metadata of the Biodiversity Exploratories, and contribute the first implementation of semantic units.
We evaluate how they can be used in SPARQL queries to answer sets of competency questions of different complexities, show how semantic units can be visualized to support the cognitive interoperability for users, and outline how the visualizations may be utilized for knowledge graph applications.

Additionally, we implemented two applications using large language models on the source data.
The first investigates how and which structured metadata categories can be constructed from publication and dataset titles and abstracts to alleviate the workloads of this time consuming task.
The second utilizes embeddings to find latent concepts in titles and abstracts, clusters publications and datasets around anchor concepts, and assigns entries to the research goals of the Biodiversity Exploratories to further enrich the metadata.

The results of our competency question evaluation suggest that semantic unit query complexity is comparable to traditional queries for competency questions in which information about a known entity must be retrieved.
For queries that retrieve all entities that fit a specific pattern (environment type, experiment type, etc.) however, query complexity with this semantic unit implementation seems to grow considerably, leading us to conclude that further user evaluation and implementation changes are necessary.

Results for the LLM extraction task suggest that LLMs bear potential to extract structured metadata categories from titles and abstracts when fine-tuned correctly, especially for metadata categories that only have two possible values.
Multi-label categories however, are more difficult to predict.
Further, the results for embedding tasks suggest that there is potential for enriching BE metadata using anchor clustering and research goal linking approaches, while more tuning is necessary for latent topic discovery.

In conclusion, this work advances research on semantic units and paves the way for further investigations into their potential to enhance cognitive interoperability and to reduce query difficulty for non-domain experts.
We also demonstrate the potential of LLMs for metadata curation and latent information extraction, pointing to promising directions for future work.
Overall, this thesis contributes to FAIRer research data representation and brings knowledge graphs closer to becoming tools to support biodiversity research.
\clearpage

\clearpage
\chapter*{\centering Appendix A\\ Use of Generative AI}
Generative AI (model: \textit{OpenAI GPT-5 Thinking}) was used as a programming assistant for the code and figures underlying Chapter \ref{chapter7}, \textit{\nameref{chapter7}}, the third part of the method section.
Below, we supply a list of scripts and figures that were created with the use of generative AI.
We also provide a link to the GitHub repository that contains a text file with conversation links and a mapping to which scripts were generated in which conversation\footnote{\url{https://github.com/fusion-jena/Biodiversity-Exploratories-Knowledge-Graph/tree/main/GenAI}}.
The raw scripts can also be found in the GitHub repository.

\textbf{Figures:}
\begin{itemize}
    \item Fig \ref{fig:top1vsmargin}: Plot of cosine similarity scores (x-axis) of the top candidates and the margin between top1  and top2 candidates (y-axis).
  Instead of dots, colored hexes are used to represent the number of documents per coordinate, with darker colors (purple) representing lower number of documents, and lighter colors (yellow) representing higher number of documents. The light blue dotted line on the left side of the plot represents the cosine similarity threshold set for our runs of the scripts, accepting candidates at a score of $>=$0.3.
    \item Fig \ref{fig:allgoals}: Collection of plots for the four research goals of the Biodiversity Exploratories. Plots show the distribution of cosine similarity scores (x-axis) and total number of documents per score (y-axis).
    \item Fig \ref{fig:cov}: Plot showing the tradeoff between threshold of cosine similarity scores to assign candidates to research goals (x-axis) and the overall document coverage this results in (y-axis). The blue vertical line represents the threshold of $>=$ 0.3 used in our runs.   
\end{itemize}

    





\textbf{Scripts:}
\begin{itemize}
    \item SLURM scripts 
    \item Scripts for metadata reconstruction
    \item Script to generate figures
    \item Embedding and Clustering workflow scripts
\end{itemize} 


\bibliography{reference}

\listoffigures
\lstlistoflistings              
\clearpage

\listoftables

\clearpage
\chapter*{\centering Eigenständigkeitserklärung}
\small
\begin{enumerate}[leftmargin=*, label=\arabic*., itemsep=0.25em]
    
\item Hiermit versichere ich, dass ich die vorliegende Arbeit selbstständig verfasst und keine anderen als die angegebenen Quellen und Hilfsmittel benutzt habe.
Ich trage die Verantwortung für die Qualität des Textes sowie die Auswahl aller Inhalte und habe sichergestellt, dass Informationen und Argumente mit geeigneten wissenschaftlichen Quellen belegt bzw. gestützt werden. Die aus fremden oder auch eigenen, älteren Quellen wörtlich oder sinngemäß übernommenen Textstellen, Gedankengänge, Konzepte, Grafiken etc. in meinen Ausführungen habe ich als solche eindeutig gekennzeichnet und mit vollständigen Verweisen auf die jeweilige Quelle versehen. Alle weiteren Inhalte dieser Arbeit ohne entsprechende Verweise stammen im urheberrechtlichen Sinn von mir.
\item Ich weiß, dass meine Eigenständigkeitserklärung sich auch auf nicht zitierfähige, generierende KI Anwendungen (nachfolgend „generierende KI“) bezieht.
Mir ist bewusst, dass die Verwendung von generierender KI unzulässig ist, sofern nicht deren Nutzung von der prüfenden Person ausdrücklich freigegeben wurde (Freigabeerklärung). Sofern eine Zulassung als Hilfsmittel erfolgt ist, versichere ich, dass ich mich generierender KI lediglich als Hilfsmittel bedient habe und in der vorliegenden Arbeit mein gestalterischer Einfluss deutlich überwiegt. Ich verantworte die Übernahme der von mir verwendeten maschinell generierten Passagen in meiner Arbeit vollumfänglich selbst.
Für den Fall der Freigabe der Verwendung von generierender KI für die Erstellung der vorliegenden Arbeit wird eine Verwendung in einem gesonderten Anhang meiner Arbeit kenntlich gemacht.Dieser Anhang enthält eine Angabe oder eine detaillierte Dokumentation über die Verwendung generierender KI gemäß den Vorgaben in der Freigabeerklärung der prüfenden Person.
Die Details zum Gebrauch generierender KI bei der Erstellung der vorliegenden Arbeit inklusive Art, Ziel und Umfang der Verwendung sowie die Art der Nachweispflicht habe ich der Freigabeerklärung der prüfenden Person entnommen.
\item Ich versichere des Weiteren, dass die vorliegende Arbeit bisher weder im In- noch im Ausland in gleicher oder ähnlicher Form einer anderen Prüfungsbehörde vorgelegt wurde oder in deutscher oder einer anderen Sprache als Veröffentlichung erschienen ist.
\item Mir ist bekannt, dass ein Verstoß gegen die vorbenannten Punkte prüfungsrechtliche Konsequenzen haben und insbesondere dazu führen kann, dass meine Prüfungsleistung als Täuschung und damit als mit „nicht bestanden“ bewertet werden kann. Bei mehrfachem oder schwerwiegendem Täuschungsversuch kann ich befristet oder sogar dauerhaft von der Erbringung weiterer Prüfungsleistungen in meinem Studiengang ausgeschlossen werden.
\end{enumerate}
\small

 \noindent\makebox[0.45\linewidth]{\dotfill}\\[-0.5ex]
 Ort und Datum
 \vfill
 \noindent\makebox[0.45\linewidth]{\dotfill}\\[-0.5ex]
 Unterschrift




\end{document}